\documentclass[twoside,12pt]{article}
\usepackage{times,graphics}
\newcommand{\bc}{\begin{center}}
\newcommand{\ec}{\end{center}}
\newcommand{\bq}{\begin{quotation}}
\newcommand{\eq}{\end{quotation}} 
\newcommand{\np}{\vfill\eject}
\begin{document}

\begin{center}

{\bf \huge The Most Metal-Poor Galaxies}

\bigskip
\bigskip
\bigskip

{\bf \Large Daniel Kunth\footnote{email: kunth@iap.fr} and 
G\"oran \"Ostlin\footnote{Present address: Stockholm Observatory, SE-133~36 Saltsj\"obaden, Sweden. email: ostlin@astro.su.se}}

\medskip

Institut d'Astrophysique de Paris (IAP) \\
98bis Boulevard Arago \\
75014 Paris, France \\

\bigskip

{\small Received August 19, 1999} \\

\normalfont

\bigskip

\end{center}

\begin{abstract}

Metallicity is a key parameter that controls many aspects in the formation
 and evolution of stars and galaxies. 
In this  review we focus on the  metal deficient galaxies, in particular
the most metal-poor ones, because they play a crucial r\^ole in the cosmic scenery.
We
first set the stage by discussing the difficult problem of defining a 
global 
metallicity and how this quantity can be measured for a given galaxy. The
mechanisms that control the metallicity in a galaxy are  reviewed in
 detail and
involve many aspects of modern astrophysics: galaxy formation and evolution,
massive star formation,
 stellar winds, chemical yields, outflows and inflows etc. Because metallicity
 roughly
scales as the galactic mass,  it is among the dwarfs that the most
 metal-poor
galaxies are found. The core of our paper reviews the considerable
 progress
made in our understanding of the properties and the physical processes
 that are
at work in these objects. The question on how they are related and may
 evolve
from one class of objects to another is discussed.   While discussing
 metal-poor
galaxies in general, we  present a more detailed discussion of a few
 very
metal-poor blue compact dwarf galaxies like IZw18.
Although most of what is known relates to our local universe, we show 
that it
pertains to our quest for primeval galaxies and is connected to the 
question of
the origin of structure in the universe.  We discuss what QSO absorption 
lines  and
known distant galaxies tell us already? We illustrate the importance
of
star-forming metal-poor galaxies for the determination of the primordial 
helium abundance,
their use as distance indicator and discuss the possibility to detect nearly
metal-free galaxies at high redshift from  Ly$\alpha$ emission. \\

{\bf Keywords:} galaxies: abundances -- galaxies: compact  -- galaxies: dwarf
 -- galaxies: individual: IZw18 -- galaxies: individual: SBS~0335-052
 -- galaxies: evolution  

\end{abstract}

\vspace{0.91cm}
To appear in: the Astronomy and Astrophysics Review (A\&AR)

\np 

%%%%%%%%%%%% The command \tableofcontents creates a table of contents.
%%%%%%%%%%%% NB that latex has to be run twice for the table to be updated
%%%%%%%%%%%%

\tableofcontents \np
%\listoffigures \np 
%\listoftables \np 

\normalfont

%%%%%%%%%%%%%%%%%%%%%%%%%% The text of the paper stars here %%%%%%%%%%%%%%%%%%%%%%%%%%%%%

\section{Introduction ~ -- What is a metal-poor galaxy?}

The  discovery of  extragalactic objects with very low heavy element abundance was made by
Searle and Sargent (1972) who reported on the 
properties of two intriguing  galaxies, IZw18 and IIZw40.
They emphasised that they could be genuinely young galaxies in the
process of formation, because of their extreme metal under-abundance, more than
10 times less than solar, and even more extreme than that of H{\sc ii} regions found in the
outskirts of spiral galaxies. At the time of this discovery the general wisdom 
that most galaxies (in particular the ellipticals) had been 
formed over a short period during a dynamical 
free fall time of few $10^{7}$
years (Eggen et al. 1962) started to be challenged (e.g. Searle and Zinn 1978).
It is also during the 70s that the first hierarchical models of galaxy formation were
constructed (Press and Schechter 1974). Because dwarf galaxies condense from smaller 
perturbations than giants, the Cold Dark Matter models (CDM) predict that low mass 
galaxies could still be forming at the present epoch. The discovery by Searle and 
Sargent (1972) has been an impressive stroke, since one of these two galaxies (IZw18) 
is still in the book of records, as we shall later elaborate more on. These two objects gave 
rise to many  
systematic searches for more objects in a quest for local genuine  young galaxies or 
``unevolved galaxies", depending upon the alternative viewpoints that some galaxies could be caught 
in the process of formation or that they  simply were the result of a very mild evolution over the 
Hubble time. These galaxies had the advantage of being gas-rich, with spectra 
dominated by strong emission lines (see Fig. 6) favouring their detection. Many 
techniques have been 
employed to find them, sometimes at large distances despite their intrinsic low luminosity. 

The last nearly three decades have brought a wealth of data from numerous studies 
on dwarf galaxies, including information on their chemical composition. It 
became clear that ``metal-poor'' would be analogous of ``low mass'' galaxies 
(Lequeux et al. 1979). 
For this reason, our review will largely focus on dwarf galaxies, but we shall address the
question of  the  existence of large and massive proto-galaxies 
- essentially devoid of metals - at large redshift in the last section. The dwarf galaxies are 
not only interesting for understanding the process of galaxy formation. For the gas-rich ones with 
active  star formation, one motivation to study them has been the hope to better understand 
the processes of massive star formation in low metallicity gas. They turn out to be test cases for chemical 
evolutionary models and offer the possibility to approach the primordial 
helium abundance with a minimum of extrapolation to early conditions.
Many galaxies of different kinds can be identified as metal-poor and it is an interesting 
question 
to find out about the connections that bridge them together. Finally, a lot of new studies 
concentrate on the impact of massive star formation onto the interstellar medium (ISM) in 
star-bursting dwarf objects, 
that in turn can lead to constrain the supernovae rate, the stellar initial mass function (IMF), 
the metal dependence of the winds 
in Wolf-Rayet (WR) stars etc. In fact there are many issues in astronomy  where it is
essential  to  understand dwarf galaxies (actively 
forming stars or not).

A definition of what is a metal-poor galaxy is indeed necessary at this
point. To define the  metallicity ($Z$, i.e. the mass fraction
elements other than hydrogen and helium) of a galaxy requires some words of caution
because in a given galaxy, depending on where one looks, this quantity may 
vary substantially. For example in our Galaxy, the  bulge, the solar 
neighbourhood and the halo differ in metallicity.
The most metal-poor  halo stars have heavy element abundances 10$^{-4}$
times that of the Sun (Cayrel 1996) while stars in the Galactic centre may be three
times more metal rich than the Sun. On the other hand, the large 
ionised complexes in the ISM show a narrower range down to only 1/10 the 
solar value. Thus metallicity depends on what one looks at: stars or gas, 
and if one considers gas -- what phase: neutral atomic, molecular or ionised? 
Moreover the metallicity of stars is found to depend on their age, and
 depends on which elements are investigated. That a star or nebula is deficient of a 
certain element does not automatically mean that the overall 
chemical abundances are low.

Thus, one must be careful in defining what  metallicity means for a
given galaxy before comparing observations and looking for trends. Another
complication stems from the different techniques in use for determining chemical abundances. 
Metallicities of Local Group dwarf spheroidals (dSph) have largely been investigated 
through photometry of their resolved stellar populations, which are dominated by old stars, 
and they are found to be metal-poor, as measured by [Fe/H]. In general dSph 
galaxies contain little or no gas, and no H{\sc ii} regions. 
Dwarf irregular (dI) galaxies on the other hand usually have plenty of gas and 
ongoing star formation  as witnessed by the presence of H{\sc ii} regions. 
It is relatively easy to derive metallicities, especially oxygen (O) of the ionised gas 
in H{\sc ii} regions explaining why most investigations are based on such. Hence the 
``metallicity'' of dSphs and dIs measures quantities
that are not directly comparable. For galaxies which have no 
H{\sc ii} regions and very low surface brightness,  the only available spectroscopic method 
hitherto, has been to study individual bright  stars which are within reach in nearby galaxies
only. In more distant galaxies, individual stars are not observable and the 
metallicity has to be inferred from either absorption line spectra of an 
integrated stellar population (not necessarily homogeneous in age or 
composition) or from the nebular emission lines, if any. Again, the different 
methods in general measure different elements.

So what useful definitions of ``metallicity'' do we have at hand? 
The  possibility of using the metallicity of the H{\sc ii} regions  has the advantage of
providing an ``up to date'' metallicity, while stars provide the
metallicity of the cloud from where they were born, perhaps a Hubble time ago. 
Nebular abundances show large spatial abundance variations and gradients in 
some  galaxies, e.g. giant spiral galaxies, like the Milky Way. 
Although  measurable abundance inhomogeneities could be 
expected, most dwarf galaxies  seem rather 
well mixed. A possible problem with metal-poor H{\sc ii} regions is self-pollution 
of fresh metals by winds from young massive stars, so that the abundance 
inferred from the nebular emission lines may not be representative of that 
in the local ISM (Kunth \& Sargent 1986).
On the other hand, the metallicity of the stars in a galaxy depends on which 
stellar population is studied.  Thus it is not surprising, in particular 
for galaxies which have experienced continuous star formation, that stars have 
different abundances. Integrated spectra of galaxies provide a luminosity-weighted 
average of the metallicity. This average  
metallicity  will change with time due to the 
photometric evolution of the stellar population, even if no new stars are formed.

A good, well defined, metallicity indicator would be the fraction of baryonic 
matter that 
has been converted into heavier elements by means of stellar nucleosynthesis. 
This material may have been returned to the ISM or may still be
 locked up in stars. 
Such a definition would indicate that our main interest goes beyond
 the element abundances themselves from the fact that they provide information about the 
history of a galaxy. The relative abundances and gas mass fractions  
might  unveil different histories among  galaxies with 
the same metallicities. However, it is clear from he above discussion that such
 a definition of metallicity remains impractical, since  not directly measurable. 
However, we would like to keep this ideal definition in mind for the rest of the 
discussion.

Now we wish to go back to the question -- What is a metal-poor galaxy?
Under the assumption that various tracers 
give a plausible picture about the ``ideal metallicity'' we can now 
try to compare different kinds of galaxies.
One of the most fundamental parameters for a galaxy is of course its mass. 
This mass, which may consist of stars, gas, dust, baryonic- and non-baryonic 
dark matter, is more difficult to measure, than e.g. the  luminosity, but to the 
first approximation, mass and luminosity correlate. Based on their luminosity 
and size, galaxies can be divided into dwarfs and giants. It has been found 
that the metallicity of a galaxy in the local Universe
correlates positively with its luminosity (although with a large  scatter), thus also reflecting a positive 
correlation with mass. The reason for this behaviour is a fundamental issue to 
understand. 
One explanation could be that dwarfs evolve more slowly because of smaller 
mass densities, which to the first order fits with the
observation that dwarfs, except dwarf elliptical/spheroidals, are more gas-rich
 than giants.  Another possible explanation is that dwarfs have weaker
gravitational potentials hence are more susceptible to loose metal enriched 
material from supernova driven winds.
 
A natural reference 
for element abundances and the ratio between them, could be the Sun. 
Thus a starting point could be that  ``metal-poor'' means anything which
has sub-solar abundances and vice versa for   ``metal rich'', which implies 
 that basically all local galaxies fainter than our Galaxy are metal-poor.
High redshift neutral gas clouds, which may be the building blocks of today's 
galaxies, are observed to have  metallicities down to
$0.001 Z_{\odot}$. Thus, there is a large range of metallicity 
to explore, and it is meaningful to distinguish between metal-poor, very metal-poor
 and extremely metal-poor. Since this review is called  ``the most metal-poor galaxies'' 
it is natural that we focus on the latter two subclasses. What do we see locally? 
Among dSph we find metallicities extending down to 1/100 $Z_{\odot}$, while the LMC 
and SMC are at roughly 1/3 $Z_{\odot}$ and 1/8 $Z_{\odot}$ respectively. 
Dwarf irregulars have sub-solar abundances, ranging down to 1/40 $Z_{\odot}$. In addition 
there are many blue compact galaxies (BCGs)  in the 
range 1/10 to 1/50 $Z_{\odot}$,
 with, as we shall show later, IZw18 at the extreme.

A more workable definition could use the minimum enrichment  one 
predicts for a single burst of star formation, using the instantaneous recycling  closed box 
model. Kunth and Sargent (1986) found that such minimal expected metallicity increment
in a pristine H{\sc ii} region would be higher than or equal to the metallicity of IZw18.
Similarly one finds that even  converting only on the order of two 
percent of  pristine gas in a galaxy to stars, results in a metallicity 
 of 1/50 $Z_{\odot}$, i.e. the metallicity of IZw18. There are of course a lot 
of uncertainties that go into these calculations,  but they can be a useful guide. 
Another, more practical guide, is to consider the
O/H distribution of star forming dwarf galaxies studied over the last 30 years. This shows 
a peak  around 1/10 $Z_{\odot}$ and  drops sharply below that value. 
Moreover, for most models: of  stellar winds, evolutionary tracks, WR-stars,
star formation etc., a critical dependence on metallicity is seen  around 1/10 $Z_{\odot}$. 
This is why we have adopted throughout this review the working hypothesis that galaxies with 
metallicity below 1/10 $Z_{\odot}$ will be considered as very metal deficient.
Therefore galaxies like the Magellanic clouds will not be our main interest in this paper.
Moreover, such a limit means that this review will be biased towards dwarf galaxies.
In particular we will focus on blue compact  galaxies (BCGs). The reason is partly
due to selection effects: since blue compact galaxies have bright emission lines and 
high surface brightness, it is fairly easy to discover them and derive their metal content.
Thus, there exists a lot of high quality data on BCGs, but we should  keep in mind that
 very metal-poor galaxies may
be as common among other types of dwarf galaxies.

Metallicities can be studied at great distances under special conditions. 
Observations of high redshift  QSOs and radio galaxies (e.g. Dunlop et al. 1994), reveal
the presence of dust and  metal rich gas, suggesting that prior stellar nucleosynthesis 
has already taken place.  High redshift QSO absorption line systems show 
a wide range of metallicities, from one thousandth solar up to 1/3 solar. Thus, while the 
average metallicity of the Universe certainly must have increased since the early epochs, 
the situation is  more complex than a simple picture where high redshift means metal-poor, 
and low redshift metal-rich. Objects with high and low metallicities are found at all 
redshifts. Surely we expect objects that in the local Universe appear as metal 
deficient to be even more deficient at high redshift, if we could observe their
precursors. Also the ancestors of the local metal rich galaxy population, i.e. the giant 
spirals and ellipticals,  should have started out with very low abundances  unless they 
were gradually built up by merging smaller galaxies. Currently, 
both the theoretical and observational pictures suggest  that the latter is an important
mechanism. Dwarf galaxies, the survivors who form the local metal-poor galaxy
population, may thus be the principal building blocks of the Universe on large scales.

The structure of the rest of paper is as follows:
In Section 2 we discuss how metallicities are measured and in Sect. 3  the 
physical mechanisms that control the metallicity of a galaxy. In Section 4 we review
the physics of metal-poor galaxies in the local Universe, while in Sect. 5 
turning to some key objects like IZw18. In Sect. 6 we discuss survey techniques, 
and the distribution in space and luminosity of metal-poor galaxies.  In Sect. 7
we examine observed trends in the metal-poor galaxy population and various possible
evolutionary  links. In Sect. 8 we  focus on cosmology and  the high redshift Universe, 
and in Sect. 9 we conclude.

Throughout this paper we adopt 12+log(O/H)=8.91 as the solar oxygen abundance.
As customary, element ratios given in square brackets represent logarithmic 
values with respect to solar values, 
e.g. [Fe/H] $=$ log(Fe/H) $-$ log(Fe/H)$_{\odot}$.
We use H$_0 = 75$km/s/Mpc, and rescale results from the litterature based
on other values of H$_0$ when necessary. A list of commonly used abbreviations
and acronyms are given at the end of the paper.

%%%%%%%%%%%%%%%%%%%%%%%%%%%%%%%%%%%%%%%%%%%%%%%%%%%%%%%%%%%%%%%%%%%%%%%%%%%%%%%%%%%%%%%%%%%%%%
%%%%%%%%%%%%%%%%%%%%%%%%%%%%%%%%%%%%%%%%%%%%%%%%%%%%%%%%%%%%%%%%%%%%%%%%%%%%%%%%%%%%%%%%%%%%%%
%%%%%%%%%%%%%%%%%%%%%%%%%%%%%%%%%%%%%%%%%%%%%%%%%%%%%%%%%%%%%%%%%%%%%%%%%%%%%%%%%%%%%%%%%%%%%%

\np

\section{How are metallicities measured?}

Below we  list the main ways that have been used to derive metallicities for 
metal-poor galaxies. One has to be aware that these different abundance estimators
sample not only different elements, but also look-back times and population mixes.

\subsection{H{\sc ii} regions}

Abundances are relatively easy to measure in star-forming dwarf galaxies
because they contain gas clouds in which large numbers of hot stars are
embedded.  Their spectra are dominated by nebular emission lines similar to
those of high-excitation giant H{\sc ii} regions in late type spiral galaxies.
What is observed in the optical are narrow emission lines superimposed on a
blue stellar continuum (see Fig 6.). They are identified as helium and hydrogen
recombination lines and several forbidden lines. Methods used
in determining abundances are well understood and generally more reliable
than those based on stellar absorption line data because transfer problems
become of minor importance. 
From the optical, O, N, S, Ne, Ar and He lines are currently measured. With modern
detectors fainter lines such as lines of Fe have been studied  (Izotov and Thuan, 1999).
 The ultraviolet (UV) region is dominated by the hot stellar continuum and shows relatively 
weak emission lines except for those that originate in stellar winds. However 
there are a few notable exceptions and
owing to the International Ultraviolet Explorer (IUE) and more recently the Hubble 
Space Telescope (HST) nebular carbon
and
silicon abundances have been determined.

 Oxygen is the most reliably determined element, since the most important
 ionisation stages
 can all be  observed. Moreover the [O{\sc iii}]${ \lambda  4363}$ line allows an accurate 
determination of the electron temperature. The intrinsic uncertainty 
in this method (reflecting a simplified conception of the H{\sc ii} region physics, possible
problems with temperature fluctuations etc.)
is of the order of $\sim$0.1 dex (Pagel 1997). Furthermore, when the electron temperature 
cannot be determined, empirical relations (cf. Pagel 1997) between the 
oxygen abundance and the [O{\sc ii}]${\lambda 3727}$ and [O{\sc iii}]${\lambda\lambda 4959,5007}$ strength relative
to H$\beta$ \ are used, though with lower accuracy (0.2 dex or worse). 
For other species, in general,
one does not observe all the ionisation stages expected to be present in
the photoionisation region and an ionisation correction factor  
must be applied to derive the total abundance of the element in question.

One important aspect of H{\sc ii} region abundances is that they can be obtained 
also at great distances. This makes them powerful tools also for studying 
high redshift galaxies, with the price that our view will be biased towards
actively star-forming systems. For a discussion on possible problems associated
with deriving abundances in very distant galaxies, see Kobulnicky et al. (1999).

\subsection{Planetary nebulae}

Planetary nebulae (PNe) enable a spectroscopic abundance determination,
using the same technique as for H{\sc ii} regions. 
The intrinsic uncertainty due to temperature fluctuations (Peimbert 1967) 
and gradients is
probably at least  0.1 dex (Pagel 1997).
The derived oxygen abundances
reflect the initial composition of the star since the ejected gas that forms
the PN has not been enriched in oxygen by the central nucleosynthesis
(however see remark below) and that
moreover, the hot ejecta has not mixed with the surrounding colder ISM.
The progenitor stars of PNe have masses up to several times that of the Sun, meaning 
that the derived O abundance traces the ISM abundance 0.1 to several Gyr ago,
depending on the mass of the progenitor. This is a large range in lookback
time over which significant chemical evolution may have occurred, and thus
it is important to assess the mass of the progenitor star. A model that 
relates the oxygen abundance  as derived from PNe and H{\sc ii} regions (i.e. 
the present ISM abundance)  has been presented by Richer et al. (1997). They find that
the abundance gap, the difference between these two abundance estimators,
is a function of the star formation history and metallicity. The more extended
the epoch of star formation has been and the higher the ISM metallicity,
the larger the abundance gap.

While the PNe abundances of oxygen traces the ISM abundance at the birth
of the progenitor star, this is not true for carbon and nitrogen which have
been enriched during the stellar evolution (Pagel  
1997, p. 199). Moreover for some types (PNe type I, cf. 
Peimbert and Serrano 1980)  there might have been also enrichment of oxygen.

\subsection{Photometry of resolved stellar populations}

Colour-magnitude diagrams (CMDs) provide a photometric estimate of
the stellar abundances, usually  assumed to be represented by iron, from the 
colour of the red giant branch (RGB). Stars more massive than 
approximately 2 solar masses never ascend it and  thus stars on the RGB have 
ages from approximately
1 to 15 Gyr, and their abundance reflects the ISM abundance of Fe when
the stars were born. The width of the RGB provides a measure of the 
metallicity spread, of course convolved with the broadening from age dispersion 
and the photometric error  function (see e.g. Lee et al. 1993).

The calibration of this method rests upon the comparison of the RGBs 
with those of old  Galactic globular clusters with metallicity
estimated from integrated spectroscopy or spectroscopy of individual 
giant stars (see Da Costa and Armandroff 1990). Therefore the  RGB colour 
of an old population measures an  average metallicity and is 
unable to reveal abundance ratios. 
Moreover, comparison with old Galactic globular clusters rests upon
some degree of similarity between the two kinds of objects. However since
it is now evident that the bulk of the stellar population in many dEs is  younger
than that of Galactic globular clusters, and  has a considerable
abundance spread, one must  be aware that physically different properties
of globular clusters and galaxies may possibly cause systematic errors.

Other  metallicity indicators include the colour and morphology of the 
horizontal branch (HB) in the CMD (Grebel 1998), although this cannot be uniquely
translated into metallicities due to poorly understood ``second parameter
effects'', also known from the study of HBs in globular clusters.
Intermediate in character between photometric and spectroscopic abundance 
determinations is the fraction of carbon (C) stars among the late type 
giants, and the number of carbon rich (WC) to nitrogen rich (WN) Wolf-Rayet (WR)
stars. Such indicators give only very crude guesses of the 
the metallicity. Moreover, each population  sample stellar populations of 
different age and are not directly comparable.

A major drawback in using CMDs for abundance determinations is that it is
limited to  nearby galaxies, basically the Local Group and its immediate
surroundings. Photometric abundances can also be derived using  
Str\"omgren photometry. However the relatively narrow filters employed, limit the
usefulness of Str\"omgren photometry for extragalactic objects, a situation that could change with the 
new generation of large telescopes.
There are also other photometric systems designed for metallicity sensitivity,
e.g. the intermediate wide Washington system.

\subsection{Stellar spectroscopy}

 In the nearest galaxies it is possible to 
spectroscopically determine the abundances of individual stars. 
While this method should yield  accurate abundance determinations,
it is not directly translatable into global values, since there are
star to star variations and poor statistics; in practice,
only few stars, and in general the most luminous, can be observed for each galaxy.
The Magellanic Clouds have been  accessible since more than a decade (Spite et al. 1986; 
Russell and Bessell 1989; see also Haser et al. 1998) as well as the 
most nearby dwarf satellites of the Milky Way  with present instrumentation 
(e.g. Shetrone et al. 1998).  Spectroscopic observations of
young metal-poor stars would also be of great importance for improving models 
of stellar evolution and population synthesis.

With the new generation of 8m class telescopes, the distance to which individual
luminous stars can be studied has been increased, and it is now possible to reach 
outside the Local Group. An outline of the merits of spectroscopy of luminous 
early type stars and on some work in progress can be found in Kudritzki (1998).
In the near future the situation will change when 8m class telescopes are
equipped with spectrographs that allow  observations of many stars 
simultaneously with sufficient spectral resolution.

\subsection{Estimates from spectroscopy and photometry of  integrated light}

In remote galaxies without H{\sc ii} regions, integrated spectroscopy
may be used to derive metal abundances. This method 
have been widely employed e.g. for giant ellipticals. Absorption line features, such as 
the Mg$_2$ band, are compared to  populations of  observed or
model stars to estimate metallicities (cf. Mould 1978; Worthey 1994).
This method is difficult to use since the interplay between stellar population mixes, 
star formation history and stellar initial mass function (IMF),  which influences
the spectral shape and the absorption line strengths, is not known a priori;
and it has turned out to be non trivial to transform the observed Mg$_2$
strengths into metallicity. In addition, there are indications of 
nonsolar [Mg/Fe] values (e.g. Worthey et al. 1992).
Since, in general,  the surface brightnesses of  dwarf  galaxies  decrease with 
decreasing luminosity, even integrated spectroscopy is difficult and time 
consuming, and practically impossible for the really faint dwarfs.

Integrated photometry is a poor-man's tool.
 When no spectroscopic 
information may be obtained and the galaxy is too distant to resolve the stellar
 population, 
some constraints may be put on the metallicity (cf. Sect. 4.2.2) from 
integrated 
photometry under certain assumptions. However, as for the case of integrated 
spectroscopy  
(but to a much greater extent), the assumed population mix, star formation 
history, IMF, 
internal extinction etc., influence the broad-band colours and make the derived metallicity 
very uncertain. The infamous age--metallicity degeneracy is here acting in its full power. 
If some of these parameters  can be constrained, this method may be used in a 
statistical
sense for large samples.

\subsection{Other  methods}

Another possible metallicity indicator is that of the cold (neutral) ISM in a 
galaxy, where problems with self pollution
 should be small or absent. By using background continuum sources, absorption lines 
from the neutral ISM may be used to estimate chemical abundances.   
 Kunth et al. (1994) used UV-absorption lines arising in the neutral 
gas in the line of sight towards the young stellar 
association in IZw18 to derive oxygen abundances of  the gaseous halo 
of IZw18.
The method is similar to the one used for deriving abundances in QSO absorption 
line system. This method is rather unexplored for galaxies, but may prove to be of 
future use if the  difficulties due to line saturation can be tackled or circumvented 
(Pettini and Lipman 1995)

X-ray observations have been used to derive metallicity of the hot intra cluster 
medium in rich galaxy clusters, and has recently been applied also to starburst
galaxies (Persic et al. 1998). X-ray observations are interesting because they
 may provide a means of observing ``hot'' metals produced in the current star 
formation event that  have not yet mixed with the photoionised gas, and to study 
abundances in gas  expelled by ``superwinds''.

In principle, abundances may be derived also from radio observations. This has been 
done in the local ISM in our galaxy. For external galaxies they have not been 
applied much due to sensitivity problems. CO observations of dwarf galaxies are
complicated by the fact that for low metallicities the CO flux depends both on the
mass and metallicity of the molecular gas complex. These questions may be further
addressed by space based infrared spectroscopy.

The chemical composition of the ISM  has been studied in the ultraviolet (UV) with e.g. 
the IUE and HST satellite observatories. A relatively unexplored 
wavelength region is however the far UV. New space borne instruments 
like FUSE (far ultraviolet spectroscopic explorer) may be of importance here.
Other methods include observations of supernova remnants (see Pagel 1997) and 
the concept of dynamical metallicity, cf.  Haser et al. (1998).

%%%%%%%%%%%%%%%%%%%%%%%%%%%%%%%%%%%%%%%%%%%%%%%%%%%%%%%%%%%%%%%%%%%%%%%%%%%%%%%%%%%%%%%%
%%%%%%%%%%%%%%%%%%%%%%%%%%%%%%%%%%%%%%%%%%%%%%%%%%%%%%%%%%%%%%%%%%%%%%%%%%%%%%%%%%%%%%%%
%%%%%%%%%%%%%%%%%%%%%%%%%%%%%%%%%%%%%%%%%%%%%%%%%%%%%%%%%%%%%%%%%%%%%%%%%%%%%%%%%%%%%%%%

\np

\section{What controls the metallicity of a galaxy?}

Here we will discuss some mechanisms of importance for understanding
why some galaxies are more metal-poor than others. It is unavoidable that this 
will include a discussion on
nucleosynthesis and chemical evolution models, however this is not the
main scope of this paper and excellent reviews can be found elsewhere, e.g. 
Pagel (1997), Matteucci (1996), Maeder (1992), Prantzos (1998).

\subsection{Stellar evolution and nucleosynthesis}

As the IGM condenses into galaxies it contains the  primordial
abundances of H, D, $^3$He, $^4$He and $^7$Li. One of the beauties of some 
star-forming  dwarf galaxies is that they can be so  metal-poor that their
abundance analysis bear strong cosmological implications.  
The current wisdom is that most of the element production in the Universe,
apart from the early ``Big Bang'' nucleosynthesis, occurs in stellar interiors.
Part of the products of stellar nucleosynthesis are released when stars die or
in stellar winds from evolved stars, while another part remains locked up in
stellar remnants: white dwarfs, neutron stars and black holes. These remnants
may not be completely sterile since in a binary system, accretion onto a white 
dwarf may lead to a type Ia supernova. 

The end products that enrich the ISM can be predicted from models of stellar evolution
and nucleosynthesis. The returned mass in metals as a function of the initial mass
is referred to as the stellar yield (Maeder 1992). The absolute and relative yield for 
different elements depend on stellar mass in a non-linear fashion. Moreover, the yields 
for the so called ``secondary'' elements depend themselves on the initial composition of the 
star. An example is  $^{14}$N, although observations strongly indicate the need also for  
primary production. Furthermore the mass-loss 
history of stars needs to be accounted for, since  elements that would have been
subject to further nucleosynthesis might have been ejected, and also the effects of 
stellar binarity.
Unfortunately, the yields for many elements are still very uncertain, also for important
elements like C and N (Prantzos, 1998), and this has to be kept in mind when 
interpreting abundances and abundance ratios. The oxygen yield is among the best
determined ones, still the uncertainty is around  a factor of two.

Low mass stars ($M < 1 M_{\odot}$) are long lived and simply lock up part of the gas, 
since their lifetimes are longer than or comparable to the age of the Universe. On the 
other hand they contribute to light and have imprinted in their chemical composition the 
conditions of the ISM at the time and place they were formed. 
Intermediate mass stars (between 1 and 8 $M_{\odot}$) undergo dredge-up processes 
that  significantly affect the C and N abundances and even the $^4$He abundance, and 
such stars are important contributors to these elements.
Massive stars ($ M  > 8 M_{\odot}$) 
are short-lived and complete their evolution in less than $5\times 10^7$ years. As such a star
collapses and becomes a neutron star its envelope is ejected in a supernova explosion, 
carrying away the earlier nucleosynthesis products and the ones resulting 
from explosive nucleosynthesis in the inner layers. This picture is still uncertain since 
the initial masses for which this is valid are not well known and it has been argued
that some of the most massive stars may collapse into black holes, without an associated
SN explosion. Before their
dramatic end, massive stars undergo mass-loss processes via stellar winds as exemplified 
by Wolf-Rayet stars and other variable stars (e.g. luminous blue variables) that
can carry away some CNO processed material, reducing the yield of O and increasing that 
of C, N and He. The efficiency of stellar winds depend
strongly on their chemical composition (Maeder 1992) since metals increase the opacity.
At low metallicity the mass loss is small and the resulting He, C and O production is 
insensitive to the initial stellar composition, although it can be affected by the tendency 
of mass loss to increase with metallicity. 
Hence one expects the yields of these elements to be metallicity dependent.
In addition to stars losing part of their mass, 
a galaxy as
a whole may be subject to mass loss, which will influence the ISM abundances, (see Sect. 3.3).
 Since different elements are produced in stars
of different mass, they enrich the ISM on different timescales. Massive stars constitute the 
main source of oxygen and other $\alpha$-elements, thus these elements are ejected on short 
timescales. Also significant amounts of carbon and nitrogen are produced in massive stars. For iron, 
massive stars  dominate, but on long timescales the contribution from SN-Ia produced in 
binary systems  may be important.

The initial mass function (IMF, the distribution of stellar masses in a population of newly borne stars) 
is a critical issue. To predict the element production of a
population of stars, the stellar yields have to be convolved with the IMF to form the 
{\it net} yield, defined as the mass of newly synthesised elements per unit mass 
locked up in remnants and long lived stars.  Unfortunately, the IMF is not yet 
well determined even locally, and it is uncertain whether it is universal or 
depends on environment and metallicity. In most extragalactic studies the IMF is assumed 
constant in time. This 
assumption needs to be examined with care although no observational evidence
has convincingly contradicted it by now. The strongest claim by Terlevich
and Melnick (1981) for IMF variations with metal content of gas has never been 
compelling. There have been some suggestions that starburst galaxies should have an 
IMF biased towards massive stars or deficient in low mass stars (cf. Scalo 1990). 
However there is no direct evidence for this or for a low mass cut-off in giant H{\sc ii} 
regions like 30 Doradus or in globular clusters.
Marconi et al. (1994) argue that the chemical evolution of  starburst galaxies
is well understood  with a normal Salpeter-like IMF. Currently it seems that, for massive 
and intermediate massive stars, the IMF is reasonably well described with a power-law 
and a slope close to Salpeter's (1955) original value while it flattens, but does not cut off, 
at masses below 1 $M_{\odot}$. For a discussion on the IMF, its derivation, and possible
variations,  see  Scalo (1998). If the total mass involved in star formation at each instant
is modest, the high mass part of the IMF will be badly populated and consequently  
predictions for nucleosynthesis and spectral evolution will be sensitive to statistical 
fluctuations.

\subsection{Star formation history} 

The most important parameter of the chemical enrichment is expected to be the 
star formation history (SFH). Since massive stars dominate  the production
of most elements,  the metal
production rate will to a first approximation be directly proportional to the
star formation rate. Likewise, in a simple chemical evolution picture, the
metallicity is  a function of the fraction of gas turned into stars.

Star formation (SF) seems to occur in different modes: one being relatively 
undramatic with continuous star formation at a regular pace while 
in ``bursting'' galaxies, star formation may be dominated by a small number of  
short intense bursts of star formation separated by extremely long intervals 
of time. While the latter scenario is mainly observed  in blue compact galaxies it 
is believed to occur also in dwarf and giant elliptical galaxies. An intermediate 
picture is that of ``gasping'' SF, characterised by extended star formation 
episodes separated by moderately long periods of less active SF, which
probably is the most realistic picture for many dwarf irregular 
galaxies. While a starburst produces a lot of metals, the  
supernovae and stellar winds may eject  gas into the intergalactic 
medium on short time scales. If, on the other hand,
the SF is continuous, the energetic feedback from dying stars will have less influence on the ISM.
Different scenarii for SF regulation have been implemented with stochastic 
self-propagating star formation, self regulated star formation and also a gas density 
threshold. Moreover  interactions, mergers and stripping may play an important r\^ole
in regulating the SF. 
 
Discrete star-bursting behaviour in dwarf galaxies may strongly affect the abundance
ratio of elements such as N, Fe and C which partly come  from longer lived
intermediate mass stars as compared to O (Garnett 1990,  Gilmore and Wyse 1991,
Richer and McCall 1995). 
In particular, this could explain the tendency of C/O and Fe/O to increase with O/H 
especially at low O/H, but this awaits more accurate measurements of 
these ratios in extremely metal-poor galaxies. Moreover, since the net yield for many 
elements appears to be metallicity dependent (Maeder 1992), element production 
by subsequent generations of stars will depend on how fast the ISM is enriched
and hence the star formation history. In addition, variations in element ratios
could affect nucleosynthesis and thus the net yields for a given stellar generation.

A last note concerns ``starbursts'':  This notion is frequently used in the literature 
to describe regions/galaxies with varying degree of active star formation. A proper 
definition of a starburst  is that it involves 
an unsustainably high star 
formation rate (SFR) in terms of the gas consumption timescale or the timescale to 
build up the observed stellar mass (i.e.  the time averaged SFR is much lower than 
the present). Many galaxies have SFRs fluctuating with 
time, but this does not necessarily imply that the SFR is unsustainable over a 
Hubble time.

\subsection{Outflows and Inflows}
When  massive stars are about to end their lives  they explode as a 
supernovae (SNe). The energy output from a SN is over a short period,
comparable to that of a whole galaxy. 
In a galaxy with a high local star formation rate, the collective action of 
supernovae may lead to a galactic superwind, which may cause loss of gas. 
Stellar winds can also contribute to the energetics of the ISM
at the very early stage of a starburst (Leitherer et al. 1992). The relative
importance of winds  compared to SNe increases with metallicity.

A continuous wind  proportional to the star formation rate  has
been applied in models predicting the evolution of starburst galaxies. But since 
different elements are produced on different timescales, it has
been  proposed that only certain elements are lost (or in different
proportions) hence reducing the effective net yield of those metals as 
compared to a
simple chemical evolution model (Matteucci and Chiosi 1983, Edmunds 1990). 
The SNe involved in such a wind are likely to be of type II because type Ia SNe 
explode in isolation and will less likely trigger chimneys from which
metals can be ejected out of the plane of a galaxy. In this framework O and
 part of Fe  are lost while He and N 
(largely produced by intermediate stars) are not. This would result in a cosmic 
dispersion in element ratios such as N/O between galaxies that have experienced mass 
loss and those that have not.

In a dwarf galaxy which has a weaker gravitational potential, these effects 
may result in gas loss from the galaxy. 
Recently galactic winds have been observationally investigated in dwarf galaxies
(e.g. Israel and van Driel 1990, 
Meurer et al. 1992; Marlowe et al. 1995; Martin 1996,1998).
In VII~Zw403 Papaderos et al. (1994) detected extended X-ray plumes which they 
interpreted as the result of outflows of hot gas. Lequeux et al. (1995) and Kunth et al.
(1998) have shown that the escape of the Ly$\alpha$ photons  in star-forming
galaxies strongly depends on the dynamical properties of  their interstellar medium.
The Lyman alpha profile in the BCG Haro2 indicates a superwind of at least 200 km/s, 
carrying a mass of $\sim 10^7 M_{\odot}$, which can
be independently traced from the H$\alpha$ component (Legrand et al. 1997a).
However,  high speed winds do not necessary carry a lot of mass.
Martin (1996)  argues that a bubble seen in IZw18 (see also Petrosian et al.
1997) will ultimately blow-out together with its hot gas component. 
Although little is known  about the interactions between the evolving supernova
remnants, massive stellar bubbles and the ISM it is possible that an outflow takes 
the fresh metals with it and in some cases leaves a galaxy totally cleaned of gas.

Will  the gas leave a galaxy or simply
stay around in the halo? Tenorio-Tagle et al. (1999) point out
that superbubbles may initially expand with speeds that well exceed the local
escape velocity of the galaxy but their motion into the gaseous halo causes
a continuous deceleration lowering the velocity to values well below the escape
speed. In such a case, ejecta condense into a cold phase, forming droplets that
fall back and settle down onto the disc of the galaxy hence
changing the composition of the ISM (the ``Galactic fountain'' model). Similarly in chemodynamical models 
(Hensler and Rieschick 1998, and references therein), the gas cools and falls back.
Modelling the effect of SNe feedback on the ISM, De Young and Heckman (1994) 
suggested  that the smallest dwarfs could have their entire ISM removed by a superwind.
However, using models including dark matter, MacLow and Ferrara (1999) and Ferrara and Tolstoy (1999)
conclude that winds are not very efficient in ejecting the ISM. Outflows are in most cases 
confined to the galaxy
and ``blow-away'' occurs only for the smallest (luminous mass $M_{lum} < 10^7 
M_{\odot}$) galaxies considered, while in other cases the mass loss is very modest. 
Winds may however still be  efficient in ejecting fresh metals. Ferrara and Tolstoy (1999) 
nevertheless argue that outflows are not likely to be more metal rich than the average ISM value.
Moreover, in their model, the SFR is a function of mass density which results in a 
mass--metallicity relation. Since  the least massive dwarfs  loose their 
entire ISM after the first star formation event, this results in a minimum  expected 
ISM metallicity for a gas rich dwarf of 12+log(O/H)=7.2, i.e. the abundance of IZw18.
Many assumptions go into these calculations, which must be further examined. Murakami 
and Babul (1999) showed that in high density environments the IGM pressure could confine 
outflows to the parent galaxies, inhibiting  mass loss (cf. Babul and Rees 1992) .

It is clear that the r\^ole of galactic winds 
in regulating the chemical evolution is not a settled
issue yet. If the metallicity--luminosity relation (cf. Sect. 4.1, 4.2 and 7.1) holds 
from gas-rich to
gas-poor systems then the loss of metals due to galactic winds should be a
second order effect (Skillman 1997).

It is also possible that a galaxy is subject to infall of gas, although evidence 
for this is scarce.  Infalling gas may come from external galaxies, stolen in the 
process of interaction or from an  external origin, perhaps via isolated pristine 
 H{\sc i} clouds (if such clouds exist). 
There is  evidence that blue compact galaxies and low surface brightness galaxies (LSBGs) 
are sometimes associated 
with  H{\sc i} clouds (Taylor 1997), which in general have optical counterparts (Taylor, 
private communication). A third possibility is that gas expelled 
in a previous superwind falls back on its host galaxy. There are indications
that infall of metal-poor gas, perhaps in the form of gassy dwarf galaxies,
may have had major impact on the chemical evolution of the disc of our Galaxy
(Edvardsson et al. 1993).
However one should recall that the
existence of infalling gas on the Galactic disc is well established, as inferred
from high velocity clouds (Mirabel 1989). Infall of unpolluted gas could act as to 
lower the ISM abundance.
Generally, models of Galactic chemical evolution including infall  assume pristine 
gas, but results with pre-enriched matter do not differ as long
as the metallicity does not exceed $0.1 Z_{\odot}$ (Tosi 1988).

\subsection{Mergers and interactions}

It has become evident that interactions and merging between galaxies is a 
major driver in the evolution of  galaxies, affecting the number 
evolution and morphological mix of the galaxy population.
Mergers and interactions will also affect the  star formation activity in galaxies,
and in this respect be important for chemical evolution. Gas flow or accretion may lead
to increased star formation (SF) activity or starbursts, whereas stripping or harassment in rich
environments may inhibit SF. Gas accretion may also affect ISM abundances even if new 
star formation is not triggered, and may also provide a means for large scale mixing 
of the ISM.

Are some dwarfs  ejected debris from interacting disc galaxies, 
and what consequences for their chemical evolution would this have? One would
 expect rather high metallicities, but if the process occurred 
long ago, before the interacting giants were enriched, this is not necessarily true.
Evidence that dwarf-like objects are being formed within the gaseous tails of the
encounters is now quite well established (e.g. Duc and Mirabel 1994, 1998), although
their ultimate fate is unknown. Abundance 
analysis shows that their metal 
content is comparable to that of the parent galaxies re-enforcing the view that
the gas from which they originate comes from the central regions of their parents. 
Another aspect is that many metal-poor dwarf 
galaxies in clustered environments may have been  swallowed or torn apart by massive 
galaxies. The  Sagittarius dwarf galaxy, a recently discovered Galactic satellite 
(Ibata et al. 1994), may perhaps be such a case.

\subsection{Mixing}

Measured abundances in interstellar gas will of course depend on how well,
and on what timescale, the ISM is mixed and on what timescale fresh metal cools
and becomes visible. Kunth and Sargent (1986) argue that H{\sc ii} regions are
self-polluted within the ongoing burst providing  the ejected oxygen can
recombine fast enough to be observed in the H{\sc ii} zone while some can become
neutral and be observed in the H{\sc i} cloud. Pantelaki and Clayton (1987) dismiss 
this possibility from the fact that most of the ejecta should remain for a long
time in the hot gas generated by SN events. Spiral galaxies display radial
abundance variations, indicating that radial mixing is inefficient. On the 
other hand, barred spirals display smaller abundance gradients, since bar 
perturbations induce radial gas flows. Roy and Kunth (1995) discuss mixing
processes in the ISM of gas rich galaxies, and conclude that dwarf galaxies
are expected to show kpc scale abundance inhomogeneities. 
On the other hand, chemodynamical models (Hensler and Rieschick 1998), 
predict that the ISM will be well mixed and chemically homogeneous through 
cloud evaporation.

The observational situation is still not completely clear and rather few 
dwarfs have been subject to high quality studies of their chemical homogeneity.
Most dwarf irregulars seem rather homogeneous (Kobulnicky and Skillman 1996, 
Kobulnicky 1998) with the exception 
of NGC~5253 where local N/H overabundances has been attributed to localised 
pollution from  WR stars (Kobulnicky et al. 1997). There is also marginal evidence 
for a weak abundance gradient in the LMC (on the scale of several kpc, Kobulnicky 1998).
The situation is less clear in BCGs: 
In IIZw40, Walsh and Roy (1993) found a factor two variation in 
the oxygen abundance. IZw18 appears to be rather homogeneous (e.g. Skillman and Kennicutt 1993, 
Vilchez and Iglesias-P\'aramo 1998, Legrand et al. 1999) while recent spectroscopy of 
SBS0335-052 (Izotov et al. 1999b) 
reveals small but significant variations in accordance (though to a much lesser 
extent) with previous results (Melnick et al. 1992).

The possibility that metallicities in the neutral gas phase are 
orders of magnitude below the H{\sc ii} region abundances  would be
an ultimate test of large scale inhomogeneities. 
Recent O/H abundance determination in the H{\sc i} envelope of the very metal-poor 
compact dwarf IZw18 (Kunth et al. 1994) suggests the  possibility of a 
striking discontinuity between the H{\sc i} and H{\sc ii} gas phases: the measured O/H in the 
cold gas appears to be 30 times lower (i.e. $\sim$1/1000 $Z_{\odot}$) than that 
of the associated H{\sc ii} region.
Note however that Pettini and Lipman (1995) have strongly warned against the use of the O{\sc i} 
interstellar lines in  deriving O/H for the neutral gas, mainly because these lines are
saturated and the velocity dispersion is unknown (see also van Zee et al. 1998).
A further HST observation of O and S lines in IZw18 has unfortunately not produced 
consistent results (unpublished). 
Nevertheless Thuan et al. (1997) circumvented these problems in the case of SBS0335-052 although
their result awaits independent measurement of unsaturated lines such as the SII${\lambda 1256}$  multiplet. 

A crucial question would be  how to interpret the presence of metals in the H{\sc i} zone.
If indeed IZw18  experiences its very first episode of star formation,  the 
oxygen present in the ionised gas should originate from the ongoing burst and one
can  speculate that metals in the H{\sc i} were produced at an earlier epoch from 
population III stars prior to the collapse of the proto galaxy. On the other hand
the enrichment in the neutral gas could originate in a previous burst allowing for 
a time scale long enough  to homogenise a cold cloud of 1 kpc diameter as discussed by Roy and 
Kunth (1995), but see Tenorio-Tagle (1996).  
To circumvent the problem, Legrand (1998) has conjectured that in between bursts, IZw18 had 
maintained a minimum continuous star formation rate of only $10^{-4} M_{\odot}$/yr over the last 14 Gyrs. 
Such a SFR is comparable to the lowest SFR observed in low surface brightness galaxies. This scenario 
nicely explains the lack of galaxies with metallicities below IZw18, the absence of H{\sc i} clouds without
 optical counterparts and the homogeneity of the metal abundances in dwarf galaxies.
The question of the metallicity of the cold neutral gas is indeed very important for
understanding how much enrichment  has really 
occurred, since for many galaxies, a considerable fraction of the total baryonic
mass is in the form of neutral hydrogen.
If some dwarfs are not well mixed on large scales (e.g. LSBGs which have large H{\sc i}-discs) 
they would appear more metal-rich after converting a given fraction of gas into stars,
because one would be observationally biased towards the star-forming, more metal rich 
regions. On the other hand, if some dwarfs can mix their whole ISM on not too long 
timescales, the fresh metals will be efficiently diluted and the galaxy appear more
metal-poor. Galaxies with very turbulent ISM and those involved in mergers could 
possibly mix more easily on large scales.

Once ejected into the ISM, part of the metals may be locked up into dust grains.
This is observed in the local ISM for some elements (see e.g. Pagel 1997) and may result 
in strange element ratios and apparent under-abundances in extragalactic H{\sc ii} regions, 
although the effect is believed to be small for H{\sc ii} regions, due to grain destruction. 
In fact Bautista and Pradhan (1995) find that in the Orion nebula,  the depletion of 
iron into dust grains is probably a minor effect.
In the colder gas
associated with damped Ly$\alpha$ systems (see Sect. 8.2) depletion onto dust grains 
may be important for some 
elements.

\subsection{Chemical evolution models}

With assumptions about stellar yields, IMF and star formation history, models of the chemical evolution 
of galaxies can be constructed. Various assumptions such as instantaneous recycling
and closed box (i.e. the total mass of the system is conserved and perfectly mixed
at all times) allows one to construct simple analytic models (cf. Pagel 1997). 
In the simplest case, with a constant net yield, the ISM abundance 
will be a simple function of gas mass fraction ($\mu_{\rm gas} = M_{\rm gas}/M_{\rm gas+stars}$).
However, many of the effects mentioned above are likely to complicate the real picture.

Many dwarfs are believed to undergo short bursts of star formation separated by
long quiescent epochs.
Several additional ingredients were added to the closed-box models  when it was  
realised that  they could not account for the observed $Z-\mu_{\rm gas}$  
distribution by simply changing the number of bursts from galaxy to galaxy.
Models with normal or differential  winds  (selectively enriched in heavy elements)
have been applied to
starburst galaxies (see Matteucci 1996 for a review). 
They seem to be successful in reproducing the observed He/H vs. O/H 
distribution with a number of bursts between 7 and 10 in general, differential
winds and various amounts of primary and secondary nitrogen from intermediate
stars. This is in agreement with Pilyugin's conclusions (Pilyugin 1992, 1993) 
although other possibilities 
have been explored. For instance, one can vary the IMF from
galaxy to galaxy by changing the slope of the lower mass cut-off 
(Marconi et al. 1994) while Olofsson (1995)  proposed that different
behaviour of N/O versus O/H can be attributed to an effect of mass loss as a
function of metallicity. On the other hand individual galaxies do not fit 
into these schemes whenever one enters into details. A few galaxies for instance 
tend to have large N/O for their their O/H. 
The galaxy IZw18 falls  into this 
category and one is forced to assume that N is produced as a primary element
in massive stars. Even this assumption does not relax the need for a strong
star formation efficiency (expressed as the inverse of the time scale of star 
formation) and strong O-enriched winds (Kunth et al. 1995).
Note that recent work, e.g. by Izotov and Thuan (1999) on low metallicity BCGs,
shows a reduced scatter for N/O and C/O at low abundances, to some extent removing 
the need for all the  mechanisms originally invoked to produce a scatter.
Clearly, the way that chemical evolution in metal-poor galaxies proceeds 
is far from beeing a settled issue.

More sophisticated ``chemodynamical'' models attempt to describe both the star
formation history, its 
impact on the abundances and the interaction with the ISM in a self consistent way. Lately, 
these models have been applied to dwarf galaxies with some success (Hensler and Rieschick
1998). Recent models of the chemical evolution versus redshift in the Universe  by
Cen and Ostriker (1999) predict that  metallicity shows a stronger dependence on
the local density (i.e. galaxy mass) than on redshift, hence objects with high and low abundances 
are found at all $z$. In the local Universe, these models predict that any region  denser 
than the cosmic average (a minimum requirement for a galaxy) should have a metallicity
of ~$0.01 Z_{\odot}$ or higher.

%%%%%%%%%%%%%%%%%%%%%%%%%%%%%%%%%%%%%%%%%%%%%%%%%%%%%%%%%%%%%%%%%%%%%%%%%%%%%%%%%%%%%%%%
%%%%%%%%%%%%%%%%%%%%%%%%%%%%%%%%%%%%%%%%%%%%%%%%%%%%%%%%%%%%%%%%%%%%%%%%%%%%%%%%%%%%%%%%
%%%%%%%%%%%%%%%%%%%%%%%%%%%%%%%%%%%%%%%%%%%%%%%%%%%%%%%%%%%%%%%%%%%%%%%%%%%%%%%%%%%%%%%%

\np

\section{The general properties of metal-poor galaxies}

Here we will give an overview of the observed  properties and chemical abundances of
metal-poor galaxies, while in Sect. 5 we will discuss a few of these in more detail.  
As a metal-poor galaxy is in general a dwarf galaxy, this
section will be biased towards dwarf galaxies.
 
Dwarf galaxies come in different kinds, with different properties and metallicities.
There is a partial overlap in the classification and physics of metal-poor galaxies.
At faint absolute magnitudes the dwarf irregular (dI), dwarf elliptical (dE) and low 
surface brightness galaxy (LSBG) classes converge. Moreover blue compact galaxies (BCGs)
to some extent overlap with actively star forming dIs. The possible connections between 
different types of galaxies will be further discussed in Sect. 7.2. 
A lot of the data on dI and dE/dSph galaxies are from the Local Group. It is not our 
intent here to give a general description of the Local Group dwarfs, recent and excellent 
reviews can be found in e.g. Mateo (1998, which also contains a compilation  of very
useful data) and Grebel (1998); see also the book ``Stellar astrophysics for the Local 
Group'' (eds. Aparicio et al. 1998) and the proceedings of the recent IAU symposium
``The Stellar Content of Local Group Galaxies'' (eds. Whitelock and Cannon 1999). Here we will only focus on 
those points that are relevant for the understanding of the most metal-poor galaxies.

As we have mentioned, metallicity correlates with luminosity for galaxies, at least in 
the local Universe. More metal rich galaxies are on average more luminous and have in
addition higher 
surface brightness. This produces a  bias against detecting metal-poor galaxies. Moreover, 
except for H{\sc ii} region spectroscopy, most ways of measuring metallicity discriminate against low 
luminosity galaxies. 

\subsection{Dwarf irregular  galaxies}

Dwarf irregulars are in general well described by exponential 
surface brightness profiles. They contain fair amounts of neutral and perhaps molecular 
gas and in general show evidence for star formation, at low (or moderate) rates. 
There are some examples of galaxies  of intermediate
type between dI and dE, e.g. the Local Group dwarfs Pegasus and Phoenix. 
Such galaxies have no signs of ongoing star formation, but  contain gas (Mateo 1998).
Dwarf irregulars partially overlap BCGs
in the classification criteria, i.e. many BCGs
are indeed irregular, and dIs often contain bright H{\sc ii} regions which may 
be picked up in emission line surveys. Of course the transition is gradual and 
to some extent  arbitrary. 
There is also some overlap with the LSBG class of galaxies discussed in Sect. 4.3.

The Local Group contains around a dozen (depending on the magnitude 
limit) dIs  most of which are rather metal-poor. Dwarf irregulars are 
also found in 
the local field and in nearby clusters and groups, but metallicity determinations 
become rarer with increasing distance.

Dwarf irregulars in general show ongoing star formation and H{\sc ii} regions.
The star formation history (SFH) of Local Group dIs has been reviewed recently by 
Grebel (1998), and Mateo (1998). Most dIs seem to have experienced more or less
continuous star formation over a Hubble time. Many Local Group dIs show evidence for 
an episodic star formation history to a varying degree (e.g. Leo A, Tolstoy et al. 1998), 
but it is difficult to say if the past SFH should be described in terms of short bursts  or 
more extended periods of increased SF as compared to the time averaged value. 
There seems to be no clear pattern when comparing different galaxies,
 i.e. no support for co-ordinated bursts.
Star formation in dIs is probably governed by different mechanisms
than in spiral galaxies, i.e. there are no density waves available 
for triggering star formation. Despite their closeness, it is quite fair 
to say that the regulation of star formation in the Local Group dIs 
(and dEs too) is not yet well understood. Interactions, ram-pressure
stripping, superwinds and other recipes have been suggested as  processes
 governing
the star formation history and morphological segregation, but results are 
still inconclusive.

Young and Lo (1996, 1997b) find that neutral hydrogen in dIs exists in two 
phases and that the presence  of cold H{\sc i} may be the key factor by which
 a dI is presently forming stars or not.
Intermediate dE/dI types do in some cases have gas, but not in the form of 
cold H{\sc i}, which could explain the lack of present star formation (Young and 
Lo 1997b). The H{\sc i} content of dIs in the nearby Universe has been investigated
by e.g. Thuan and Seitzer (1979), Schneider et al. (1992, 1998) and C\^ot\'e
et al. (1997); see Skillman (1996) for a review. The dynamics of dIs can 
be investigated through H{\sc i} rotation curves, and dIs (as most other dwarfs) 
turn out to be dominated by dark matter (cf. Mateo 1998).

\subsubsection{Chemical abundances of dwarf irregular galaxies}
 
Dwarf irregular galaxies are in general metal-poor. Nebular abundances 
in Local Group dIs range between roughly one third of solar (e.g. LMC) and $\sim$ 1/40 of solar. 
Abundances of dIs are usually derived from H{\sc ii} regions. For some dIs there
exist also metallicity estimates of [Fe/H] using the same 
technique as generally used for dEs, i.e. colour--magnitude diagrams (CMDs). 
 
We present in  Table 1 some basic 
properties for the most metal-poor (oxygen abundance less than or equal to 1/10
of the solar value) Local Group dIs.  Some dIs (Leo A, SagDIG and Sextans A)
do indeed have oxygen abundances comparable to the most metal-poor
blue compact galaxies. If available, we present also [Fe/H] values
based on stellar photometry, absolute B-magnitudes and the logarithm
of the estimated total (dynamical) mass expressed in solar masses.

\begin{table}[h]
\caption{{\protect\small The most metal-poor Local Group dIs (data from Mateo 1998). The second column
gives 12+log(O/H), and the third column [Fe/H]. The fourth and fifth columns give the integrated absolute B-band 
magnitude and the logarithm of the derived dynamical mass (in solar masses), respectively.}}
\begin{center}
\begin{tabular}{ccccc}
\hline
Galaxy name 	& 12+log(O/H) 	& [Fe/H] & $M_B$ 	& log($\cal M$)	\\ 
\hline
Leo A		&  7.3 		&  	& -11.3		& 7.0	\\
SagDIG		&  7.42		& 	& -12.1		& 7.0	\\
Sextans A 	&  7.49  	& -1.9 	& -14.2		& 8.6	\\
Gr 8 		&  7.62  	&  	& -11.2		& 7.6	\\
EGB 0427+63	&  7.62		& 	& -11.6		& 	\\
WLM		&  7.75  	& -1.5	& -13.9		& 8.2	\\
IC 1613		&  7.8		& -1.3	& -14.2		& 8.9	\\
Sextans B	&  7.84		& -1.2	& -13.8		& 8.9	\\
Pegasus		&  7.93 	& -1.0	& -12.3		& 7.8	\\
\hline
\end{tabular}
\end{center}
\end{table}

Only for a few dIs have abundances
of individual stars been determined. 
For the Magellanic Clouds, Pagel (1992) concludes that the agreement between
stellar and gaseous heavy element abundances is about as good as can be expected in view of 
the uncertainties.
However, for NGC 6822, metal lines of  
early type stars are considerably weaker than
expected from the measured  nebular oxygen abundance (Massey et al. 1995). 
If there is a real discrepancy, this  is worrying
since one argues that ISM oxygen abundances can be obtained with high accuracy.
Direct spectroscopic observations of individual stars in more dIs are much needed.

Skillman et al. (1989) confirmed in a seminal paper  a strong correlation between absolute magnitude 
and metallicity for local dIs in the sense that the fainter systems have lower metallicities. 
This was hinted already in some earlier papers: e.g. Lequeux et
al. (1979), Kinman and Davidson (1981). 
 The existence of a relation between metallicity 
and luminosity constitutes a fundamental observation that we need to understand to address the 
chemical evolution of dwarf galaxies. The metallicity--luminosity relation for dIs and other 
galaxies will be further addressed in Sect. 7.1. More recent work in this area include that of 
Richer and McCall (1995) and Hidalgo-G\'amez and Olofsson (1998). The comparison of dIs and dEs 
is not straightforward, as will be discussed in 
next section, since different chemical elements  are sampled in general.

The best studied dIs reside in the Local Group, but plenty of observations of relatively 
nearby dIs outside the Local Group can be found in the literature (e.g. Skillman et al. 1989). 
The use of H{\sc ii} regions for determining oxygen abundances makes even relatively 
distant dIs accessible for chemical investigation of their ISM. 
Dwarf irregulars in the M81 group have been studied by Skillman et al. (1994) and  
Miller \& Hodge (1996),  and dIs in the Sculptor group by Miller (1996). 
Abundances are similar to Local Group dIs and follow the same metallicity--luminosity 
relationship (Skillman et al. 1989, Skillman 1998).   
A sample of quiescent dwarfs (dIs/LSBGs) studied by van Zee et al. (1997b,c) shows 
again typical Local Group abundances and adheres to the metallicity--luminosity  relation; 
see also the section on LSBGs (4.3.1). The most metal-poor dIs are included in Table 3.

Vilchez (1995) presented a spectrophotometric study of star forming dwarf galaxies 
(dIs  and BCDs) in different environments, from low density to the core of the Virgo
cluster. For a given luminosity, the high density environment dwarfs are systematically
overabundant ($\sim$0.5 dex) with respect to dwarfs in low density environments 
and the Skillman (1989) relation. However,  metallicities especially for dwarfs 
in Virgo are uncertain and the observed over-metallicity may be spurious.
Later work by Lee et al. (1998) finds O-abundances of 12+log(O/H) = 8.0 to 8.3 for 
absolute magnitudes ~$ -15 > M_B > -16$~ indicating that the systematic overabundance 
of Virgo dIs with respect to local ones is small or non-existent.  Some cluster dwarfs seem to
be anomalously metal rich (perhaps related to the ``tidal dwarfs'', see below), while 
others appear to be rather normal (Vilchez 1999; private communication).
It now appears that dwarf galaxies can be spawned in the process of violent galaxy
interactions. These ``tidal dwarfs''  will 
be discussed separately in Sect. 4.5.

Carbon abundances have been determined for a few dIs by 
Garnett et al. (1995) and Kobulnicky and Skillman (1998), although most galaxies
in these investigations are rather BCGs, revealing a strong positive correlation 
between [O/H] and [C/O].  
Nitrogen abundances exist for more dIs (e.g. Garnett 1990), but rarely  for the most 
metal-poor Local Group dIs. The typical [N/O] values for metal-poor dIs are of the same order 
as those of BCGs. The [C/O] and [N/O] nebular abundance ratios will be 
further discussed in the BCG section.

\subsection{Dwarf elliptical/spheroidal galaxies}

Let us first say a few words about the nomenclature.
Luminous Elliptical galaxies follow a well defined relation between luminosity
and central surface brightness  in the sense that more luminous objects have lower 
central surface brightness (a projection of the fundamental plane), whereas the
opposite relation appears to hold for ``diffuse'' early type dwarfs, and late type 
galaxies (e.g. Binggeli 1994). 
Thus the kind of galaxy classically termed dwarf elliptical (like Fornax) appears to 
be physically distinct from non-dwarf elliptical galaxies (like M87), see Binggeli 
(1994) and Kormendy and Bender (1994). This has created some debate on what to call 
low mass ellipsoidal galaxies.
Ferguson and  Binggeli (1994) proposed to call objects with $r^{1/4}$ profiles 
(e.g. M32 and giant elliptical galaxies)  {\it elliptical} (E) and those
with more exponential profiles   {\it dwarf  elliptical} (dE). Kormendy and 
Bender (1994) calls the latter class {\it spheroidal} and uses the prefix dwarf 
to indicate the low luminosity galaxies in each class. 
Lately, the clear structural distinction between these two classes of objects has begun to
be smeared out: Jerjen and Binggeli (1997) show that, as  luminosity increases,   
the luminosity profiles of low mass 
ellipsoids approach those of giant ellipticals, if the central 0.3 kpc is excluded.
In this review we will be primarily
interested in the low luminosity low metallicity systems, which we will refer to as dEs, or
occasionally dSph (when speaking specifically about the satellites of our Galaxy), which have nearly 
exponential profiles.

Dwarf elliptical galaxies were long thought to be made up of
exclusively old stars, but it has now become evident that many seem to have experienced 
several star formation episodes, and in many cases quite recently (Grebel 1998). 
This makes the distinction between dEs and dIs less clear. The SFH in Local Group dEs are
mainly explored through their resolved stellar populations, e.g. by means of colour magnitude 
diagrams (CMDs). A famous example of a CMD revealing distinct and well separated episodes of star 
formation is that of the Carina dwarf galaxy (Smecker-Hane et al. 1994).

Masses have been derived for many local dEs using measured velocity
dispersions of their stars and they tend to be  dark matter dominated (cf. Mateo 1998). 
Most dEs are very gas poor. In fact, the gas content is sometimes lower than what one 
would expect from mass loss from the old stellar population alone. The best 
known example is NGC~147 where neither H{\sc i} (Young and Lo 1997a) nor CO (Sage et 
al. 1998) has been found. Thus there might be a mechanism that removes gas, 
or the gas is in a phase which is not observable. In other dEs, the gas 
content is in rough agreement with expectations from stellar mass loss 
(Young and Lo 1997a). An interesting case is  the Sculptor dwarf
that has no H{\sc i} within its optical extent, but some well outside in the halo, 
perhaps moved there by a  superwind (Carignan et al. 1998).

Most data on dEs come from the Local Group population, but dEs also
exist in large number in galaxy clusters like Virgo, Fornax and Leo. 
There appear to be very few pure field dEs (i.e. not accompanying a giant 
galaxy, Binggeli et al. 1990), but their luminosity function is not well
known since faint dEs  have in general low surface brightness hence may be 
absent in many surveys (cf. Sect. 6).  In the Local
Group, new dEs are still beeing discovered  (e.g. Armandroff et al. 1998),
implying that not even locally do we have a complete picture about
the dE population. 

\subsubsection{Chemical abundances of Local Group dE/dSph galaxies}

Aaronson (1986) showed that dEs seemed to follow a tight relation
between metallicity ([Fe/H]) and optical luminosity. This has later
been confirmed and a rather tight  luminosity metallicity relation for 
dwarf ellipticals is now quite well established (Caldwell 1998).   
Metallicity data for dEs based on CMD analysis, can be found at several places
in the literature (see Mateo 1998 for a recent compilation). These data 
also reveal that most dEs have a considerable spread in metallicity as
measured from the width of the RGB.

Stellar spectroscopic abundances exist for several of the Milky Way
satellites, e.g. the Sextans dSph (Suntzeff et al. 1993).
Another well studied galaxy is Draco, where Shetrone  et al. (1998) found ~$-3.0 <$ [Fe/H]$ < -1.5$
confirming earlier findings of a substantial  metallicity spread (Lehnert et al. 1992). 
The  metallicity derived from a CMD is [Fe/H] $=-2.0 \pm 0.15$.

Richer and McCall (1995) made a spectroscopic study of planetary
nebulae (PNe) in  local dEs, including data from the literature on the  Fornax dwarf. 
They also revisit the metallicity--luminosity relation for dwarf irregulars 
by Skillman et al. (1989), using  new determinations of the distance and metallicity.
Their PN spectra have rather low signal to noise requiring empirical relations 
between [O{\sc iii}]${\lambda 5007}$ and H$\beta$ to estimate lower bounds on the oxygen abundances 
in the three dEs, from which they make a transformation
to a ``mean oxygen abundance''. They claim these to be systematically
higher in dEs than in dIs of the same luminosity. Although based on a sound
line of reasoning their derivation of the ISM mean oxygen abundance
involves several steps, and must be regarded as quite uncertain. Thus, while 
tentatively very important, these results need to be quantitatively confirmed for a 
larger sample of PNe in more galaxies, with higher S/N and better spectral 
coverage. Recently, PNe were studied also in the Sagittarius dwarf (Walsh et al. 1997).

Richer and McCall (1995) also claim that [O/Fe] is systematically higher in dEs 
than dIs, where the latter class  was represented by the Magellanic Clouds.  
The Fe abundance adopted for LMC and SMC are based on young supergiants, 
and thus do not measure the same stellar generation as in the dEs. To compensate 
for this, they modify the abundance ratios.

If one instead compares the Magellanic Cloud O-abundances with [Fe/H] estimates
from field stars with ages comparable to those of the red giants in the 
dEs (e.g. Hilker et al. 1995) the Magellanic Cloud [O/Fe] values increase 
and thus the difference compared to the dEs decreases. In Fig. 1 we plot [Fe/H] vs. 
[O/H] for all nearby dwarfs with
both elements measured. It is clear that the overabundance over oxygen 
with respect to iron seems to be a general feature of local dwarfs (see also Fig. 7 
in Mateo 1998) as it is for giant halo stars (Barbuy 1988) and metal-poor unevolved 
halo stars in our Galaxy (Israelian et al. 1998). There is no indication that  [O/Fe] 
is  significantly higher in dEs than dIs. In a later work, Richer et al. (1998) 
find [O/Fe] to be  systematically  higher in  dEs than e.g. in  M32 and 
the Galactic bulge, but similar to the galactic halo, and they suggest that the star formation timescale 
has been shorter in dEs than ellipticals and spiral bulges.

\begin{figure}
\resizebox{14cm}{!}{\includegraphics{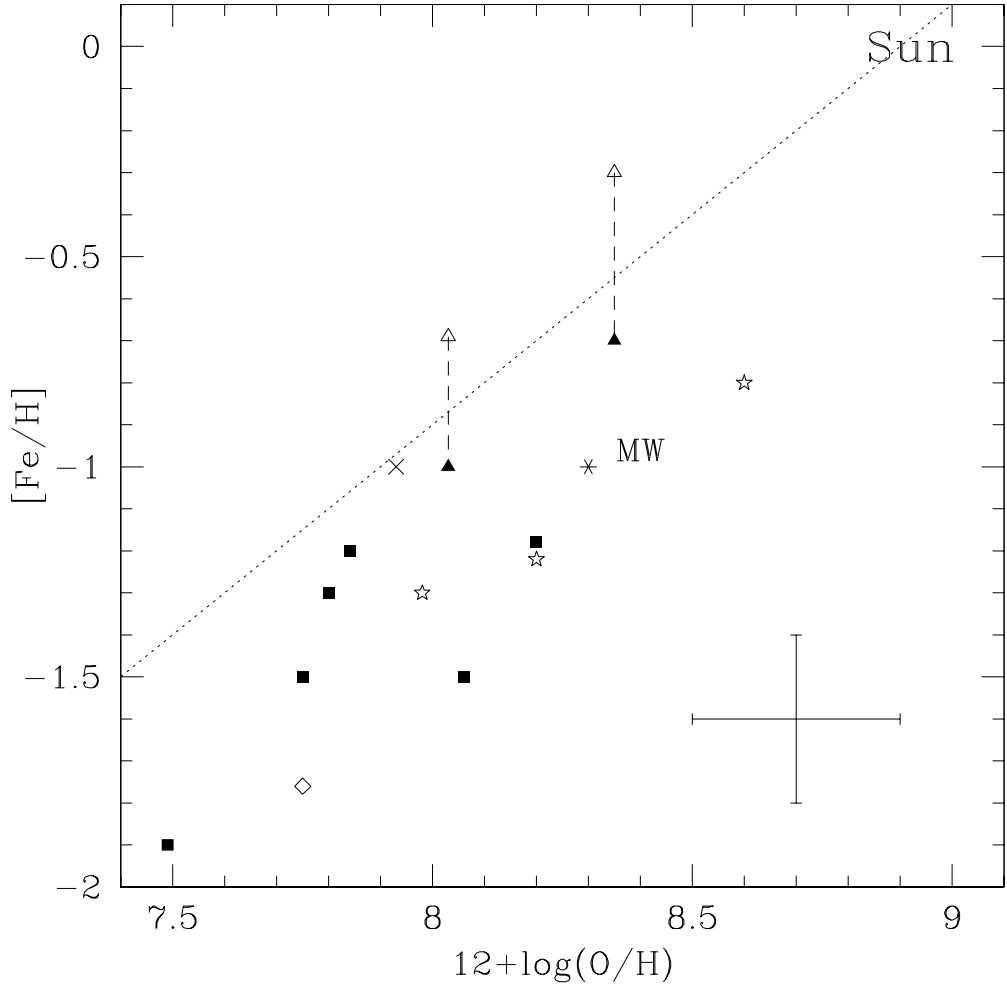}}
\caption{{\protect\small [Fe/H] vs. 12+log(O/H) for all local dwarf galaxies  where
both iron and oxygen abundances have been obtained (data taken from Mateo 1998), plus the Magellanic
clouds, a BCG, the Sun and the Milky Way halo. dIs are denoted by filled squares, dI/dE intermediate 
type (Pegasus) by a cross, and dE (Sagittarius) by an asterisk; stars are the dEs studied by 
Richer \& McCall with
oxygen abundances from planetary nebulae. 
The triangles joined by dashed line represent the Magellanic 
clouds, where for the open symbols the [Fe/H] are based on young supergiants, while
for the filled symbols the [Fe/H] have been determined from Str\"omgren photometry of
the field population. The open diamond indicates the location of the nearby BCG VII~Zw403 
(see Sect. 4.4.7).
The location of  Milky Way halo stars is indicated with ``MW'' (from Richer and McCall 1995), 
and that of the Sun with ``Sun''. 
The dotted line corresponds to solar [O/Fe]. In general (except for the open triangles)  
Fe/H traces
an older stellar population than O/H to some extent explaining the apparent general deficiency of 
iron, with respect to oxygen. The typical errors are around 0.2 dex for both quantities, as 
illustrated by the error bars in the lower right corner. }}
\end{figure}

Now, how metal-poor are the Local Group dEs? As was discussed above, [Fe/H] measured
from CMDs tends to be systematically smaller than [O/H] measured from H{\sc ii} regions with 
up to 0.5 dex, making these two quantities difficult to compare.
Of the Local Group dEs, a dozen have [Fe/H] $\le -1.5$, and the lowest value, [Fe/H] $ = -2.2$, 
is found for the Ursa Minor dSph, see Table 2. In summary, many Local Group dEs are very metal 
poor, which probably is related to their intrinsic faintness. Taking the 
average offset between [Fe/H] and [O/H] into account, Mateo (1998) concludes that dEs appear to 
be more metal rich in view of their luminosities,
in agreement with Richer and McCall (1995).

\begin{table}[h]
\caption{{\protect\small The most metal-poor  dE/dSph galaxies in the Local Group, 
including all galaxies with [Fe/H]$\le -1.0$.   The second column
gives [Fe/H], and the third column gives nebular oxygen abundances, when available. 
As discussed in the text,
there is a tendency for the measured abundance of oxygen to be higher than that of  iron, with 
respect to the solar values. The fourth and fifth 
columns give the integrated absolute V-band 
magnitude and the logarithm of the  dynamical mass (in solar masses), 
respectively. All data taken from Mateo (1998).}}
\begin{center}
\begin{tabular}{ccccc}
\hline
Galaxy name 	&	[Fe/H]	& 12+log(O/H)	&	$M_V$	& 	log($\cal M$)	\\ 
\hline
Ursa Minor	&	-2.2	& 		&	-8.9	& 	7.36	\\
Draco		&  	-2.0	&		&	-8.8	&	7.34	\\
Carina		&	-2.0	&		&	-9.3	&	7.11	\\
Andromeda III	&	-2.0	&		&	-10.3	&		\\
Leo II		&	-1.9	&		&	-9.6	&	6.99	\\		
Phoenix		&	-1.9	&		&	-10.1	&	7.52	\\
LGS 3		&	-1.8	&		&	-10.5	&	7.11	\\
Antlia		&	-1.8	&		&	-10.8	&	7.08	\\
Sculptor	&	-1.8	&		&	-11.1	&	6.81	\\
Sextans dSph	&	-1.7	&		&	-9.5	&	7.28	\\
Tucana		&	-1.7	&		&	-9.6	&		\\
Andromeda II	&	-1.6	&		&	-11.1	&		\\
Andromeda I	&	-1.5	&		&	-11.9	&		\\
Leo I		&	-1.5	&		&	-11.9	&	7.34	\\	
Fornax		&	-1.3	&	8.0	&	-13.2	&	7.83	\\
NGC 185		&	-1.2	&	8.2	&	-15.5	&	8.11	\\
NGC 147		&	-1.1	&		&	-15.5	&	8.04	\\
M32		&	-1.1	&		&	-16.7	&	9.33	\\
Sagittarius	&	-1.0	&	8.3	&	-13.4	&		\\
\hline
\end{tabular}
\end{center}
\end{table}

\subsubsection{Abundances of dEs outside the Local Group}

Unfortunately, very little is known about the chemical abundances of dEs outside 
the Local Group.
The M81 group is near enough that
the study of colour magnitude diagrams is feasible
with powerful telescopes. Caldwell et al. (1998) report on HST photometry
of two dEs in the M81 group, concluding that their metallicities are similar
to those of Local Group dEs with the same luminosity. Of course two 
data points is far too little to assess whether the Local Group dE population 
is typical of dEs in general, but this investigation nicely shows that it is 
feasible to obtain CMDs of dEs outside the Local Group.

Held and Mould (1994) obtained integrated spectra of 10 {\it nucleated} dEs in
the Fornax cluster, and derived metallicities in the range [Fe/H] $= -1.4$~ to ~$-0.7$.
They found that the metallicities are tightly correlated with the UBV-colours.
The  range in luminosity is too small to deduce any relationship with 
the metallicity, but interestingly the median metallicity, $\langle$[Fe/H]$\rangle = -1.1$,
and luminosity, $\langle M_V\rangle = -16.1$ (assuming a distance of 18.6 Mpc for the
Fornax cluster, Madore et al. 1999), 
is in perfect agreement with the 
metallicity--luminosity relation for local dEs of Caldwell et al. (1998). The tight 
correlation between metallicity and colour  is quite remarkable since the large
metallicity spread at a given luminosity indicates very different enrichment 
histories. The colour-metallicity relation also offers some hope to obtain metallicity
information from photometry of distant galaxy populations. One should, however, 
 keep in mind that the relation was established from a biased 
sample of  nuclei of galaxies with uniform luminosity and morphology.

Although the Virgo cluster is known to be very rich in dEs, their metallicities
are to a large extent unknown.
Based on optical and near infrared colours  Thuan (1985) concluded 
that Virgo cluster dEs had metallicities in the range: $1/3 Z_{\odot}$ to $Z_{\odot}$. 
These conclusions were based on comparisons 
with  models that are now outdated. Since the possibility of using photometry
for metallicity estimates is of interest for the study of distant faint galaxies
we decided to reassess this finding. To do this we have compared 
the photometric data by Thuan (1985), Bothun and Caldwell (1984), James (1991,1994), 
and Zinnecker and Cannon (1986) with a new set of models by Bruzual and 
Charlot (2000). Special care was taken when modelling galaxies with inhomogeneous 
data (e.g. different aperture sizes).
When using simple stellar populations (where all stars have the same age and metallicity) 
and a standard Salpeter or Scalo IMF, the 
resulting metallicities are in the range from less than $1/10 Z_{\odot}$ to $Z_{\odot}$, 
with a median around $1/3 Z_{\odot}$, and a typical age of 3 Gyr. 
Models with exponentially decreasing SFR (e-folding time 3 Gyr) produced slightly better 
fits with similar metallicity and a median age of 11 Gyr.
Assuming a distance of 16.2 Mpc to the Virgo cluster (Macri et al. 1999)
we find a median luminosity $\langle M_V\rangle = -15.3$, which with a median 
metallicity of $\langle$[Fe/H]$\rangle \approx -0.5$ means that the Virgo dEs are more metal
rich by $\approx 0.5$ dex  as compared to the ~$M_V - Z$~ relation for local dEs
by Caldwell et al. (1998). Thus, either the Virgo cluster dEs are overabundant,
or perhaps more likely, the photometrically derived metallicities are too high.
We find no clear trend between metallicity and luminosity in this analysis.

In conclusion, the photometric data on Virgo dEs indicate that these galaxies
have  metallicities typically around 0.3 $Z_{\odot}$ (in rough agreement with 
Thuan's earlier result). Today it would be possible to probe 
fainter systems  using CCDs and modern near-IR arrays, although it is not yet clear how 
powerful a tool optical/near-IR colours are in deriving metallicities. 
Note that recent spectroscopy for six dEs in the Virgo cluster
gives metallicities ranging from a few tenths of solar to solar, with small
radial gradients as compared to giant ellipticals (Gorgas et al. 1997). 
Their median inferred [Fe/H] and luminosity is in rough agreement with 
the relation for local dEs by Caldwell et al. (1998).
Spectroscopic metallicity determination 
of intrinsically faint dEs outside the Local Group is now 
feasible with 8-10m class telescopes, and should be pursued.

\subsection{Low surface brightness galaxies}

A low surface brightness galaxy (LSBG) is, as the name implies, a galaxy with 
surface brightness fainter than  $\mu_{B,0} = 23$ mag/arcsec$^2$
(the central B-band surface brightness). This class spans from tiny dwarfs,
sometimes similar to dSph galaxies, to luminous giants like Malin~1 (Bothun et al. 1987).
The latter are in general not metal-poor by our standards, and will not 
be considered here. Instead, we will be interested in the LSBGs with low
integrated luminosities (the dwarf LSBGs). Many dEs could in  fact classify
as LSBG, but we will reserve this name for late type galaxies, often
gas rich, keeping in mind that dEs in general have comparable
surface brightness.  Low surface brightness gas poor galaxies (as observed 
e.g. in Virgo, e.g. Binggeli et al. 1984) will be covered by the dE class. 
Dwarf irregulars with low surface brightness, will 
be included in our LSBG class. The appearance of a typical LSBG on the digitised
sky survey is shown in Fig. 2.

The LSBGs pose a severe problem for the general understanding of the 
galaxy population. This is because their nature make
them difficult to detect with most techniques that have been applied
to survey the galaxy content of the Universe. A striking example is
the not too distant giant ($M_B = - 21.1$) Malin 2 (Bothun  et al. 1990). 
With an integrated apparent magnitude of 14.65 it is bright enough to be
included in most galaxy catalogues. The fact that it is not illustrates 
that despite the claims, galaxy catalogues are not magnitude limited,
but rather surface brightness limited (cf. Disney and Phillipps 1983). 
This has the consequence that
 our view of the LSBG population is largely incomplete, and since 
dwarfs tend to be of low surface brightness, it is evident that we 
are lacking many local metal-poor LSBGs and dEs. See e.g. Bothun et al.
(1997) for a general discussion on the problems of finding LSBGs.
In view of these difficulties, the true space density of LSBGs is badly known, 
although some recent progress has been made.

Van der Hulst et al. (1993), studying H{\sc i} properties of LSBGs,  found that while
the gas mass and gas mass fractions are high, the column density of H{\sc i} is low due
to very extended H{\sc i} morphologies. The neutral gas densities are found to be lower 
than the empirical threshold for star formation found for normal high surface brightness galaxies (Kennicutt 
1989) explaining the low star formation rates. A similar study was performed by van Zee
et al. (1997a) who compared dwarf LSBGs with ``normal'' gas-rich dwarfs, finding no
qualitative difference however and noting that both types were equally inefficient star formers. 
De Blok et al. (1996) found further support for low gas densities as compared to
normal high surface brightness galaxies. Moreover the gas mass fraction is higher in LSBGs
than other types of galaxies
when measured with respect to the absolute blue luminosity (typically $M_{HI}/L_B = 1$, in solar units)
 or the dynamical mass (McGaugh and de Blok 1997).
They suggest that the total mass density of a galaxy regulates its rate of evolution. The dynamical 
masses are of the order of $10^{10} M_{\odot}$ and the H{\sc i} masses an order of magnitude smaller. 
It is likely that LSBGs have experienced continuous star formation at low rate (Bergvall \& 
R\"onnback 1994). Bell et al. (1999) suggest that the principal difference between blue and redder
LSBGs is the e-folding timescale for star formation. 
The general conclusion seems to be that LSBGs are gas rich and unevolved, both chemically and
photometrically, which may be related to their low mass densities.

In view of their probably high space density and large gas mass fractions, the LSBGs 
constitute an important fuel reservoir for star formation in the Universe. Gas in LSBGs can be
made accessible to other galaxies e.g. in collisions. Likewise, if the extended low 
density gas somehow can be forced to collapse, LSBGs may be sites of future starbursts.  
The possible physical relations 
with other galaxies will be addressed in Sect. 7.2. Moreover, there have been suggestions 
that LSBGs may be responsible for some of the metal line absorption systems seen in QSO spectra 
(e.g. Phillipps et al. 1993).

\subsubsection{Chemical abundances of LSBGs}

Despite their faintness, many LSBG contain H{\sc ii} regions and the metallicity of 
LSBGs have been investigated through the derivation of nebular oxygen abundances. McGaugh 
(1994)  used an empirical determination
of the oxygen abundance to find values of ~$12+$log(O/H) ranging from 
7.3 to 8.8, for individual H{\sc ii} regions with a strong peak around 8.4.
 There was often considerable scatter between H{\sc ii} regions in the same galaxies.
Taking weighted averages,  eight galaxies have O abundances less than 1/10 of the solar value,
ranging down to $\sim$ 1/15, and the absolute B magnitudes of the 
metal-poor ($\le 10$\% of solar) subsample range from $-15.8$ to  $-20.3$ (for H$_0$=75 km/s/Mpc).

Similarly,
R\"onnback \& Bergvall (1995) found that a sample of LSBGs with ~$M_B = -14$~ to ~$-18.5$,  
and selected to have blue colours, all had low metallicities. Typical oxygen abundances are 
around 1/10 of solar, extending down below 1/20. They also derive N/O values. Five out of 
13 galaxies had an oxygen
abundance below 10\%  of solar, e.g. the remarkable edge-on galaxy ESO~146-G14 ($M_B=-16.6$) 
with 12+log(O/H)=7.6 (see also Bergvall and R\"onnback 1995). 
Several of the galaxies in these two samples fall below the metallicity--luminosity relation for dIs, 
see Fig. 10.

Van Zee et al. (1997b,c), for a sample of quiescent LSBGs/dIs derive oxygen abundances ranging
from 12+log(O/H)=7.6 to 8.3, matching perfectly  the $M_B - Z$ relation  
for dIs (Fig. 10).  
A massive neutral hydrogen cloud was found in the Virgo cluster,
by Giovanelli and Haynes (1989).
It has an optical low surface brightness
counterpart, with some central faint H{\sc ii} regions from which Salzer et al. (1991)  
derived 12+log(O/H)=7.66.
De Blok and van der Hulst (1998) found  no very metal-poor LSBGs in their investigation, but
three galaxies examined for abundance gradients  were found to have none.
This together with small or non existent colour gradients (R\"onnback \& Bergvall
1994, de Blok et al. 1995, Patterson and Thuan 1996) suggests that the stellar 
populations are spatially homogeneous in age and metallicity. This is supported also 
by recent  near infrared surface photometry (Bergvall et al. 1999).

There is only a weak correlation between metallicity and absolute blue 
luminosity (McGaugh 1994, and Bergvall \& R\"onnback 1994), see however Fig. 10.
Metallicity and surface brightness are related in the sense
that LSBGs have lower than average abundances for their 
absolute magnitudes as compared to high surface brightness galaxies.
However, neither investigation found any strong 
correlation between surface brightness and oxygen abundance.

Some  LSBGs have a too low surface brightness to be observable spectroscopically 
at the time of these studies. The new generation of 8-10m class telescopes may be 
used to investigate if late type galaxies with extremely low surface brightness
are even more metal-poor. Also, new surveys for extremely low surface brightness
galaxies may yield many new interesting metal-poor candidates, though abundances 
may  be very difficult to obtain if the galaxies do not have H{\sc ii} regions.
The most metal-poor LSBGs are included in Table 3.

\subsection{Blue compact and H{\sc ii} galaxies }

The concept of ``compact galaxies'' was introduced by Zwicky (1965) to denote
galaxies barely distinguishable from stars on the Palomar Sky Survey plates.
Originally, most studies of blue compact galaxies (BCGs) concerned objects from
the lists of compact/emission line/UV-excess galaxies produced by Zwicky (1966),
Haro (1956) and Markarian (1967). However, only a fraction of the objects
in these lists are BCGs, the others being AGNs, normal spirals with nuclear star 
formation, H{\sc ii} regions in the outskirts of nearby spirals etc. Later, many apparently 
similar objects have been added, mostly from
emission line surveys (cf. Sect. 6). This type of galaxy is sometimes 
also referred to as H{\sc ii} galaxies (Melnick et al. 1985b, Hazard 1986, Terlevich et 
al. 1991), since they have spectra reminiscent of Galactic H{\sc ii} regions (and were often
discovered because of  this property). Other types  include ``blue amorphous galaxies'' 
(Sandage and Brucato 1979, Gallagher and Hunter 1987). 
Different notation reflects a focus on different physical aspects meaning that the 
classifications do not necessary overlap completely, and this loosely classified 
group may contain objects with different evolutionary  history.
Without arguing that any name is better than the other, we shall henceforth  
use the name BCGs, unless  we want to draw attention to differences.
Since many BCGs have been discovered by means of objective 
prism surveys, which 
are not very sensitive to the properties of the host galaxy, ``ordinary''
 dIs with bright H{\sc ii} regions may ``contaminate'' samples selected in this way. 
Note also that not all galaxies denoted as BCGs in the literature are strictly compact
according to a surface brightness criterion.
BCGs have luminosities in the range $M_B \approx -12$ to $M_B \approx -21$. 
Those BCGs which are less luminous than $M_B \approx -17$ are commonly 
referred to as blue compact dwarfs (BCDs). Interestingly, even some  luminous
BCGs may be very metal-poor. In Fig. 2 we show the appearance of BCGs on the
Digitized Sky Survey, and a LSBG and a normal spiral for comparison. In Fig. 5
and 9 we show high resolution images of the two most metal-poor BCGs, and in 
Fig. 6 a spectrum of one of them (IZw18).

The blue compact galaxies are the class of objects where 
most galaxies with low abundances as derived from H{\sc ii} regions have been 
found. The over-representation of BCGs is probably related to selection effects,
since  high surface brightness and prominent emission line spectra
make it relatively  easy to determine nebular abundances. 
Early studies (e.g. Arp 1965, Zwicky 1966) hinted that BCGs had dramatically
different properties compared to ``normal'' dwarf and giant galaxies.
The wide interest in low metallicity BCGs was triggered by the work
of Sargent and Searle (1970) and Searle and Sargent (1972) where they showed
two BCGs to be metal-poor and forming stars at high rates. They concluded
that either these galaxies were young, now forming their first generation of 
stars, or that star formation occurred in short bursts separated by long
quiescent periods (Searle et al. 1973). Today the latter explanation 
seems the correct one for the majority  of BCGs, if not all.

\begin{figure}[h]
\hspace{1.0cm} \resizebox{5.6cm}{!}{\includegraphics{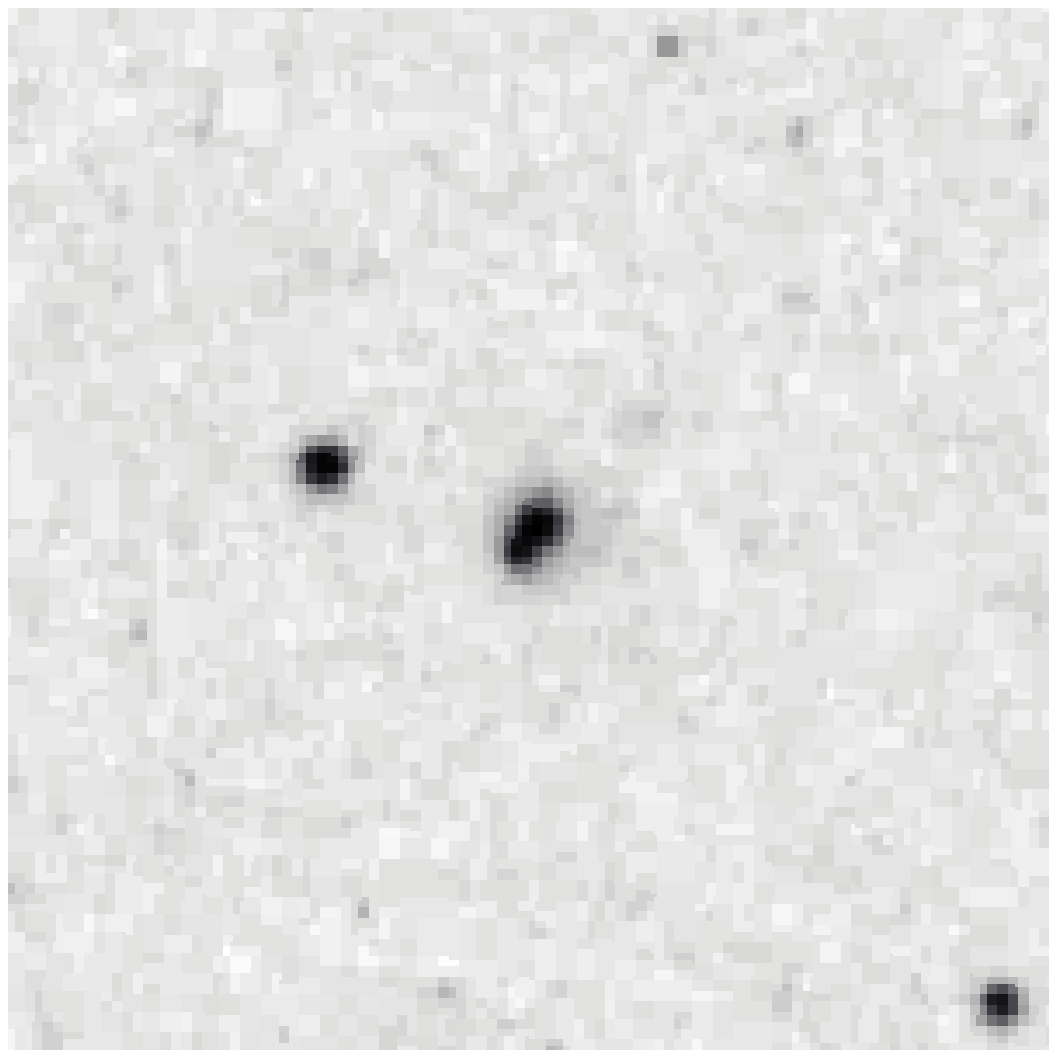}}
\hspace{0.1cm} \resizebox{5.6cm}{!}{\includegraphics{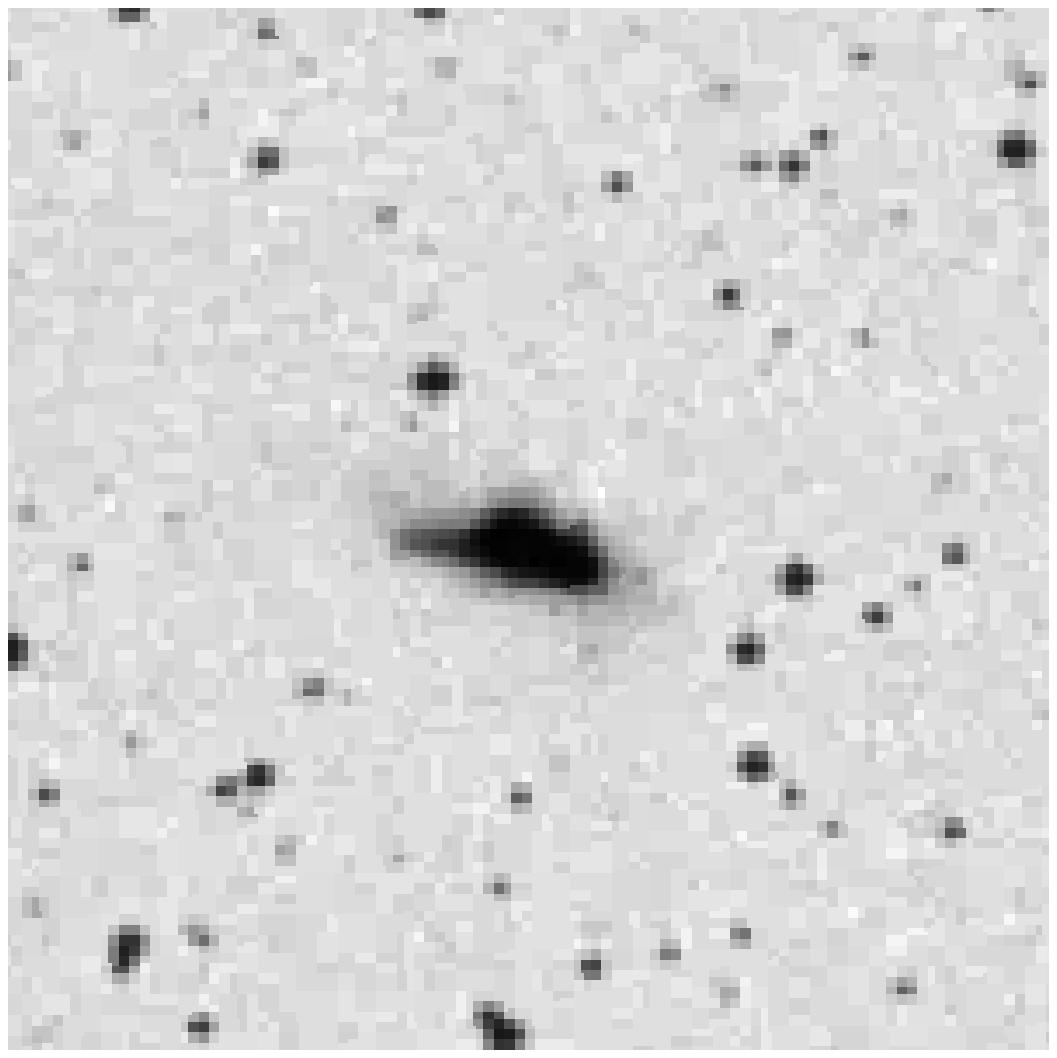}}

\vspace{0.5cm}

\hspace{1.0cm} \resizebox{5.6cm}{!}{\includegraphics{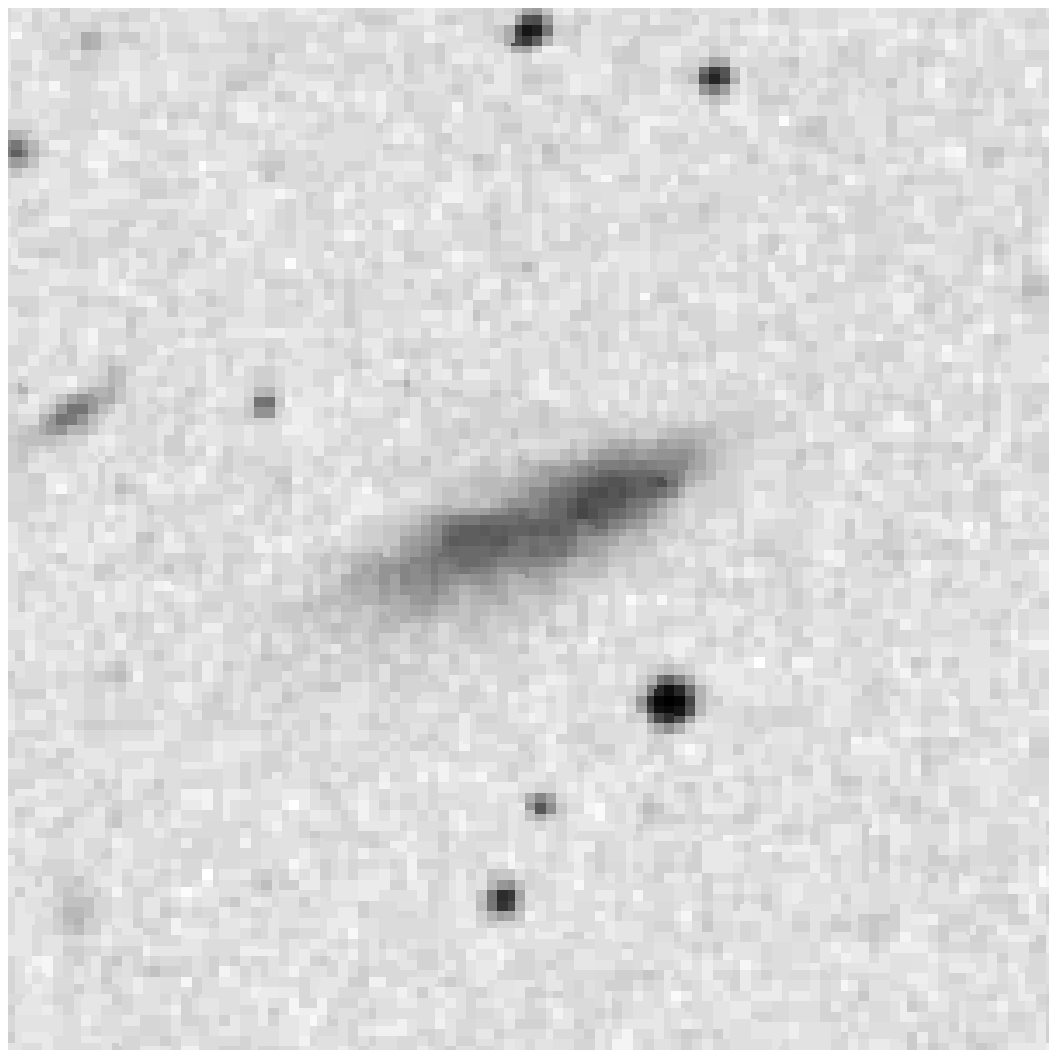}}
\hspace{0.1cm} \resizebox{5.6cm}{!}{\includegraphics{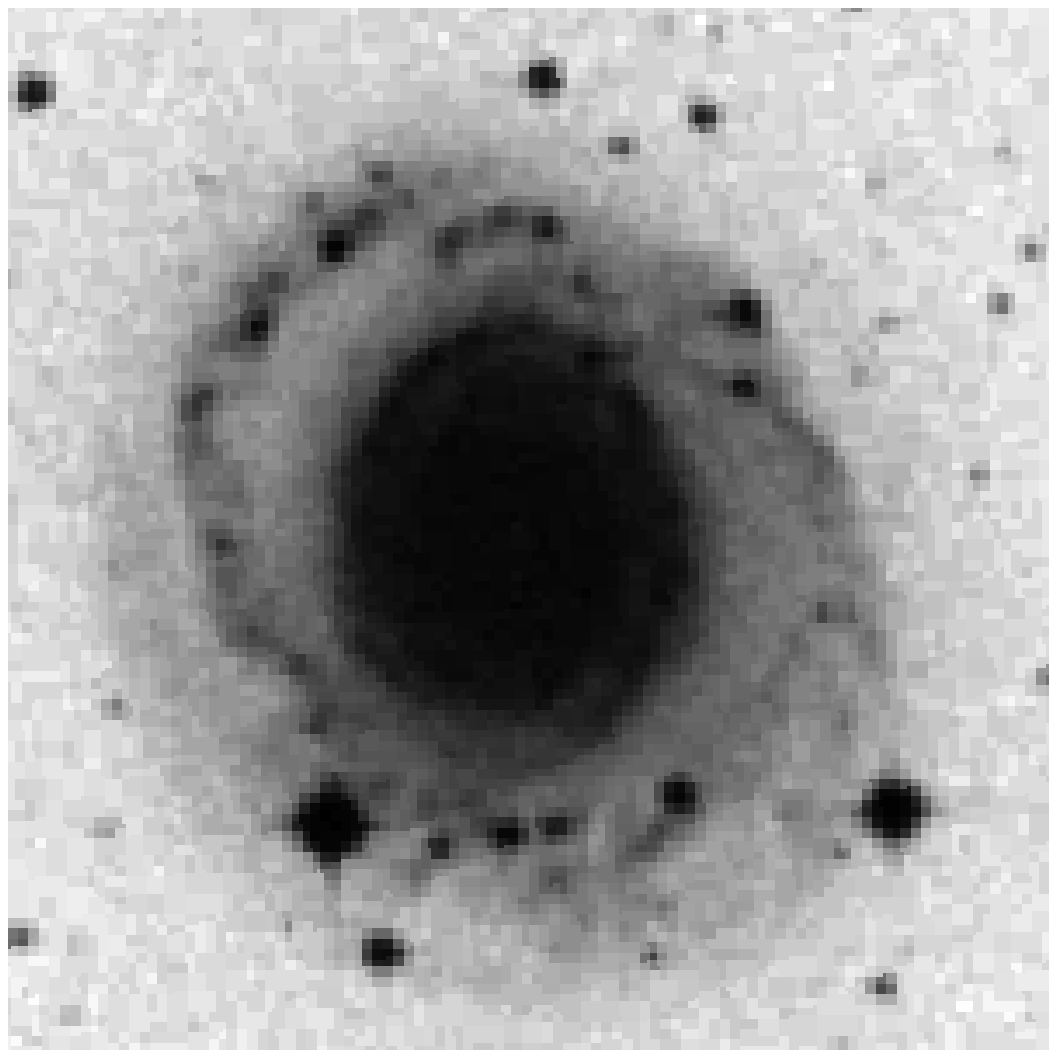}}

\caption{{\protect\small Some galaxies as seen in the Digitized Sky Survey (obtained from {\it Sky View}).
All images have dimension $3 \times 3$ arcminutes, north is up east is left. 
Top left: The ``prototypical'' BCG IZw18, distance $(D) = 10$ Mpc, 
$M_B=-14$, $Z \approx 1/50 Z_{\odot}$. Top right: ESO~338-IG04 (=Tol1924--416), a luminous metal-poor BCG, 
$D=38$ Mpc, $M_B=-19$, $Z \approx 1/10 Z_{\odot}$. Bottom left: ESO~546-34, a metal-poor LSBG, $D=20$ Mpc, $M_B=-16$, $Z\approx 1/20 Z_{\odot}$. Bottom right: The giant ``normal'' spiral NGC~6753 shown for comparison, $D=40$ Mpc, $M_B=-21.5$}
 }
\end{figure}

\subsubsection{Morphology and structure of BCGs}

BCGs are smallish galaxies with high central surface brightness.
The detailed morphology and surface photometry of BCGs have been studied by many  investigators. 
The central morphology
is often irregular due to the presence of active star formation,
but this contains very little information on the extended light distribution which
likely traces the dominant stellar mass.

Loose and Thuan (1986a) 
 defined four subclasses depending on the morphology of the central star forming 
region and the surrounding host galaxy. Kunth et al. (1988) reached the similar conclusion 
that BCGs constituted a ``mixed bag'' of morphologies, including objects that appeared
to be isolated extragalactic H{\sc ii} regions (e.g. Pox186), irregular morphologies, as well as more common cases
with symmetric outer envelopes suggesting the presence of an old population.
Salzer et al. (1989b) classified the emission line galaxies in the UM survey according
to morphology and emission line properties. Telles et al. (1997) divided H{\sc ii} galaxies
into two classes: Type I which have irregular morphology and are more luminous, while
Type II have symmetric and regular outer structure.
The existence of regular haloes, if corresponding to stellar emission, is in itself 
suggesting  fairly high ages since relaxation times are of the order of a few 
times $10^8$ years. Of course the stars may be younger if they formed later on 
in an already relaxed gaseous disc.

Quantitative surface photometry reaching faint isophotal levels began with the development
of CCD detectors. Bergvall (1985) found that ESO~338-IG04 (=Tol1924--416) followed an exponential
like surface brightness distribution in the I-band, suggesting the presence of an
old stellar disc. Loose and Thuan (1986b) on the contrary found Haro2 to follow a
more elliptical like light distribution, with some suggestion of redder colours in 
the halo. Similar results were obtained by Kunth et al. (1988).

The shape of surface brightness profiles in BCGs has been subject to some debate
and both exponential and $r^{1/4}$ laws have  been claimed to best match the data.
Papaderos et al. (1996a) have proposed that  profiles in general can be fit with a 
three-component model, with an exponential light profile in the outer parts. Exponential
outer profiles were also found by Telles and Terlevich (1997). On the other hand 
Doublier et al. (1997) find  $r^{1/4}$ profiles in a substantial fraction of objects, 
while the rest have exponential light profiles. We note however that these studies do 
not compare easily because only a few objects are in common.
For most BCGs the shape of the  profile changes with 
radius, meaning that the  fitting shape will be uncertain and critically
depend on the sensitivity limits. Bergvall and \"Ostlin (1999) go deeper than other
published studies (to levels fainter than $\mu_B=28$ mag/arcsec$^2$), and claim
that $r^{1/4}$ types laws are favoured when using deep red (R or I band) data,
but that discs provide decent fits to the outer parts, and are favoured for B-band 
data.  The shape of the luminosity profiles
depends on how one constructs them, and especially the amount of central excess 
(identified with the  ``starburst'' component) depends  critically on the chosen method
(Marlowe et al. 1997). 
Several investigators have found high underlying surface brightness ($\mu_B 
= 20$ to $23$  mag/arcsec$^2$) and short scale lengths in BCGs as compared to other dwarf 
galaxies (Papaderos et al. 1996a, Telles and Terlevich 1997, Marlowe et al. 1997).
Bergvall and \"Ostlin (1999) find much lower central surface brightness values 
when using deeper data for a sample of luminous BCGs. Thus there might be differences
arising from the different nature of the objects, but also from
different observational methodology.
Now, one can ask  what meaning the shape of the luminosity profile really 
has?  Most dIs and LSBGs, and faint dEs, are well described by exponential like laws,
while $r^{1/4}$ laws are found in ellipticals of high and low (e.g. M32) luminosity. 
Recently, Jerjen and Binggeli (1997) showed that brighter dEs gradually approach
the luminosity profiles of Ellipticals, although they never become as curved as  $r^{1/4}$.
A systematic homogeneous survey at faint isophotal levels, including all known types of 
low luminosity galaxies  would be illuminating and moreover useful in understanding 
relations between dwarfs. For an example, see Fig. 8 where we show a luminosity profile
of IZw18.

In general, the central parts of BCGs contain one or a few star forming knots 
(often found to be composed of many individual bright star clusters), which may 
be identified with the ``starburst''
region. The central knots in most cases give rise to excess surface brightness.
To quantify the strength of the starburst the excess light may be integrated
and compared to the underlying galaxy light. This gives rather modest
starbursts in most BCGs, amounting to a brightening with typically less than
one magnitude in the blue (Marlowe et al. 1999, Papaderos et al. 1996b). Given
that it  must have a low mass to light ratio, this suggests 
that the starburst contains only a minor fraction 
of the integrated stellar mass, and that the subsequent fading in luminosity will
be very moderate. Of course, some of these galaxies may already have passed their SFR peak  
and be in the process of fading. However,  the amount of central
excess depends on how luminosity profiles were constructed and to what depth 
the profiles are fitted, and moreover depends on an a priori assumption of 
 the true shape of the underlying galaxy.
If whatever creates the increased star formation does so, not only in the centre, but
throughout the galaxy in question, the ``burst strength'' will be underestimated.
Colour profiles can here yield useful additional information.

\subsubsection{The age of the underlying population}
To detect underlying old populations it is essential to reach faint isophotal levels,
and moreover to take into account the possible contamination from gas ionised by
UV-photons leaking out from the central starburst. Thuan (1983) attempted to constrain 
the star formation histories by NIR aperture photometry and concluded that  BCGs 
were old. However, this was based on central colours and old models, and the data
do no longer allow for an unambiguous conclusion. Surface
photometry or resolved star photometry would be the preferred method to investigate
ages of BCGs. With modern detectors, it is feasible to make surface photometry
of metal-poor BCGs at faint levels in the near IR. Bergvall and \"Ostlin
(1999) found a very clear signal in $V-J$ of old stars formed on rather short 
timescales in the haloes of some luminous BCGs, while e.g. 
$B-V$ remained inconclusive, illustrating the power of NIR observations.

Loose and Thuan (1986b) and Kunth et al. (1988) found  $B-R$ colours to redden with increasing 
radius, suggesting the presence of underlying old populations.
Optical surface photometry has continued to unveil red haloes in most BCGs studied in detail
(Papaderos et al. 1996,  Doublier et al. 1997,1999; Marlowe et al. 1999) and
Telles and Terlevich (1997) found the underlying colours to be consistent with those
of blue  LSBGs and amorphous galaxies.
However ages and star formation histories are not yet well constrained, since model
predictions  are degenerate. In many cases, the most influential parameter is the assumed shape
of the star formation history which is what one wants to determine ultimately. Depending on how 
deep the halo is to be probed, and what pass band is used,   different
 ages are obtained for the underlying population, indicating that composite stellar populations
are present, or that nebular emission contributes to the colours, or both. For an example of
colour profiles revealing a redder halo, see Fig. 8.

Nevertheless, all these studies demonstrate that the majority of BCGs are not young. Cases were no 
underlying populations have yet been found exist, but as more detailed studies are 
performed, old populations turn up in most young galaxy candidates, as e.g. in the previous 
young galaxy candidates Pox186 (Kunth et al. 1988; Doublier et al. 1999) and ESO~400-G43 
(Bergvall and J\"ors\"ater 1988; Bergvall and \"Ostlin 1999). 

\subsubsection{Ongoing star formation, starbursts and star clusters}

The present star formation process is another important aspect of BCGs, since they offer
fine laboratories for studying vigorous  star formation in metal-poor environments with
relatively small extinction problems. The ongoing star formation has been studied by  
means of spectral synthesis of either integrated colours (Thuan 1983, Bergvall 1985)
or spectral energy distributions (Lequeux 1981, Thuan 1986, Fanelli et al. 1988, etc.) or both. 
This gives information on the current star formation event and in general short bursts are 
found to best match the data (e.g. Mas-Hesse and Kunth 1999). However this does not exclude
longer star formation episodes as the following example illustrates: if star formation 
propagates through a galaxy, but typically takes place in individual luminous short lived 
H{\sc ii} regions, from studying the most luminous H{\sc ii} region one might get the 
illusion to witness a sudden starburst event 
although the average SFR might have been continuous. With this in mind, we can use spectral
synthesis to investigate the duration of individual star-forming events, the initial mass 
function (IMF), the stellar content (e.g. WR stars), and other interesting properties. 

That most BCGs are not necessarily efficient star formers was shown by Sage et al. 
(1992) who investigated the neutral and molecular gas, infrared and optical properties 
of a small sample of BCGs and dIs. Most galaxies were found to be not more efficient than normal
spirals in forming stars, the  star formation efficiencies  being high only when compared
to other dwarfs. However compact star clusters (see below) require high SFRs to be gravitationally bound
and there are indeed BCGs which appear to be very efficient star formers, and true 
dwarf analogues of giant starbursts.

Detailed studies have revealed that star formation in BCGs and dwarf starbursts often take 
place in dense ``super star clusters'' (e.g. NGC1569, Arp and Sandage 1985; NGC1705, Melnick 
et al. 1985a).  Super star clusters (SSCs) are comparable in luminosity to R136 
(the central cluster of 30 Doradus in the LMC), and are
sometimes much more luminous. It has been 
proposed that these may be newly born globular clusters, although it is still quite uncertain 
whether they are massive enough and if they are gravitationally bound.
Conti and Vacca (1994) studied He2-10 and Meurer et al. (1995) nine starbursts galaxies, both using the HST/FOC, to find bright SSCs with UV luminosities 
close to those expected for proto globular clusters. \"Ostlin et al. (1998) 
studied the luminous metal-poor BCG ESO~338-IG04 (=Tol1924--416) with the HST/WFPC2 and found more than hundred
luminous star clusters whose ages and masses were estimated from multicolour photometry.  
It would  be interesting to know to what extent it is a general feature of 
BGCs to reveal young star clusters when studied with enough spatial resolution. Given that
special conditions may be required for SSCs to form, their presence and abundance can yield
important insights into the starburst mechanism.  Similar objects are found in massive 
starbursts (e.g. M82, O'Connell et al. 1995), merging galaxies (e.g. the {\it Antennae}, Whitmore 
\& Schweizer 1995), and in the circum-nuclear regions of giant barred spirals (e.g. Barth 
et al. 1995). Ho and Filipenko (1996) managed to determine the velocity dispersion of 
SSCs in NCG1569 and NGC1705 and showed that they have masses on the order of $10^5 M_{\odot}$,
comparable to old Galactic GCs.

Bound massive star clusters offer an alternative way to probe the SFH in BCGs (and other galaxies). 
Thuan et al. (1996) found old GCs around Mrk996, and \"Ostlin et al. (1998) in addition 
to old ones a rich population of intermediate age ($\sim$ 2.5 Gyr) GCs in ESO~338-IG04, 
revealing a former starburst event. In view of the many young SSCs at least a fraction could 
survive to become GCs.  
Moreover, young SSCs may be used to investigate the recent evolution of the starburst because 
they represent true ``simple stellar populations'' (coeval on a time scale of around one 
million year) and should be chemically homogeneous. In the case of ESO~338-IG04 it is clear 
from the age distribution of young SSCs that the present burst has been active for at least 
30 Myrs (\"Ostlin et al. 1998, 1999c).

\subsubsection{The Wolf-Rayet galaxies}
Wolf-Rayet (WR) stars are considered to be highly evolved descendants of the
most massive O-stars. They are extreme Population~I stars and have spectra
characterised by broad emission lines resulting from dense, high-velocity
winds. These stars are detectable in external galaxies by their prominent
emission lines at around 4650-4690~\AA\ (the ``Wolf-Rayet bump''). This
bump has been detected in many emission line galaxies (Allen et al. 1976,
Kunth \& Sargent 1981, Kunth \& Joubert 1985, Conti 1991,
  Vacca \& Conti
1992; see Schaerer et al. 1999 for the latest updated catalogue), providing a new insight on
the process of massive star formation in metal-poor galaxies.
 Arnault et al. (1989),
Cervi\~no \& Mas-Hesse (1994), Meynet (1995) and Schaerer \& Vacca (1998)
have discussed the dependence of the WR bump strength on the parameters
that define the star-forming episodes (metallicity, age, IMF
slope, etc.). The two most interesting properties of the Wolf-Rayet bump is its
strong dependence on metallicity and the constraints it can impose on the
age of the cluster. Since the WR phenomenon is tightly coupled to the
generation of strong stellar winds, its incidence decreases significantly
with decreasing metallicity, so that at $Z=1/20~ Z_{\odot}$  only very massive stars 
(initial
mass $>$ 80 {M$_\odot$} ) might become WR stars. This small mass range implies
 that the
detection of the WR bump in low metallicity galaxies can provide important information
on the upper mass limit of the IMF. 

The relative population of WR to O stars is usually measured through the
{$L(WR)/L(H\beta)$} ratio, the luminosity of the WR-bump over the $H\beta$-luminosity. 
To compare with model predictions it is  
necessary to 
integrate over the whole ionised region which poses some observational technical problems. 
 The measurements of this ratio might also be strongly affected by
differential extinction. Since $L(WR)$ is of stellar origin, it
should be affected by the same extinction as the stellar continuum. On the
other hand, $L(H\beta)$ is of nebular origin and might suffer from
a larger amount of extinction. Ignoring this effect may lead to a
significant overestimation of the {$L(WR)/L(H\beta)$} ratio. Schmutz and Vacca (1999)
 have 
questioned the use of the 4640\AA\ emission feature which may not entirely be due to WR stars
but to large numbers of O stars or contamination from other nebular lines.
Observations show that very short bursts are compatible in general with a
Salpeter IMF and a large upper mass cut-off.
Recent results by Cervi\~no (1998) show that if a
significant fraction of massive stars are formed in binary systems, mass
transfer episodes can lead to the formation of WR stars during longer
periods of time than predicted by models based on the evolution of single
stars alone. Therefore, age calibrations through  WR features has to be 
taken with caution. 

Schaerer (1996) has evaluated the effect of the evolution of the WR stars
population on the He{\sc ii} narrow emission line at 4686~\AA , by combining
model atmospheres accounting for stellar winds with evolutionary tracks.
He concludes that for metallicities in the range
 $Z=1/5 ~ Z_{\odot}$ to $Z_{\odot}$, a
strong nebular He{\sc ii} emission line should  originate
 in early WR phases
 when WC stars begin to appear. The He{\sc ii} emission line is
indeed detected in a few objects with very young stellar populations,
below 3~Myr, and therefore starting to produce WR stars (NGC~2363 and
Mrk~36), in good agreement with the scenario proposed by Schaerer.
This is also supported by the detection of WR stars in IZw18 (that came as a surprise
 in view of its very low metallicity) (see Fig 7; Legrand et al.,
1997b, Izotov et al. 1997a). A last argument is the spatial distribution
of the nebular HeII lines that follows that of the WR features (Legrand et
al. 1997b, Maiz-Apellaniz et al. 1998, and De  Mello et al. 1998). 
Other broad emission lines from C{\sc iv} at 5805\AA\ 
 originating from WC stars (representing more evolved phases than WN stars)
 are currently observed (Schaerer et al. 1997,1999). 
The observations of WR features in low metallicity objects is indeed a
challenge to our understanding of the WR phases and forces to reassess  the metallicity
 dependence of stellar winds, the binary channel for WR production and the effect of 
rotation onto the evolution of massive stars (Maeder and Meynet, private communication).

\subsubsection{Gas content and dynamics of BCGs}

The kinematics of galaxies may be investigated by means of the motion of their stars or
gas. In BCGs the former is in general not possible, due to the absence of strong absorption 
lines, and studies have been restricted to the gas phases. 

Chamaraux et al. (1970) were the first to detect neutral hydrogen in a BCG, namely
IIZw40, and more BCGs were detected in various surveys (not only targeting BCGs) :
 Lauqu\'e (1973), Bottinelli et al. (1973, 1975), Chamaraux (1977). 
Gordon and Gottesman (1981) and Thuan and Martin (1981) conducted the first extensive 
systematic surveys of neutral hydrogen in BCGs, and in total more than 200 BCGs, 
predominantly in the northern hemisphere, were observed and the majority detected. 
BCGs have been found to be gas rich. 
The ratio of neutral hydrogen mass to integrated blue luminosity is typically in the 
range $0.1 \le M_{HI}/L_B \le 1.0$ (in solar units, Thuan and Martin 1981, Gordon and 
Gottesman 1980). Thus BCGs are as gas rich as spirals and dIs, but less than LSBGs 
(Staveley-Smith et al. 1992).

HI interferometry of IIZw40 was reported by Gottesman and Weliachew in 1972, providing
the first spatially resolved investigation of H{\sc i} in a BCG. This was followed by studies
of IIZw70+71 by Balkowski et al. (1978) and  IZw18 by Lequeux and Viallefond (1980) and
Viallefond et al. (1987) who suggested the existence of dark matter (perhaps molecules)
to account for the dynamical mass.
Bergvall and J\"ors\"ater (1988) provided a detailed mass model for  ESO~400-G43 and also
found evidence for dark matter dominated dynamics in agreement with recent findings by 
Meurer et al. (1996, 1998) and  van Zee et al. (1998c).
A substantial fraction of BCGs seems to have H{\sc i} companions (Taylor 1997),  probably 
representing uncatalogued gas rich LSBGs. 

There is a current view suggesting that the formation of stars depends
primarily on the amount of molecular gas. However the situation in low
metallicity gas  is still under debate. Many attempts to detect CO in BCGs
 galaxies have been reported so far (Combes 1986, Young et al. 1986, Arnault et al. 
1988, Sage et al. 1992, Israel et al. 1995, Gondhalekar et al. 1998) but  the
 CO luminosity  of  BCGs  is very
 low, in comparison with their observed star formation rate, mostly yielding only upper limits. This lack of
 detection may be because
 the low metallicity of BCGs  hides the true molecular phase by 
a low CO to H$_2$ conversion factor. However other explanations may be invoked 
as well:
the CO excitation could be lower than for molecular clouds in our Galaxy, or
the molecular clouds in BCGs could be UV-photodissociated as a result of high star formation rates.
Gondhalekar et al. (1998) conclude that
 the CO luminosity correlates rather weakly with the FIR luminosity, i.e. FIR
 luminosity may not be a good tracer of molecular gas. 
 Obviously the lack of CO detection does not preclude the
presence of H$_2$ molecules in these gas-rich galaxies.
 In fact there are mechanisms by
which molecular hydrogen can be formed in absence of grains from hydrogen atoms
 in gaseous phase via a reaction involving
 negative hydrogen ions (Lequeux and Viallefond 1980).
It is therefore possible that in  star-forming galaxies with well localised
 massive star formation surrounded by huge H{\sc i} gaseous envelopes that the
 molecular hydrogen is abundant and makes up a significant fraction of the dark
 matter dynamically detected (Lequeux and Viallefond 1980,
 Lo et al. 1993). New capabilities such as the FUSE mission
will offer  the unique possibility to detect for the 
first time cold H$_2$ in absorption against the stellar continuum of blue massive
 stellar clusters. IZw18 will be the first target to be  searched for H$_2$.

\"Ostlin et al. (1999a,b) investigated the H$\alpha$ velocity fields of a sample of
luminous BCGs utilising scanning Fabry-Perot interferometry. The velocity fields were 
found to be complex, and in many cases showed evidence for dynamically distinct components,
e.g. counter rotating features. Their analysis suggests that mergers involving gas rich dwarfs
are the best explanations for the starbursts in these systems. Masses were modelled
both dynamically and photometrically, and some galaxies showed apparent rotational mass
deficiencies which could be explained if the studied BCGs are not primarily supported by 
rotation,  if stars and gas are dynamically decoupled
(e.g. due to gas flows) or if the galaxies are not in dynamical equilibrium.
There are also indications that the width of emission lines in BCGs is related to virial
motions and may provide dynamical mass estimates (see Sect. 7.1, and Melnick et al. 1987).

Flows in the ionised gas have been detected in several BCGs (Marlowe et al. 1995; Martin 
1996, 1998; Meurer et al. 1997), and suggested by X-ray observations in VIIZw403 (Papaderos et al. 1994). 
Flows have also been found from studies of the Ly$\alpha$ emission.
 Although  Ly$\alpha$ emission in starbursts is expected to be strong, it turns out that dust 
is very effective  in suppressing this line because the effects of resonant scattering in 
a gas-rich medium  dramatically reduce the effective mean free path of the  Ly$\alpha$ photons. 
On the other hand this mechanism does not explain why in
 many galaxies with little dust content such as IZw18 (Kunth et al. 1994)  Ly$\alpha$ 
is seen in absorption whereas in dustier ones such as Haro2 (Lequeux et al. 1995) 
the line is seen in emission but with a clear P-Cygni profile. Kunth et al. (1998) found that the 
strength of Ly$\alpha$ emission is in fact only weakly correlated with metallicity and suggested that 
the dynamical state of ISM is also a major regulating mechanism. A new model explains Ly$\alpha$ 
profiles in starburst galaxies by the hydrodynamics of superbubbles powered by massive stars 
(Tenorio-Tagle et al. 1999).

\subsubsection{Starburst triggers in BCGs}

Searle and Sargent (1972) had already  speculated on the reasons for a sudden increase 
of the star formation rate in BCGs, followed by long quiescent phases. The  regulation of
their star formation histories in general, is of course a very important question, intimately 
linking to the physical relations between BCGs and other dwarfs, which will be addressed
in Sect. 7.2. Moreover, given the heterogeneity of BCGs as a class  the possibility that 
different mechanisms operate is not excluded. The star formation rate in BCGs (even
the mass averaged value, cf. 3.2) varies considerably, and some are rather star forming dwarfs
than starburst ones. Given the increased central H{\sc i} densities (van Zee et al. 1998c) it
is evident that some mechanism that can concentrate gas in the centres is needed.

A  scenario with statistical fluctuations proposed by Searle et al. (1973), 
has been further elaborated by
Gerola et al. (1980). The basic ingredients are  a positive feedback of star formation which 
in combination with the small masses of BCGs give rise to large fluctuations in the SFRs.
A popular explanation is that supernova driven winds halt star formation by expelling the 
gas. Later, the lost gas might accrete back on the galaxy and create a new starburst 
(Tayler 1975, Dekel and Silk 1986, Silk et al. 1987,  Babul and Rees 1992). A problem
with models where the precursor accretes gas continuously is the time scale for onset
of rapid star formation, -- why would the galaxy wait to form stars for a long time, then 
suddenly (on time scales of the order of $10^7$ years) burst into an unsustainable star formation rate?

It is well known that mergers between giant galaxies can produce impressive starbursts 
(Sanders et al. 1988), and it is possible that mergers  are key mechanism in building
up the galaxy population and regulating starburst activity (Lacey et al. 1993). Therefore
this possibility is attractive also for explaining dwarf starbursts, i.e. BCGs. It is 
evident that there are at least a few BCGs that seem to be interacting/merging in some form 
(e.g. IIZw40, He2-10, and galaxies in \"Ostlin et al. 1999a,b). Moreover, many of these 
BCGs show Super Star Clusters (SSCs) of the same kind as those found in giant mergers.

On the other hand several studies indicate that many BCGs are pretty isolated (cf. Sect. 6.3).
However there might be low surface brightness or pure H{\sc i} companions missing in
present catalogues. A related question is whether tidal interactions are strong enough to
ignite bursts in dwarfs that, like BCGs (Salzer 1989) are not accompanied by giant galaxies. 
First, tidal forces between dwarfs are too modest to trigger radial gas flows unless the galaxies are 
almost in contact (Campos-Aguilar et al. 1993). Moreover it is uncertain whether dwarfs are
at all unstable against tidal perturbations (Mihos et al. 1997). Perhaps a direct contact is
needed to trigger dwarf starburst explaining why pairs are rare since one would only pick up the 
burst once the merger is already in progress. While mergers would be sufficient for producing BCGs
it is uncertain how common they are. The environments of isolated BCGs should be studied in more 
detail for this purpose.

\subsubsection{The chemical abundances of BCGs}

\begin{figure}
\vspace{-1cm}
\hspace{1cm}
\resizebox{10.0cm}{!}{\includegraphics{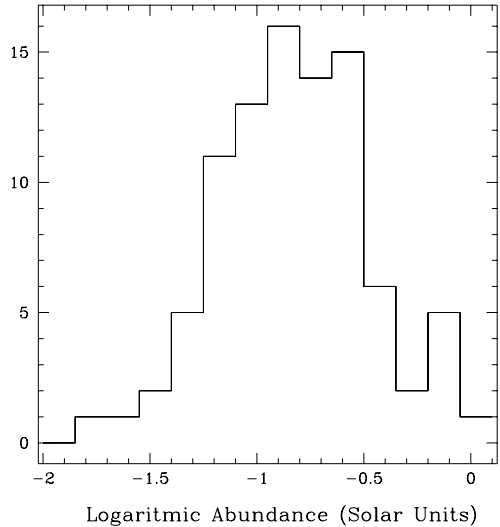}}
\caption{{\protect\small The metallicity distribution of BCGs/H{\sc ii}-galaxies from Terlevich et al. 
(1991), as measured by their oxygen abundances. Note the peak at $\sim$1/10:th $Z_{\odot}$, 
which at least partly may be due to selection effects. (Courtesy R. Terlevich)}}
\end{figure}

Blue compact  galaxies (BCGs) include  galaxies with the lowest 
measured abundances as derived from H{\sc ii} regions.
Since both BCGs and dIs in general possess bright H{\sc ii} regions, many
investigations of nebular abundances deal with objects of both classes,
which in any case are not always very distinct. Very little is known about the
metallicity of the stellar population of BCGs. Schulte-Ladbeck et al. (1999),
using a colour magnitude diagram,
derive  [Fe/H]$=-1.92$ for the old population in  VIIZw403, one of the most nearby BCGs,
which has a previously reported oxygen abundance of 12+log(O/H) $\approx 7.7$
(Tully et al. 1981, Izotov and Thuan 1999); it is included in Fig. 1.
Mas-Hesse and Kunth (1999), from the correlation between the strength
of the stellar absorption lines from massive stars (C{\sc iv} and Si{\sc iv}) and
O/H as obtained from the nebular lines, argue that the metallicity
of the young stellar populations is similar to the
gas metallicity.

The metallicity of BCGs was first addressed by  Searle and Sargent (1972),
who showed that the abundances of oxygen and neon in I~Zw18 and II~Zw40
were sub-solar. Their work has been followed by numerous investigations over
the years which have established their nature as metal-poor
galaxies, as regards oxygen: Alloin et al. (1978),
Lequeux et al. (1979), French (1980), Kinman and Davidson (1981),
Kunth and Sargent (1983), Kunth and Joubert (1985), Campbell et al. (1986), 
Izotov et al. (1991), Pe\~na et al. (1991), Pagel et al. (1992), Gallego et al. (1997) and others. 
A recent  investigation of a large sample of 80 H{\sc ii}-galaxies  can be found 
in Masegosa et al. (1994). Izotov and Thuan (1999) present O, N, Ne, S, Ar and Fe 
abundances for 50 BCGs, and in addition C and Si
abundances for 7 of these, thereby constituting the largest homogeneous high
quality source of information for  metal-poor BCGs.

The abundances of heavy elements in these objects range between  $1/2 ~Z_{\odot}$
and $1/50~Z_{\odot}$,
making them among the least chemically evolved objects in the universe. Figure 3
shows the distribution for oxygen abundance among H{\sc ii} galaxies. Oxygen is assumed
to be representative of the total metallicity of the entire galaxy, although claims have 
been reported that this may only concern the H{\sc ii} region because of incomplete 
mixing (Kunth and Sargent 1986, Kunth et al. 1994, Roy and Kunth 1995, see Sect. 3.5).
Figure 3 reveals that the oxygen abundance in this class of objects peaks slightly
above  1/10 of the solar value. This peak may be related to  selection effects 
since many surveys have selected galaxies with strong forbidden oxygen
emission lines, which happen to be strongest for an oxygen abundance of around
ten percent of the solar value. There is indeed another strong selection bias in 
Fig. 3 because it includes only galaxies with "measurable" electron
temperatures (through the use of [O{\sc iii}]$\lambda$4363) hence restricts the sample to abundances lower
than 12+log[O/H] $\approx  8.5$. The apparent low abundance 
cut-off is bounded by IZw18 (Sect. 5). The utter lack of known
galaxies with abundances smaller than  IZw18, despite concentrated observational 
efforts (see Sect. 6) has been a puzzle. Kunth and Sargent (1986) 
suggested that IZw18 could indeed be a primordial galaxy in which the observed 
H{\sc ii} regions have been self-enriched in the current burst. This idea has been tested
by several distinct approaches, but we will see (Sect. 5.1.2) that there are many indications
that IZw18 is in fact old.

According to Izotov and Thuan (1999), nine BCGs more metal-poor than 1/20
of solar exists. These are: IZw18, SBS~0335-052, SBS~0940+544, SBS~1159+545,
UGC~4483, CG~389 (Izotov and Thuan 1999); CG~1116+51 (French 1980); Tololo~65  and
Tololo~1214-277 (Pagel 1992). However another half dozen examples can be found 
in the litterature, with the most metal poor being UM~382 12+log(O/H)$=7.45$ (Masegosa et al. 1994). 
In Table 3 we list all known BCGs with oxygen abundances $\sim$1/20 solar and below.
As we discuss in Sect. 6.1.5 the number of very metal-poor galaxies will probably
increase significantly in the near future.
In total there are more than a dozen very metal-poor ($Z \le 1/20 Z_{\odot}$) BCGs 
known, while the number of BCGs with $Z \le 1/10 Z_{\odot}$ is several times larger.

\subsubsection{Abundance ratios in BCGs}

Oxygen is normally considered as representative of the metallicity of BCGs. However H{\sc ii} region 
abundance  analysis can also provide abundances of other elements. 
Especially nitrogen, helium and carbon have been investigated.
In addition  $\alpha$ elements such as argon, neon and sulphur may be studied.
Lately, iron has been added to the list. The study of helium in BCGs offers a route towards
determining the primordial He abundances, and will be discussed separately in Sect. 8.1.
In Fig. 4 we show C/O and N/O vs. O/H for BCGs.

\begin{figure}
\vspace{-1cm}
\hspace{1cm}
\resizebox{12.0cm}{!}{\includegraphics{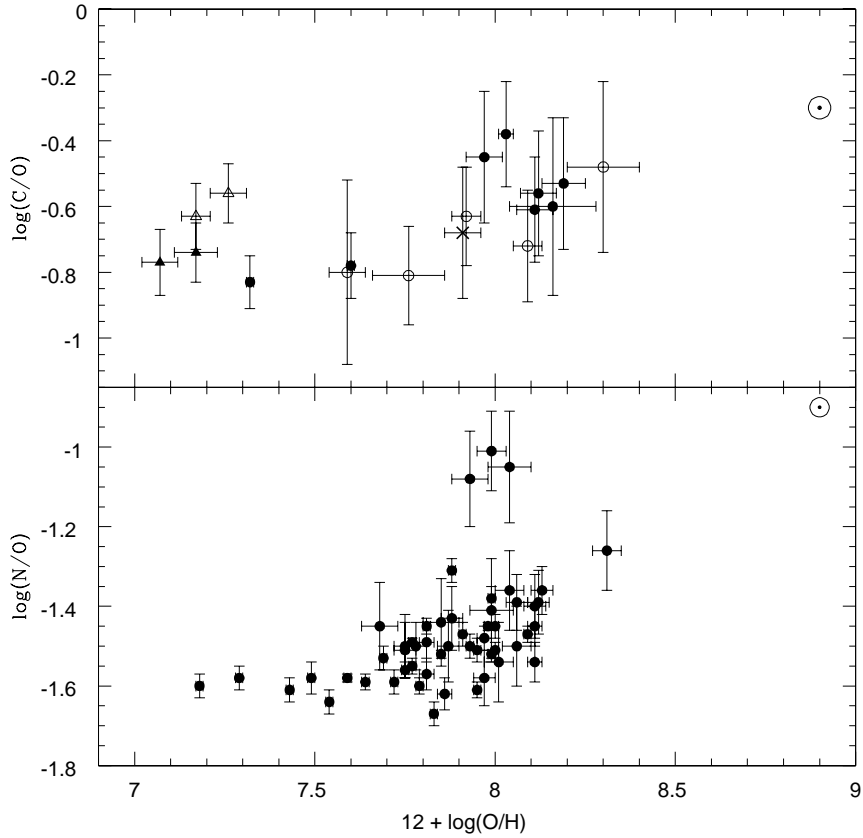}}
\caption{{\protect\small Element ratios in BCG and dIs: 
{\bf Top:} The relation between oxygen abundance and C/O. Filled symbols are from 
Izotov and Thuan 1999, which includes a reanalysis of previously published data.
The open circles are from Garnett et al. (1995). The cross is ESO~338-IG04 
(Bergvall 1986, Masegosa et al. 1994). The open triangles show the location of the 
NW and SE regions in IZw18 from Garnett et al. (1997), while the filled triangles
show the same regions as derived by Izotov and Thuan (1999).
{\bf Bottom:} The relation between oxygen abundance and N/O for BCGs, data taken 
from Izotov and Thuan (1999). 
 }}
\end{figure}

The investigation of carbon abundances in the H{\sc ii} gas poses 
some difficulties since there are no strong emission lines in the optical
regions. The investigation of carbon abundance in BCGs began with the International
Ultraviolet Explorer (IUE)
satellite and has continued with the HST. Garnett et al. (1995) presents
C/O ratios for seven galaxies, including some BCGs, and three others are
presented by Kobulnicky and Skillman (1998). Thus carbon abundances
are still not very well explored in BCGs. Garnett et al. (1995)  found that C/O
 increases with increasing oxygen abundance. The average value of this
 ratio is rather low, as compared to solar, except  possibly for IZw18 which has C/O  about a factor of two 
larger  than predicted from stellar nucleosynthesis (Garnett et al. 1997).

The relative abundance of nitrogen to oxygen increases with O/H (Pagel and Edmunds 1981,
Serrano and Peimbert 1983, Torres-Peimbert et al. 1989), implying a secondary origin 
of N in the CNO cycle. Such a behaviour was not seen at very low O/H (Lequeux
 et al. 1979; Kunth and Sargent 1983; Campbell et al. 1986) indicating
 that nitrogen is mainly a primary element in very metal-poor  gas. 
The current interpretation of this behaviour from stellar nucleosynthesis
 models is that intermediate stars produce primary nitrogen by hot-bottom
 burning. In such a phase, the third dredge-up brings carbon-rich material from
 the core onto the hydrogen burning shell (Renzini and Voli 1981;
 van der Hoek and Groenewegen 1997). The scatter of the N/O  versus O/H
 diagram has been considered as larger than the observational uncertainties 
(although they were nearly comparable two decades ago). Time delays between 
the production of oxygen due to massive stars and that of nitrogen
 is likely part of the explanation although this point of view has been
challenged by recent data from Izotov and Thuan (1999). 
Indeed their high 
 signal to noise observations  not only suggest a small intrinsic 
dispersion of log N/O ($\pm 0.02$ dex) at low metallicities but a similar behaviour 
is found for C/O and other ratios, see Fig. 4. The disagreement with Garnett et al. (1995,1997) comes mainly 
from the reassessment of  C/O  in  IZw18, the
abundances of which are thoroughly discussed in Sect. 5.1.1.
Izotov and Thuan (1999) find positive correlations between C/O and N/O
with O/H but for 12+log(O/H) $ \le 7.6$,  
C/O and N/O remain constant and independent of O/H. They  conclude that
galaxies with such low abundances are genuinely young (less than 40 Myr old), 
now making their first generation of stars. Moreover they suggest that 
all galaxies with 7.6 $\le$ 12+log(O/H) $\le$ 8.2 have ages from 100 to 500 Myr.
Thus, the question raised by Searle and Sargent almost
30 years ago would after all have a positive answer. However,
there are independent data suggesting that these galaxies do in fact contain 
old stars (see Sect. 5). Moreover, there are definitely many BCGs with  
$12+ \log({\rm O/H}) < 8.2 $ which have been convincingly shown to be much older than
500 Myr, e.g. ESO~338-IG04  from its globular 
clusters (\"Ostlin et al. 1998). Moreover, as we shall discuss below, there are alternative interpretations 
 of the abundance patterns which do not require the galaxies to be young.

H{\sc ii} regions in the outskirts of spiral galaxies may have C/O values
 as low as those of the most metal-poor galaxies, and 
H{\sc ii} regions in spiral galaxies follow the same C/O vs. O/H relation as 
dwarf galaxies (Garnett et al. 1999). This suggests that they evolve
chemically in the same manner.  Now, the discs of spiral galaxies are
several Gyr old, still the C/O ratio is as low as in the most metal-poor BCGs,
clearly indicating that C/O is not  simply a function of age.
The observed trend of C/O vs. O/H  could equally well be 
explained by a metallicity dependent yield (Maeder 1992).
Gustafsson et al. (1999) studied the carbon abundances of disc stars in
our Galaxy and concluded that the observed relation could be explained if 
carbon production occurs mainly in massive WR(WC)  stars. In this scenario,
C/O would be mainly a function of metallicity and not age.

A similar pattern is seen for N/O observations of H{\sc ii} regions in spirals.
Outlying H{\sc ii} regions appear to have N/O similar to  the most metal-poor
galaxies (van Zee et al. 1998b). Moreover, the low surface brightness galaxies
studied by R\"onnback \& Bergvall (1995) and  which are   fairly old systems 
(Bergvall et al. 1999), have N/O comparable to those of the most metal-poor BCGs.   
Pilyugin (1999) finds that if significant N production occurs in intermediate mass
stars, and the heavy element abundances have not been polluted by the
present star formation event (i.e. the time scale for cooling of fresh
metals is longer than the typical lifetime of a giant H{\sc ii} region) the 
constant N/O found at low metallicity is consistent with the presence
of previous starbursts, i.e. high ages. It is also worth commenting that
if the time scale for recycling is longer than the duration of a typical
burst of star formation, this can explain the lack of abundance gradients 
in dwarfs (Sect. 3.5). 

The elements Ne, Si, S and Ar  all shows a constant abundance relative to oxygen, 
independent of O/H as expected from stellar nucleosynthesis, since they are products 
of $\alpha$-processes (Izotov \& Thuan 1999).
Finally the Fe/O abundance ratio in BCGs is on average 2.5 times smaller than 
in the Sun with a mean [O/Fe]$  = 0.40 \pm 0.14$ with no dependence on oxygen 
abundance
(Izotov and Thuan 1999). The scatter is surprisingly small  considering the
short time scale for the production  of oxygen as compared to iron
production because different stellar masses are involved. If real,
 it would imply
that Fe could have been produced by explosive nucleosynthesis of SNe type II for 
both O and Fe at the early stage of chemically unevolved galaxies.

\subsubsection{Summary on BCG metallicities}

In summary we find that the study of heavy elements in BCGs shows that these 
systems are chemically unevolved but does not allow to infer a young age 
in terms of galaxy formation. In the range 7.1 $<$ 12 + log(O/H) $<$ 8.3 more than 100 objects
have good quality data. Alpha-elements (Ne, Si, S and Ar) have abundances relative
to  oxygen that show no dependence on oxygen abundance and are  close to 
solar values and similar to that in halo stars and distant galaxies. At low metallicities,
C/O is constant, independently of the oxygen abundance but more metal deficient galaxies should
be observed to confirm the presence or absence of a trend. The behaviour of N/O indicates a primary 
origin as anticipated already by several investigators (see Matteucci 1996, and 
references therein). The 
conclusion that N, C, Fe and O are produced only by massive stars in the most metal-poor 
systems (Izotov and Thuan 1999) needs to be checked by independent observers with larger samples. 
There is a possibility that metals observed in the most metal-poor galaxies originate 
from previous population III star enrichments, previous bursts or  continuous 
star formation at very low rate.  In this case the minimum metallicity should be 
increasing with time,   consistent with quasar absorption line data 
(Lu et al. 1996), although admittedly the connection to dwarfs is not clear.
The question of young galaxies has to be further addressed before it can be
settled, but as we have seen, most BCGs (including objects with less than 1/10 solar 
metallicity) are definitely not young.  Even the best candidates I~Zw18 and 
SBS~0335-052 seems to be old although Izotov, Thuan and their collaborators still dispute this.

While abundances and abundance ratios are the footprints of the past
chemical evolution, many factors need to be taken into account to unveil
the enrichment history and chemistry alone is of limited value in constraining
the history of galaxies. Still, the set of data presented by Izotov \& Thuan (1999) is 
probably unique with respect to data quality, homogeneity and the large number of 
atomic species included, making it a most valuable tool.

\subsection{Tidal Dwarfs}

Zwicky (1956)  proposed that dwarf galaxies could be formed from 
debris when large galaxies collide, an idea that was further addressed
by Schweizer (1978). Though not a real morphological class of its own, this group of galaxies 
have a common history which is likely to have a considerable impact on
their chemical composition. If dwarfs can form from material originating in 
giant galaxies we would expect them to be comparatively metal rich for 
a given luminosity. Some recent progress in this field (Mirabel et al. 1992,
Duc \& Mirabel 1994, 1998) suggests that, even though the final fate of 
dwarf galaxoids forming in mergers is uncertain, they have metallicities
that are systematically higher than in other dwarfs of similar luminosity.
Tidal dwarfs thus appear to be a fundamentally distinct class of galaxies.
Moreover, their future evolution is uncertain: are they stable, do they contain dark matter,
how will they be affected by future interactions with their parent galaxy?

In rich clusters, where interactions should have been frequent, explaining the 
presence of cD galaxies and the overabundance of ellipticals, we would expect
a larger fraction of tidal dwarfs. If they can survive and escape their progenitor, 
dwarfs in clusters must be more metal rich than field dwarfs. There
are preliminary indications that this may be the case (as well as indications 
that it is not, cf. Sect. 4.1.1.). Of course,  galaxies may simply evolve faster in high 
density environments. Interactions
with other galaxies and the intercluster medium  may stimulate star formation in dwarfs and
make the chemical evolution clock run faster. 

In view of their chemical abundances,
it is not likely that the local dwarfs have formed this way.
But what if this process occurred early in the
Universe, at high redshifts when the parent galaxies were not yet significantly 
enriched?

%%%%%%%%%%%%%%%%%%%%%%%%%%%%%%%%%%%%%%%%%%%%%%%%%%%%%%%%%%%%%%%%%%%%%%%%%%%%%%%%%%%%%%%%
%%%%%%%%%%%%%%%%%%%%%%%%%%%%%%%%%%%%%%%%%%%%%%%%%%%%%%%%%%%%%%%%%%%%%%%%%%%%%%%%%%%%%%%%
%%%%%%%%%%%%%%%%%%%%%%%%%%%%%%%%%%%%%%%%%%%%%%%%%%%%%%%%%%%%%%%%%%%%%%%%%%%%%%%%%%%%%%%%

\np

\section{Individual bona fide metal-poor galaxies}

\subsection{IZw18}

The blue compact galaxy IZw18 (also known as Markarian 116) was first described by Zwicky 
in 1966.  IZw18 is the galaxy with the 
lowest known metallicity as derived from the ionised gaseous component. Its oxygen 
abundance is only $\sim$1/50 of that of the Sun. Thus, often being referred to as 
``the most metal-poor galaxy known'', it is still two orders of magnitude more 
metal rich than the most metal-poor stars in the Milky way. 
It is intriguing that while IZw18 was the first BCG (together with II~Zw40) in which 
ionised gas abundances were investigated (Searle \& Sargent 1972), it remains  still the most 
metal-poor BCG known,  despite large efforts in searching for more metal-poor ones.

Neutral hydrogen in IZw18 was detected by Chamaraux (1977), and further investigated 
using  aperture synthesis by Lequeux and Viallefond (1980) who derived $M_{\rm H{\sc i}} = 
7 \cdot10^7 M_{\odot}$. Viallefond et al. (1987) mapped the galaxy in H{\sc i} with VLA and from
 the velocity field inferred a mass  $M_{dyn} \approx 9 \cdot10^8 M_{\odot}$.
Van Zee et al. (1998a) made a high resolution VLA study of IZw18 which revealed a complex
HI morphology and velocity field. A complex velocity field was also found in the ionised 
component (Martin 1996, Petrosian et al. 1996).
Molecular gas has not been detected, not surprising given the low metallicity.
Moreover, low extinction is reported in most studies  and the galaxy is not detected 
by the InfraRed Astronomical Satellite IRAS (IZw18 has not been observed with the 
Infrared Satellite Observatory, ISO), indicating a low dust content.
Deep spectra revealed that IZw18 contains Wolf-Rayet stars (Fig. 7; Legrand et al.
1997b; Izotov et al. 1997a) showing that such can exist even at very low metallicities. HST 
imaging in the optical (Hunter and Thronson 1995, Dufour et al. 1996) and near infrared (\"Ostlin 
1999a) resolves the galaxy into individual luminous stars, and shows that the star formation 
has been active for at least 30 Myrs, with indications of an even older stellar population 
(see below). In Fig. 5 we show  H$\alpha$ and  V-band images, and in Fig. 6 
an optical spectrum of IZw18.

\begin{figure}[h]
\vspace{0.5cm}
\hspace{0.6cm} \resizebox{6cm}{!}{\includegraphics{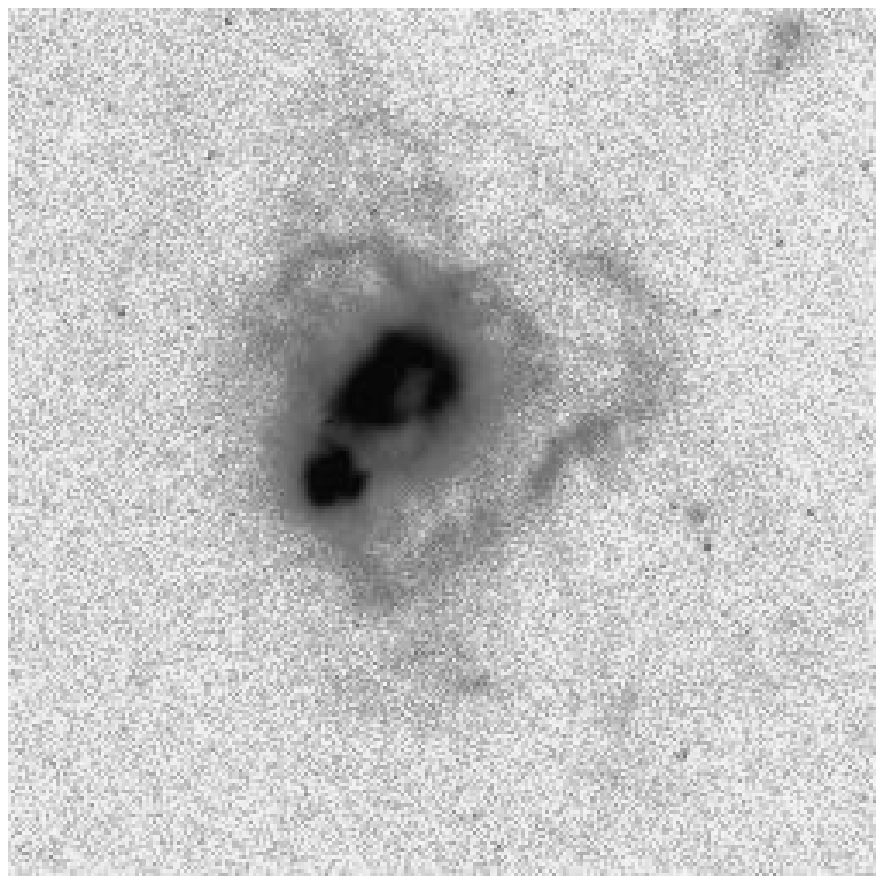}}
\hspace{1.1cm} \resizebox{6cm}{!}{\includegraphics{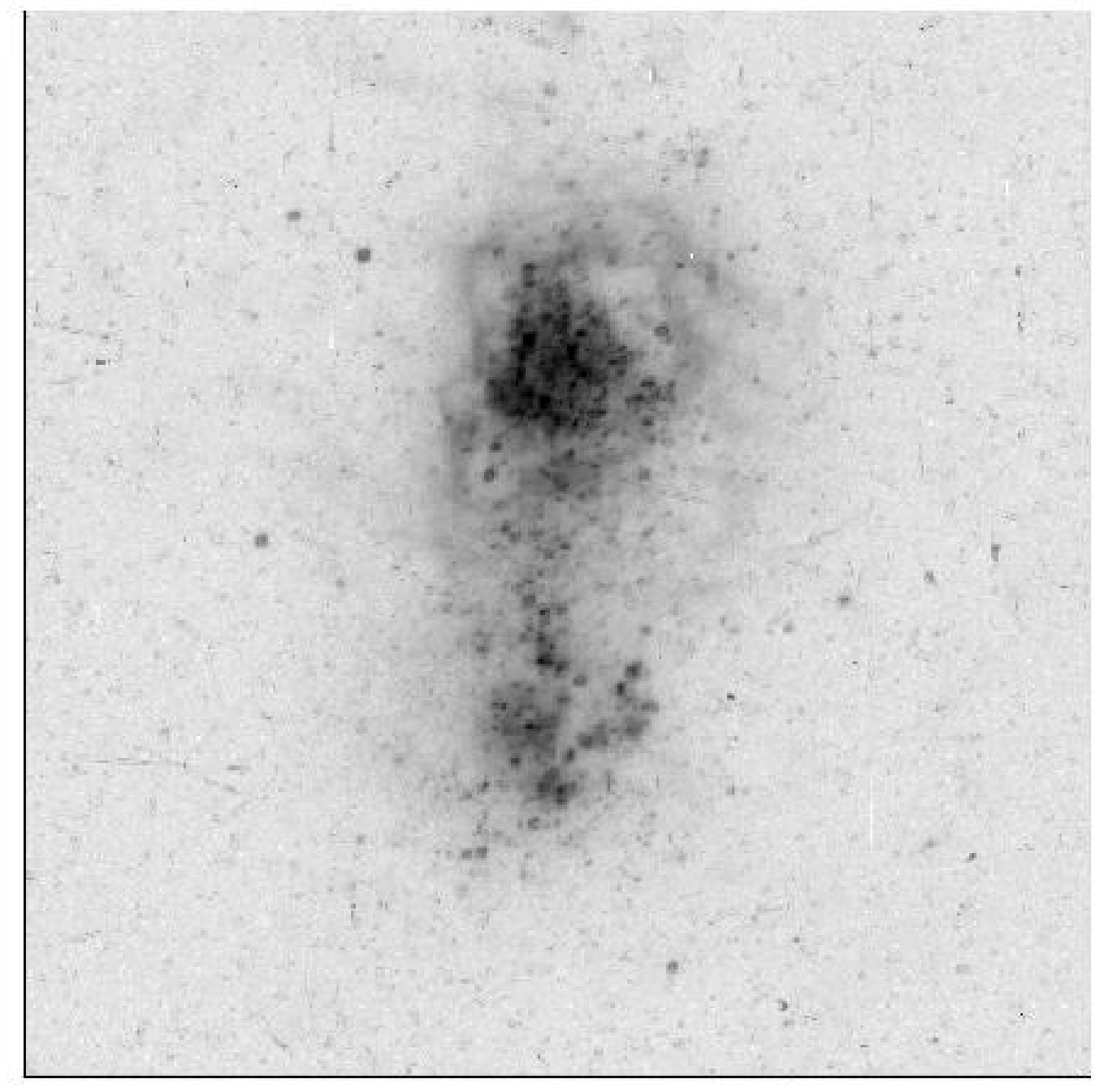}}
\caption{{\protect\small {\bf Left:} Continuum subtracted H$\alpha$ image of IZw18. Note the complex structure 
and very extended  filaments. The dimension is $45 \times 45$ arcseconds, north is up, 
east is left. In the upper right corner, the faint H$\alpha$ nebula in the companion galaxy is visible.
The image was obtained at the Nordic Optical Telescope with 0.65'' seeing (\"Ostlin).
{\bf Right:} A V-band (F555W) image of IZw18 obtained with the Planetary Camera of WFPC2 onboard
HST (cf. Hunter and Thronson 1995). The dimension is $17 \times 17$ arcseconds, North is up left,
east is down left. (Obtained from the 
HST archive.) At  a distance of 10 Mpc, 1 arcsecond corresponds to 48 pc.  }}
\end{figure}

\subsubsection{Chemical abundance of IZw18}
The first estimate of the metallicity of IZw18 came from Searle and Sargent (1972) who 
showed that the oxygen abundance was  below one tenth of solar, while the He abundance 
appeared to be normal. Alloin et al. (1978) derived 12+log(O/H)=7.2, which is close to most 
recent determinations, in particular by Izotov and Thuan (1999) who give 12+log(O/H)=7.18, 
or 1/50 of the Sun. Values of O/H in the range 1/30 to 1/60 of the solar value have been 
quoted over the last 20 years. The oxygen abundance seems to be constant, within the errors, 
over the optical face of the galaxy (e.g. Skillman and Kennicutt 1993,
Vilchez and Iglesias-P\'aramo 1998, Legrand et al. 1999). 

C/O was first derived by Dufour et al. (1988) who found [C/O]$ = -0.27$~ from IUE and 
ground based optical observations, an unexpectedly high value: higher than [C/O] 
in dwarfs like the LMC, indicating that the chemistry in 
IZw18 is anomalous. This value was later revised down by Dufour and Hester (1990) to [C/O]$=-0.5$.
Garnett et al. (1995) used the Faint Object Spectrograph (FOS) on HST but only got  a lower
limit  [C/O] $>  -1.3$. Later, they  re-observed IZw18 with the FOS and found [C/O]$=-0.63$ and 
$-0.56$ for the north-west (NW) and south-east (SE) regions respectively (Garnett et al. 1997). 
These values are again rather high when compared to other metal-poor dwarfs and
Garnett et al. (1997) proposed that a carbon enrichment from an intermediate
age population had occurred previous to the current star formation burst.
Recently, Izotov and Thuan (1999) reanalysed the HST data by Garnett et al. (1997), 
now finding [C/O]$ = -0.77$ and $-0.74$ for the NW 
and SE regions respectively. Their lower C/O follows mainly from a higher adopted electron
temperature based on new ground based spectra. Izotov and Thuan (1999) also found
indications of a temperature gradient, giving rise to the apparent abundance difference
between the NW and SE components. These last [C/O] values for IZw18 
follow nicely the trend for other very metal-poor galaxies (see Fig. 4).
 However, the fact that  the [C/O] determination has changed so much with time
is of course not satisfying. Hence  the new value adopted 
by Izotov and Thuan (1999) should perhaps not be taken at face value since part of their disagreement with other authors might 
be associated with the imperfect match between
HST and ground based apertures. A deep HST UV+optical spectrum with a single instrument like 
STIS could resolve this issue.

Similar problems were encountered with the study of the nitrogen to oxygen ratio in IZw18 that
 first yielded lower limits until 
Dufour et al. (1988) were able to derive [N/O]$=-1.36$. 
 Izotov and Thuan (1999) recently 
derived a lower [N/O] of $-1.60$ in perfect agreement with their overall trend that
[N/O] stays constant at low [O/H] in metal-poor galaxies,
 contrary to previous findings (see Fig. 4).
The helium abundance of IZw18 has been a concern for some time. Its value is of special 
importance for deriving the primordial helium abundance, since IZw18 has the lowest 
known ISM heavy element abundance of galaxies. Searle and Sargent (1972) found a rather normal He abundance, that later was revised down, e.g. $Y=0.21$ (Davidson and 
Kinman 1985), and $Y=0.226$ (Pagel et al. 1992). These values were rather low  in view of the 
standard big bang nucleosynthesis (SBBN) theory. Izotov and 
Thuan (1998a,b) find $Y=0.242$, in comfortable agreement with SBBN.

\begin{figure}
\hspace{1.0cm}
\resizebox{11cm}{!}{\includegraphics{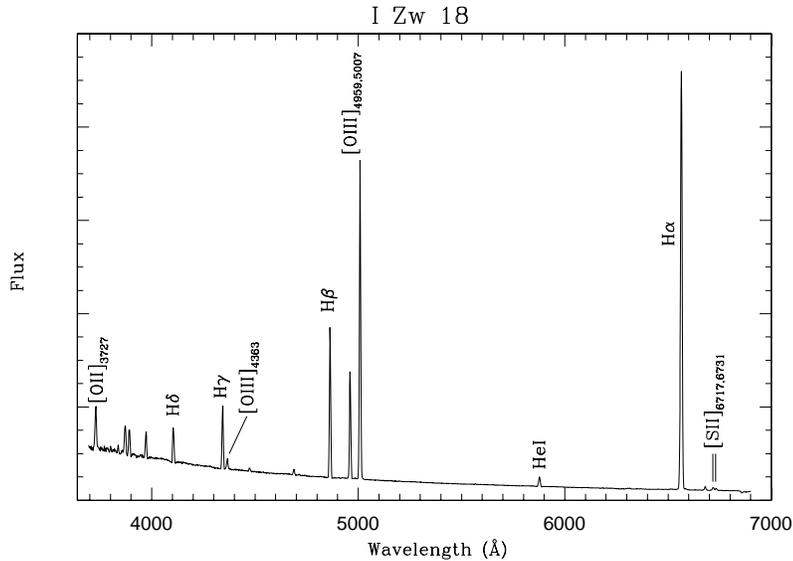}}
\caption{{\protect\small An spectrum of IZw18, with the most important emission lines labelled.
Note the very blue continuum. That IZw18 is a very low abundance object can be seen by
noting the following: [O{\sc iii}]$\lambda\lambda {4959,5007}$ and [O{\sc ii}]$\lambda {3727}$ are rather weak with
respect to H$\beta$, while [O{\sc iii}]$\lambda {4363}$ is relatively strong, and moreover the nitrogen 
and sulphur lines are very weak. (Plotted from a spectrum provided by F. Legrand).
}}
\end{figure}

\begin{figure}
\hspace{2.0cm}
\resizebox{8.6cm}{!}{\includegraphics{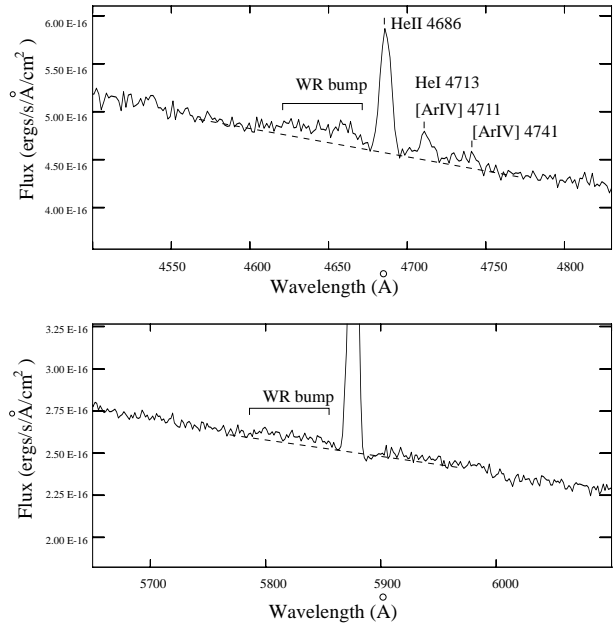}}
\caption{{\protect\small This figure shows the very weak Wolf-Rayet features detected in a
high S/N spectrum of IZw18 (Legrand et al. 1997b; see also Izotov et al. 1997a).
(Courtesy F. Legrand). }}
\end{figure}

Kunth et al. (1994) attempted to measure the oxygen abundance in the neutral gas cloud surrounding
IZw18 by using UV absorption lines, observed with the GHRS onboard HST. They found that
the oxygen abundance could be as low as 1/1000 of the solar value, indicating near pristine 
gas with abundances lower than what is found in QSO absorption line systems. However the
use of saturated lines was criticised by Pettini and Lipman (1995).
Van Zee et al. (1998) argue that the H{\sc i} velocity dispersion together with the measured OI
line of Kunth et al. (1994) imply an oxygen abundance ~$\ge$1/60 of the solar value,
suggesting that the chemical enrichment products are well mixed.

\subsubsection{On the age of IZw18}

Izotov and Thuan (1999) come to the conclusion  that all galaxies
with 12+log(O/H) $<$ 7.6 must be younger than 40 Myr, thus  in practise  newly born 
galaxies; and this would be very much so for IZw18 with its record low abundance (12+log(O/H)=7.18).
There are however other possible interpretations of the abundance ratios of metal-poor galaxies (as 
we already pointed out in Sect. 4.4.8). Moreover there is some independent evidence suggesting that IZw18 does
indeed host an old underlying population.

The age of IZw18 has been debated ever since the early seventies when the intriguing
properties of this galaxy were first realised. Being the most metal-poor galaxy, it is
of course one of the most promising candidates for a genuinely young galaxy. The absence 
of an outer regular envelope made IZw18 a good young galaxy candidate.
Thuan (1983) argued from near infrared aperture photometry that the galaxy was old, but  
comparing his colours with the recent
Bruzual and Charlot (2000) leads to a different conclusion. 
Moreover the apertures only  cover the 
central part of the galaxy which is dominated by young stars irrespective of the possible 
presence of an underlying population.

Pantelaki and Clayton (1987) claimed a high age for IZw18  based on the observed C/O and N/O ratios, 
but these were later revised downwards. A similar argument was given by Garnett et al. 
(1997) from their new C abundance, that however  was revised downwards
recently by Izotov and Thuan (1999)  weakening the conclusion. 
In view of the uncertainties on  the N/O and C/O ratios and yields, and that these ratios 
are subtle to 
interpret anyway, it is clear that they give very limited constraints on the star formation history 
in IZw18.
The chemical evolution model used by Kunth et al. (1995) and the spectro-chemical 
evolution model of Legrand  (1998) also suggest that IZw18 is not young. The latter one
predicts that IZw18 could have experienced a low but continuous star formation rate for several Gyr prior 
to the present burst.
Of course the  difficulties outlined in Sect. 3 make all constraints on galaxy ages derived 
from chemistry very uncertain.

 IZw18 has been imaged in the optical by HST by two different groups 
(Hunter and Thronson 1995; Dufour et al. 1996), both finding a resolved population of 
young massive stars. There was no evidence for old stars,
but  none against, since the observations were not sufficiently deep
to allow old stars to be detected. These data sets  were recently  reanalysed by 
Aloisi et al. (1999) and give support for an age in excess of 0.5  Gyr. \"Ostlin (1999a)
studied the resolved stellar population in the near infrared with NICMOS onboard HST, and 
found that while the NIR colour magnitude diagram was dominated by stars 10-20 Myr old,
numerous red AGB stars require a much higher age in agreement with Aloisi et al. (1999). 
The NICMOS data require stars older than 1 Gyr to be present, and an age as high as 5 Gyr 
is favoured. This holds even if  a  distance slightly higher than the customary 10 Mpc
is adopted.

In Fig. 8 we show surface brightness and colour ($B-R$ and $B-J$) profiles of IZw18
from deep CCD images obtained at the Nordic Optical Telescope and infrared images  from the 
UKIRT (\"Ostlin 1999c). The colours rise continuously 
with increasing radius and reach $B-R=0.6$ (Cousins R)
and $B-J=1.6$ at a radius of 10 arcseconds. If due to purely stellar emission, a single
stellar population model with a metallicity of $1/50 Z_{\odot}$ (Bruzual and Charlot 2000) 
indicates an age of $\log (age) = 9.1 
\pm 0.1$ for both colours, irrespective of IMF (Salpeter, Scalo or Miller-Scalo). 
With a more realistic assumption of a more or less continuous
star formation rate (e.g. for an exponentially decaying SFR with e-folding time $\tau = 3$ Gyr)  
we predict an age of at least 5 Gyr. One possible caveat is that
there might be a substantial contribution from ionised gas to the colours, but we believe 
that this is not a dominant effect. The B-band profile has an exponential shape
for radii greater than 10 arcseconds; fitting a disc to the outer parts yields a central
surface brightness of $\sim 23$ mag/arcsec$^2$ and an integrated disc luminosity 
$M_B = -11.8$. 
With a mass to light ratio for a $10^9$ years old single stellar population,  
the disc luminosity is equivalent to a stellar mass of $\sim 5 \cdot 10^6 
M_{\odot}$, but may be a factor of two higher if the SFR has been continuous (cf. the
predictions by Legrand 1998). With a mass of the young population of $\sim 10^6 M_{\odot}$ 
this means that while difficult to detect, the old population dominates the stellar mass.
This suggests that IZw18 could be hosted by a low surface brightness galaxy.

Thus, although not  foolproof, photometry (of individual stars and surface 
photometry) now indicates that the galaxy is in fact old, although Izotov et al. 
(1999a) in a last attempt to resurrect its youth, suggested that the 
distance of IZw18 has been severely underestimated.

\begin{figure}
\vspace{-1cm}
\hspace{.5cm} \resizebox{13cm}{!}{\includegraphics{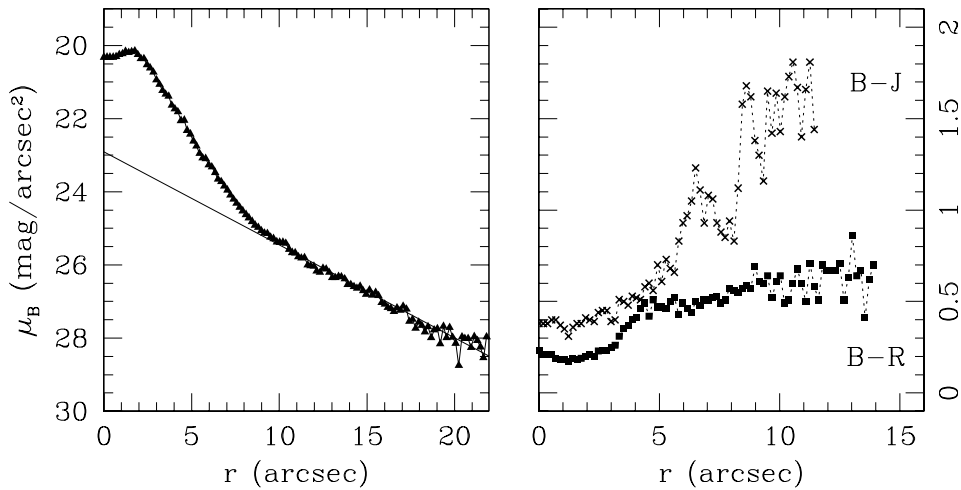}}
\caption{{\protect\small Luminosity and colour profile of IZw18. {\bf Left panel:} the radial surface brightness
profile of IZw18 in B.  Note the apparent exponential shape of the profile 
for radii greater than 10 arcseconds, indicating an underlying low surface brightness 
component with  $\mu_{B,0} = 23$ mag/arcsec$^2$. {\bf Right panel:} $B-R$ (filled squares) and $B-J$ (crosses) 
profiles of IZw18; note the red colours at large radius. From deep images obtained at the Nordic
Optical Telescope (NOT) and the UKIRT (\"Ostlin 1999c, in prep). }}
\end{figure}

\subsubsection{The companion of IZw18}

Zwicky noted what he called a ``flare'' at the north west of IZw18, which has turned out to be 
a separate galaxy.  H$\alpha$ images showed nebulosity close to the 
centre of this putative companion galaxy (Fig. 5), suggesting that it had the same
redshift as IZw18 (Dufour and Hester 1990, \"Ostlin et al. 1995). This was 
later confirmed by spectroscopy (Petrosian et al. 1996,
Dufour et al. 1996). The HST imaging study by Dufour et al. (1996) was able to
resolve several young ($\sim$ 80 Myr) luminous stars, see also Aloisi et al. (1999). 
It is located in the same H{\sc i} cloud
as IZw18 and its location coincides with a peak in the H{\sc i} column density (van Zee et al. 1998a).

In view of its position, just a kpc away from IZw18,
its chemical composition  would be of considerable interest. This galaxy is fainter than IZw18 but still very blue. However spectra have 
failed to unambiguously detect any oxygen lines, and consequently its metallicity is not known.
The problem is that the galaxy has very low central surface brightness ($\mu_{B} = 22.7$ mag/arcsec$^2$,
\"Ostlin 1999c) making spectroscopy difficult.

\subsection{SBS0335-052}

For a long time I~Zw18 seemed to play in its own league, with no other BCG
coming really close to its low oxygen abundance in the H{\sc ii}-gas. However
the entrance of SBS0335-052 on the stage changed the situation.
This galaxy was found in the Second Byurakan Survey (SBS, Markarian and Stepanian
1983). A number of papers
from 1990 and onwards have shown it to be a galaxy with an oxygen abundance
comparable to that of IZw18 (Izotov et al. 1990, 1997b; Melnick et al. 1992).
Melnick et al. (1992) and Izotov et al. (1997b) both find an oxygen abundance
of 1/40 of the solar value. The analysis by Melnick et al.  (1992) suggest that the
O abundance may be a factor of two higher in a north-western H{\sc ii} region.
Thuan et al. (1997) argues that the oxygen abundance in the neutral gas
may be even smaller by a factor 100. Mid infrared observations with ISO  revals the 
presence of dust, and a gas to dust ratio typical for more metal rich BCGs (Thuan et al. 1999).
 VLA observations have
revealed an H{\sc i} mass of $\sim 2 \times 10^9 M_{\odot}$ and a dynamical mass a factor
of a few larger (Pustilnik et al. 1999). An image of SBS0335-052 from the HST
archive (see also Thuan et al. 1997) is shown in Fig. 9 and reveals a complex
morphology. Like IZw18, SBS0335-052 has a faint companion galaxy (Sect. 5.2.1).

\begin{figure}
\resizebox{7.6cm}{!}{\includegraphics{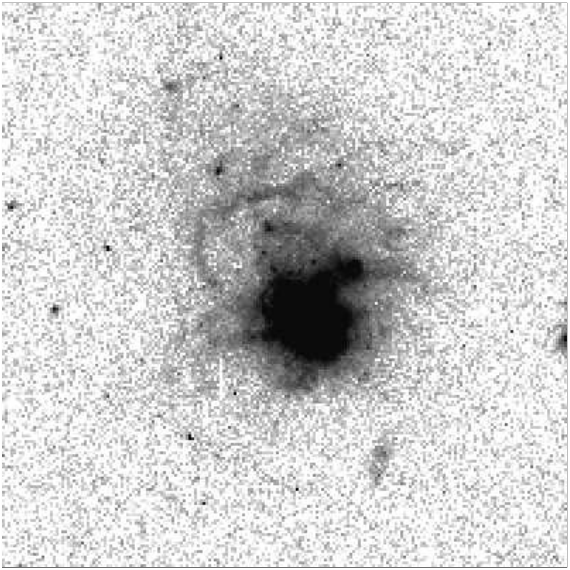}}
\caption{{\protect\small SBS0335-052 imaged with the Hubble Space Telescope in the V-band, cf. 
Thuan et al. (1997).
The size of the image is $20\times 20$ arcseconds corresponding to $5.2 
\times 5.2$ kpc. Note the extended complex filaments. (Obtained from the 
HST archive.) }}
\end{figure}

There have been several claims that this galaxy is a truly young galaxy,
not containing any underlying old population, ( Thuan et al. 1997, 
Izotov et al. 1997b, Papaderos et al. 1998). The argument put forward
is the low metallicity and the lack of any underlying population in 
surface photometric data. However Lipovetsky et al. (1999) in their study
of the companion, present also surface photometry of SBS0335-052. The
$R-I$ colours at a radius of 10 arcseconds is $(R-I)=0.4$ which indicates an
age of several Gyr for a single stellar population model with metallicity 2\%
solar and a standard Salpeter IMF (Bruzual and Charlot 2000). 
Nebular gas contamination
cannot account for this, since that would make $R-I$ bluer, leading to
an underestimate of the age. 
Even  allowing for an internal extinction of $E(B-V)=0.2$ an age of 1 Gyr
or more is required.
Changing IMF, or metallicity (to 20\% solar)
still leads to an age in excess of 1 Gyr. Relaxing the assumption of 
instantaneous star formation, and allowing for more realistic SFHs would increase 
the age to around 10 Gyr (for $\tau =3$ Gyr, cf. Sect. 5.1.2). Thus,
in view of the  surface photometry presented by Lipovetsky et al. (1999),
the age of ~$>1$ Gyr can be regarded as a conservative lower limit.
Moreover the presence of red star clusters (\"Ostlin 1999b) suggests 
an age of several Gyr. These star clusters
were discussed by Papaderos et al. (1998) but the models they used to interpret
the ages do not agree with other published models, but are systematically off-set, 
apparently  leading to an underestimate of the age of the clusters.  Thus, 
SBS0335-052, like IZw18, is probably not a genuinely young galaxy.
In some sense SBS~0335-052 is more extreme than IZw18 in that despite being 
almost as metal-poor as  IZw18, it is  more luminous ($M_B = -16.7$) and thus
lies further off the metallicity--luminosity relation for dIs, see Fig. 10.

\subsubsection{The companion of SBS0335-052}

Like IZw18, SBS~0335-052 also has a physical companion,  SBS~0335-052W at a projected distance 
of 22 kpc (Pustilnik et al. 1999). Lipovetsky et al. (1999) found an oxygen abundance of 
12+log(O/H)=7.2, even lower than that of SBS~0335-052, and fully comparable to IZw18.
 This companion, though blue is not very compact and it has 
evidently avoided inclusion in any BCG catalogue. Its central R surface 
brightness and SFR is rather low, comparable to a ``normal'' dI. The fact that both SBS0335-052
and its companion resides in the same H{\sc i} cloud but at least 22 kpc apart,   poses a problem 
for the youth hypothesis since it would require a coordinated ignition of star formation in the two clouds.

\subsection{Summary}

The most metal-poor BCGs: IZw18 and SBS0335-052 (and its companion), are probably not young
galaxies. Nevertheless they are important laboratories for studying star formation
at low abundances and the evolution of dwarf galaxies. The metallicities of 
their old  stellar populations cannot be assessed with any accuracy. There is 
nothing that requires these galaxies to be more metal-poor than the most extreme 
dSph galaxies which have [Fe/H] ${\stackrel{<^{ }}{\sim}} -2$, see Table 2 
(their oxygen abundances are still not known). 

In addition to these galaxies, there are more than a dozen other BCGs with
abundance 12+log(O/H)$\approx  7.6$ or lower, and more objects are about to 
be discovered (cf. Sect. 6.1.5). The Local Group, and its immediate surroundings, contains 
another half dozen dI with very low metallicity. Moreover there are many dE/dSph 
galaxies in the neighbourhood with [Fe/H] $\approx  -2$. In addition, LSBGs have proven
a good hunting ground for very metal poor galaxies. In Table 3 below we present
all known galaxies with 12+log(O/H)$\approx  7.6$ or lower. When several measurements
(representing different H{\sc ii} regions) for the same galaxy were available we 
took weighted linear averages; and the same was done if there were differing O/H
determinations of similar quality in the litterature (in these cases, only galaxies for 
which the weighted average was lower than the limit were included in the table).
These galaxies, taken together form an important, yet hardly homogeneous,
set of data for investigating  dwarf galaxy evolution. Another interesting class, to
which IZw18 and SBS0335-052 and its companion belong, is galaxies with low 
abundances with respect to their integrated luminosities, see Sect. 7.1.

\begin{table}[h]
\caption{{\protect\small The most metal-poor galaxies. This Table contains all galaxies
with 12+log(O/H)$ < 7.65$ that we in found the literature. The first column gives the most
common name that is also recognized by the NASA/IPAC extragalactic database (NED). 
Alternative names are given below in some cases.
A broad  galaxy classification  is given in column two. The coordinates are 
from NED (except for HS 0822+3542 and CS 0953-174). The heliocentric radial velocities 
in column 5 come mainly from NED, but
we also quote the reference given by NED. Preference was given to H{\sc i}-velocities if available.
The sixth column gives 12+log(O/H), where
uncertain values based on empirical methods (not utilising [O{\sc iii}]$\lambda 4363$) are marked with colon. 
Values marked  by asterisks represent a weighted average of values from different H{\sc ii} regions and/or authors. 
The seventh column gives the absolute B-magnitude, from direct distance measurements or from radial velocity 
and H$_0=75$ km/s/Mpc. Values for $M_B$ based on other values of H$_0$ were rescaled. A colon indicates that 
the magnitude is uncertain, either beacuse it is based on photographic data or because from other passband 
than B. The references are given in the footnotes.}}
\end{table}

\np
\addtocounter{table}{-1}

\begin{table}[h]
\caption{{\footnotesize  continued}}
\footnotesize
%\scriptsize
\vspace{-0.65cm}
\begin{center} 
\begin{tabular}{llllrlllll}
\hline
 Galaxy name 	& Type	& $\alpha_{2000}$ & $\delta_{2000}$ & $v_{\rm hel}$ & O/H &	$M_B$ \hspace{-2cm}  & \multicolumn{3}{c}{
References} \\
		& &(h~~m~~s) &$(~~^{\circ} ~~~{'} ~~~{''})$ & (km/s) & & & {\scriptsize $v_{\rm hel}$} & {\scriptsize O/H }&	{\scriptsize $M_B$ } \\
\hline
UM 283~$^a$		& BCG	& 00 51 49.5 	& $+$00 33 53	& 4510	& 7.59	& $-$16.7: 	& {\scriptsize 37}	& {\scriptsize 7}	& {\scriptsize 8}	\\
UM 133			& BCG	& 01 44 41.3 	& $+$04 53 26 	& 2098	& 7.63	& $-$16.9	& {\scriptsize 31}	& {\scriptsize 9}	& {\scriptsize 10}	 \\
UM 382			& BCG	& 01 58 09.4 	& $-$00 06 38 	& 3598	& 7.45	& $-$15.0	& {\scriptsize 31}	& {\scriptsize 9}	& {\scriptsize 11}	 \\
UM 408			& BCG	& 02 11 23.5 	& $+$02 20 31 	& 3598 	& 7.63	& $-$15.8	& {\scriptsize 31}	& {\scriptsize 9}	& {\scriptsize 11}	 \\
SBS 0335-052W		& dI/BCG & 03 37 38.4 	& $-$05 02 36	& 4017	& 7.21$^*$ & $-$14.4	& {\scriptsize 33}	& {\scriptsize 14}	& {\scriptsize 15}	 \\
SBS 0335-052E~$^b$ 	& BCG	& 03 37 44.0 	& $-$05 02 38	& 4057	& 7.29	& $-$16.7	& {\scriptsize 33} 	& {\scriptsize 4}	& {\scriptsize 15} 	\\
Tololo 0618-402 	& BCG	& 06 20 03.8 	& $-$40 16 43 	& 10493	& 7.61$^*$ & $-$17::	& {\scriptsize 31}	& {\scriptsize 9}	& {\scriptsize 26}	 \\
HS 0822+3542		& BCG	& 08 25 55.4	& $+$35 32 31	&  732	& 7.35	& $-$12.6	& {\scriptsize 1,2}	& {\scriptsize 1,2}	& {\scriptsize 1,2}	 \\
IZw18~$^c$		& BCG	& 09 34 01.9	& $+$55 14 26 	& 750	& 7.18	& $-$13.9	& {\scriptsize 32}	& {\scriptsize 4}	& {\scriptsize 16}	 \\
SBS 0940+544N~$^d$	& BCG	& 09 44 16.7 	& $+$54 11 33	& 1685 	& 7.43	& $-$14.8:	& {\scriptsize 34}	& {\scriptsize 4}	& {\scriptsize 13}	\\
CS 0953-174		& BCG	& 09 55~~~~~	& $-$17~~~ ~~	& 3540	& 7.58	&   		& {\scriptsize 30}	& {\scriptsize 9}	& {\scriptsize 27}	 \\
KUG 1013+381~$^e$	& BCG	& 10 16 24.5	& $+$37 54 44	& 1198	& 7.58	& $-$15.1:	& {\scriptsize 3}	& {\scriptsize 3}	& {\scriptsize 3}	\\
$[$RC2$]$A1116+51~$^f$	& BCG	& 11 19 33.3	& $+$51 29 40	& 1351	& 7.60$^*$ & $-$13.9:	& {\scriptsize 21} 	& {\scriptsize 17,19}	& {\scriptsize 25}	\\
SBS 1159+545		& BCG	& 12 02 02.4 	& $+$54 15 51 	& 3537	& 7.49	& $-$15.4:	& {\scriptsize 35}	& {\scriptsize 4}	& {\scriptsize 13}	 \\
Tololo 1214-277		& BCG	& 12 17 18.2 	& $-$28 01 40 	& 7795	& 7.56$^*$ & $-$17.6:	& {\scriptsize 31} 	& {\scriptsize 9}	& {\scriptsize 13}	\\ 
Tololo 65~$^g$		& BCG	& 12 25 46.9 	& $-$36 14 01 	& 2698	& 7.40$^*$ & $-$15.3	& {\scriptsize 31}	& {\scriptsize 9}	& {\scriptsize 12}	\\
$[$RC2$]$A1228+12~$^h$	& BCG	& 12 30 48.5	& $+$12 02 42	& 1254	& 7.64	& $-$13.2:	& {\scriptsize 21}	& {\scriptsize 17}	& {\scriptsize 18}	\\
CG 389			& BCG	& 14 17 01.7	& $+$43 30 13	&  649	& 7.59	& $-$14.1:	& {\scriptsize 29}	& {\scriptsize 4}	& {\scriptsize 13}	\\
\\
UGCA 20			& LSB/dI & 01 43 14.7	& $+$19 58 32	& 498	& 7.58$^*$  & $-$14.0	& {\scriptsize 41}	& {\scriptsize 42}	& {\scriptsize 28}	 \\
ESO 546-G34		& LSBG	& 02 58 37.3	& $-$18 41 57	& 1568	& 7.64:	& $-$15.5	& {\scriptsize 21} 	& {\scriptsize 5}	& {\scriptsize 5}	\\
UGC 2684		& LSB/dI & 03 20 23.2	& $+$17 17 47	& 287	& 7.64:$^*$ & $-$12.8	& {\scriptsize 41} 	& {\scriptsize 43}	& {\scriptsize 28}	\\
ESO 489-G56		& LSBG	& 06 26 17.0	& $-$26 15 56	& 495	& 7.49	& $-$13.7	& {\scriptsize 39}	& {\scriptsize 6}	& {\scriptsize 6}	 \\
ESO 577-G27		& LSBG	& 13 42 46.9	& $-$19 34 54	& 1410	& 7.57	& $-$14.4	& {\scriptsize 40} 	& {\scriptsize 6}	& {\scriptsize 6}	\\
ESO 146-G14		& LSBG	& 22 13 00.4	& $-$62 04 03	& 1680	& 7.61	& $-$16.6	& {\scriptsize 38} 	& {\scriptsize 5,36}	& {\scriptsize 5}	\\
\\
DDO 53~$^i$		& dI	& 08 34 06.8	& $+$66 10 52	& 19	& 7.62:	& $-$13.8	& {\scriptsize 21}	& {\scriptsize 22}	& {\scriptsize 22}	 \\
UGC 4483		& dI	& 08 37 03.0	& $+$69 46 50 	& 156	& 7.53$^*$ & $-$12.8:	& {\scriptsize 21}	& {\scriptsize 4,23}	& {\scriptsize 20}	\\
Leo A			& dI	& 09 59 23.9	& $+$30 44 42	& 26	& 7.30	& $-$11.3	& {\scriptsize 24}	& {\scriptsize 22,24}	& {\scriptsize 24}	\\
Sextans A		& dI	& 10 11 05.6	& $-$04 42 28	& 325	& 7.49	& $-$14.2	& {\scriptsize 24}	& {\scriptsize 22,24}	& {\scriptsize 24}	\\
Gr8			& dI	& 12 58 39.5	& $+$14 13 02	& 215	& 7.62$^*$ & $-$11.2	& {\scriptsize 24} 	& {\scriptsize 24}	& {\scriptsize 24}	\\
DDO 187~$^j$		& dI	& 14 15 55.9	& $+$23 03 13	& 154	& 7.36:	& $-$13.4	& {\scriptsize 21}	& {\scriptsize 22}	& {\scriptsize 22}	\\
Sag DIG~$^k$		& dI	& 19 29 59.0	& $-$17 40 41	& $-$79	& 7.42:	& $-$12.1	& {\scriptsize 24}	& {\scriptsize 22,24}	& {\scriptsize 24}	\\
\\
\hline
\end{tabular}
\end{center}
\end{table}

\np

\addtocounter{table}{-1}

\begin{table}[h]
\caption{ continued.}
\footnotesize
{\bf Alternative names}:
$^a$ 	UCM 0049+001;
$^b$ 	SBS~0335-052;
$^c$ 	Markarian 116;
$^d$	KUG~0940+544;
$^e$	HS 1013+3809;
$^f$	CG 1116+51;	
$^g$	Tol 1223-359, ESO~380-G27;
$^h$	RMB 132, VCC~1313;
$^i$	UGC~4459, VII Zw 238;
$^j$	UGC~912;
$^k$	ESO~594-G4;\\
{\bf References}:
1:	Kniazev et al. (1999, in prep.), Masegosa private comm.
2:	Merlino et al. (1999).
3:	Kniazev et al. (1998).
4:	Izotov and Thuan (1999).
5: 	Bergvall et al. (1999). 			% (absmag modified with 0.4 mag)
6:	R\"onnback and Bergvall (1995).		% (absmag modified with 0.5 mag)
7: 	Gallego et al. (1997). 
8: 	Vitores et al. (1996), assuming $B-R=0.6$.
9: 	Masegosa et al. (1994).
10: 	Telles and Terlevich (1997).
11:	Salzer et al. (1989a).
12:	Lauberts and Valentijn (1989).
13:	magnitude taken from NED.
14:	Lipovetsky et al. (1999).
15:	Papaderos et al. (1998).
16:	Mazarella and Boroson (1993).
17:	Kinman and Davidson (1981).
18:	Young and Currie (1998).
19:	French (1980).
20:	Tikhonov and Karachentsev (1993).
21:	RC3 (de Vaucoleurs et al. 1991). 
22: 	Skillman et al. (1989).
23: 	Skillman et al. (1994).
24:	Mateo (1998).
25:	Arp and O'Connell (1975).
26:	Estimated from continum flux in Terlevich et al. (1991), $m_B=18.5 \pm 1$, very uncertain.
27:	Galaxy from the Cambridge survey. No data in litterature on magnitude, nor accurate coordinates.
28:	van Zee et al. (1997c). 			%(adjusted to Ho=75)
29:	Velocity from the NED-team.
30:	Masegosa private communication.
31:	Terlevich et al. (1991).
32:	Gordon and Gottesman (1981).
33:	Pustilnik et al. (1999b).
34:	Augarde et al. (1994).
35:	Pustil'nik et al. (1995).
36:	Bergvall and R\"onnback (1995) shows that if effects of shocks are taken 
	into account O/H may be higher (12+log(O/H)=7.68).
37:	Kinman and  Hintzen (1981).
38:	Mathewson and  Ford (1996).
39:	Gallagher et al. (1995).
40:	Matthews and Gallagher (1996).
41:	van Zee et al. (1997a).
42:	van Zee et al. (1996).
43:	van Zee et al. (1997b).
\end{table}

\np

\small
\normalsize

%%%%%%%%%%%%%%%%%%%%%%%%%%%%%%%%%%%%%%%%%%%%%%%%%%%%%%%%%%%%%%%%%%%%%%%%%%%%%%%%%%%%%%%%
%%%%%%%%%%%%%%%%%%%%%%%%%%%%%%%%%%%%%%%%%%%%%%%%%%%%%%%%%%%%%%%%%%%%%%%%%%%%%%%%%%%%%%%%
%%%%%%%%%%%%%%%%%%%%%%%%%%%%%%%%%%%%%%%%%%%%%%%%%%%%%%%%%%%%%%%%%%%%%%%%%%%%%%%%%%%%%%%%

\np

\section{Surveys for metal-poor galaxies and their space and luminosity distributions}

\subsection{Surveys}

Haro (1956) used  multiply exposed large scale plates to search for 
emission line galaxies. He identified a number of compact galaxies with
 strong emission lines. Zwicky (1971) envisioned that galaxies would
 evolve into highly concentrated densities as for neutron stars and 
produced several lists of  ``compact galaxies''  
 selected form the Palomar sky survey. A nearly contemporary line
 of research was opened by Markarian (1967) who focused on
ultraviolet excess galaxies at the Byurakan observatory. The underlying 
assumption was completely opposite however since V.A. Ambartsumian had 
the view that galaxy nuclei originated from explosive events.

	Many metal-poor galaxies known at present  
 undergo enhanced star formation event. This type of activity has 
obviously favoured their discovery (a metal-poor galaxy not necessarily needs to be an active one!). 
Following the early discovery of two bona fide metal-poor galaxies  by Searle
 and Sargent (1972) that indeed turned out to experience an active star formation
 episode  many astronomers have embarked in the building of large samples
 of such  objects. Surveays were aimed to study their statistical
 importance in the Universe as a function of time, their relation to large
 structures and their contribution to the luminosity function.

One way to select star-forming galaxies is to rely on their ultraviolet
 excess and emission line spectrum. This property is not completely 
unambiguous since it is shared to some extent by active galactic nuclei (AGN).
There are complications of course, mainly
 because of the effect of dust that reduces the UV flux and re-radiate the
 Lyman continuum photons in the far infrared (IR excess in IRAS 60 micron 
band). 
 Follow-up observations need to distinguish the full
population of starbursts from the AGN with non thermal activity.

\subsubsection{Ground-based selection}

 Ground-based surveys have used morphological criteria, colour selection and
 emission line selection. Lists of blue compact galaxies were pioneered by 
Zwicky, followed by Fairall and others who isolated objects from their anomalous 
high surface brightness as seen on the Palomar Sky Survey. Spectroscopic 
follow ups have revealed a large proportion of  H{\sc ii} galaxies 
and AGNs (Kunth et al. 1981)

The colour selection proceeds by searching for blue  or ultraviolet excess 
objects involving various techniques such as the use of very low dispersed 
images or multiple colour direct images. Dispersed images have been used by the
First and Second Byurakan Surveys (FBS, SBS) by Markarian (1967) on IIaF emulsion
and later the University Michigan survey (UM, MacAlpine et al. 1977) and Case survey
(Pesch and Sanduleak 1983) with IIIaJ emulsion. The second method has
 been pioneered by Haro (1956) and extensively developed by the
 Kiso Observatory Survey (Takase and Miyauchi-Isobe 1984).
Low resolution slitless spectroscopy enables to detect  [O{\sc ii}]${\lambda 3727}$,
 H$\beta$, [O{\sc iii}]${\lambda\lambda 4959,5007}$  and H$\alpha$ lines  depending 
on the chosen emulsion or
 filter.
 Good seeing and excellent guiding are a requisite to avoid trailing and loss
 in detectivity. These techniques face a trade off between the dispersion
and the spectral range covered. The higher the dispersion, the easier it
 becomes to detect weak emission lines against the continuum, while a  narrow
  spectral
 range cuts significantly the sky background at the expense of the redshift
 range. The  recent surveys discussed in Gallego et al. (1997) and Salzer (1999) use the
 H$\alpha$ line which can  be bright even in low-excitation or very
 metal-poor objects. Because each technique involves specific observational
biases, modern surveys tend to combine various approaches. The use
of large CCD arrays equipped with scanning Fabry Perot interferometry or slitless spectroscopy 
 offer deeper limits at the
expense of the reduced field of view. In the future, these combinations will
probe distant H{\sc ii} galaxies populations. The most difficult problem that these
 surveys have to face is that of the follow-up observations (Terlevich et al. 1991). Getting even a 
rough oxygen abundance for faint galaxies  requires long  telescope
 time and suggests the use of multi-object-spectroscopy. 

\subsubsection{Selection effects at work}

Selection effects have to be considered in comparing objects drawn
 from different surveys and for the derivation of a luminosity function,
as first pointed out by Wasilewski (1983) and summarized by Comte (1998), who noted that  selected 
populations have different statistical
 properties that depend on their selection modes. Emission line selected samples span a broader range of 
colours than purely UV-excess objects. This is not a surprise, but Salzer et al. (1997a) 
pointed out that very blue objects  selected from their emission lines were also missed 
by the Markarian surveys. Similarly the distribution of emission line equivalent widths  strongly depends on the choice of the dispersion.  Very young
starbursts are favoured in emission line surveys while ageing bursts are more easily
picked out in ``continuum'' surveys. Finally H$\alpha$ surveys present no
 prejudice in favour of a given metallicity.

\subsubsection{How metal-poor galaxies can be found?}

	Clearly, despite their strong emission lines  metal-poor compact dwarf 
galaxies are difficult to detect simply because they are fainter than $L^{\star}$ (the 
characteristic galaxy luminosity for a Schechter type luminosity function, cf.
Schechter 1976, Binggeli et al. 1988) 
galaxies (a kind of Malmqvist bias). Only metal-poor galaxies with
 metallicity of 
the order of 0.1 solar that undergo starbursts are easy to pick out just
because  oxygen  is the major cooling species, hence the  [O{\sc iii}]  lines
 at 4959 and 5007 \AA \ are particularly strong. But as one moves to more 
deficient objects, say below 0.01 solar, forbidden lines fade and the dominant 
cooling agents are H and 
He (Kunth and Sargent 1986).Therefore a combination of H$\alpha$ objective-prism
 spectroscopy and  UV-excess searches should be promising. Comte (1998) suggests
  a mid-UV imaging survey from balloon borne or orbiting instruments. 
So far the SBS and UM surveys has given a handful of new galaxies but never 
with metallicity below 1/50 solar. Why this is so? 
It may be that extreme metal-poor star-forming galaxies are very rare or do not exist locally

\subsubsection{HI clouds}

Because it was observationally difficult to confirm the recurrent burst picture 
against the idea that blue compact dwarf galaxies are young, several H{\sc i}
 surveys have provided independent clues. In their decisive early H{\sc i} survey, Lo and Sargent (1979)
showed that the space density of protogalactic clouds required by the youth 
hypothesis must be at least 8 Mpc$^{-3}$, each cloud having a mass of about $5 \times 10^8
 M_{\odot}$ which is 2 to 3 orders of magnitude higher than what they  found.
 H{\sc i}-selected sample of clouds unavoidably turn out to have optical
 counterparts.  In all cases,  possible local H{\sc i} primeval
 clouds have been found to be associated with stars (Djorgovski 1990, Impey et al. 1990,  
McMahon et al. 1990, Salzer et al. 1991,  Chengalur et al. 1995).
Since then, other surveys have been carried out to find 
isolated H{\sc i} clouds but without success (Briggs 1997).
Some  interesting examples of H{\sc i} without 
coincident 
optical emission  
were however provided by ``off-scans'' in 21cm line studies (Schneider 1989, 
Giovanelli and Haynes 1989, Chengalur et al. 1995, Giovanelli et al. 1995) 
but they are likely to be associated with (or bridge) nearby large
 visible galaxies.
 Tyson and Scalo (1988) have suggested that  the majority of
 dwarf galaxies may be in a quiescent state, not forming young stars, and
 that they might consequently be missed by optical selection methods. A 
recent analysis  of H{\sc i}-selected galaxies (Szomoru et al. 1994)
shows that H{\sc i} searches do not yield a population of optically underluminous 
galaxies. Some faint nearby galaxies with strong emission lines remaining 
undetected in H{\sc i}  must have very little  gas or store their gas in ionised or 
molecular form. Interestingly, the recent work by Schneider et al. 
(1998) shows that the H{\sc i} mass 
function  may become steeper at faint masses, in analogy with an upturn in the optical 
luminosity function at faint luminosities. 
Detection limits for H{\sc i} surveys remain quite high ($N_{HI} \sim 10^{18} {\rm cm}^{-2}$) hence 
smaller  pockets may still be hidden. On the other hand, isolated H{\sc i} 
clouds with masses and sizes comparable to 
present dwarf galaxies would be easily seen with modern radio telescopes 
hence they must be very rare. This conclusion is reinforced by the damped  QSO
 absorption lines  that tend to occur in the haloes of bright galaxies and not in
 smaller H{\sc i} clouds
(Lanzetta et al. 1995; however see discussion in Sect. 8.2). It has been
noted that the diffuse ionising background could drive H{\sc i} clouds under the 
detection limits of current surveys (see Corbelli and Salpeter
1993a,b).

\subsubsection{New objective-prism surveys in progress} 

Evidently, the quest for finding more metal-poor galaxies is not over.
The KISS (KPNO International Spectroscopic Survey, see Salzer 1999)
is a CCD based objective prism survey. It selects candidates from their H$\alpha$
emission, an advantage in searching for low metallicity systems. 
However an H$\alpha$ survey  produces not only BCGs, but also
AGN etc. The real advantage is the usage of CCD detectors which enables a more than
fivefold increase in  limiting distance  compared to photographic surveys. KISS
finds on average 17 emission line galaxies per square degree (i.e. 170 times more than the original Markarian survey). Follow-up spectroscopy to investigate low 
metallicity candidates is under way (Salzer, private communication).

In an attempt to detect low metallicity galaxies, the Calan-Tololo survey plates 
have been searched by eye for objects without visible continuum, but with strong emission 
lines (Maza et al. 1999, in preparation).
Interesting candidates are selected for spectroscopic follow up. More than two dozen
galaxies with oxygen abundance of 1/10 $Z_{\odot}$ or less have been found so far, and of these
8 are around 1/20 $Z_{\odot}$ (Masegosa and Maza, private communication).

The HSS (Hamburg/SAO Survey) selects emission line dwarf galaxy candidates from the 
Hamburg quasar survey plates.
The HSS has already yielded many new emission line dwarf galaxies (Ugruymov et al. 1999,
Pustilnik et al. 1999a), and the follow-up study is well in progress. The dwarf ($M_B=-12.6$)
HS~0822+3542 was recently found to have an oxygen abundance of only 12+log(O/H)=7.35
(Merlino et al. 1999, Kniazev et al. 1999), i.e. not far from that of SBS~0335-052 (see also Table 3).

Other recent work includes the UCM (Universidad Complutense Madrid) survey, based on an
H$\alpha$ selection, yielding objects from tiny BCGs to AGN, but mainly relatively 
luminous ones (cf. Gallego et al. 1997). For a recent review of surveys for star
forming galaxies, see Comte (1998).

Whereas objective prism and UV-excess surveys are good at picking 
up high surface brightness dwarfs with
emission lines, we have shown that very metal-poor galaxies also come in other brands. In
particular, since metallicity, luminosity and surface brightness are positively correlated
we expect the existence of very metal-poor LSB dwarfs, like dEs and LSBGs. Surveys
for such galaxies could be rewarding in the hunt for the ``most metal-poor galaxies'',
although abundances will be more difficult to determine, especially for dEs lacking 
H{\sc ii} regions. New dEs are still found in the Local Group and its vicinity, and several surveys
for LSBGs in the local Universe have been undertaken, and should strongly benefit from 
new wide field CCD cameras and projects like the Sloan Digital Sky Survey. To get
a picture about work in progress, see the recent conference volume ``The low surface brightness
universe'' by Davies et al. (1999).

\subsection{Luminosity function}
         
How do active star forming galaxies, in particular dwarf systems contribute
to the luminosity function (LF) as compared to normal galaxies?
To some extent, although somewhat paradoxical, 
the LF of local objects is more difficult to derive. Completeness problems are
severe and combined with the need of large spatial coverage. Even
when samples become large enough, the problem remains to establish
the nature of selected objects. Low dispersion spectra must be
 used and require a diagnostic diagram to disentangle H{\sc ii} 
region-like spectra from AGN.  Most LF
 determinations use a Schmidt estimator (Schmidt 1968) combined with a V/Vm test for
completeness. One must bear in mind that such a test may not be adequate 
if large structures of our local Universe are not sufficiently averaged out.
The very faint end of the LF is naturally
the most difficult to establish. Deep CCD imaging using narrow-band filters
complemented by follow up spectroscopy (Boroson et al. 1993),  
dedicated to this problem points towards a moderate slope of the LF at low
luminosities in agreement with larger samples from the Case surveys or the UM.
These studies are not in agreement with others, showing that this
 problem will be settled only by a better understanding of selection biases.
Certainly  surveys involving CCD techniques will allow to build larger
and deeper samples.

The space density of dwarf emission line galaxies has been addressed by
Salzer (1989), who finds $\sim 0.03$ galaxies per Mpc$^3$, corresponding to $\sim$7\%
of the local field galaxy density based on the UM
survey. This number is dominated by intrinsically faint systems. A rough 
calculation of the space density of
SBS galaxies from Pustilnik et al. (1995) immediately shows these to be
an order of magnitude less abundant, probably an effect of different
selection criteria. The  Case survey is even richer than the UM survey and the
 emission line galaxies (ELGs), which it contains, may contribute up to one third of the field galaxy population
(Salzer et al. 1995). The reason is that the Case survey also is sensitive to
 galaxies with a low level of H{\sc ii} activity. A comparison of derived space 
densities of emission line/UV-excess dwarfs is given by Comte et. al. (1994)
and Gallego et al. (1997).
These differences demonstrate once more that different surveys
effectively target different types of galaxies.

\subsection{The spatial distribution of metal-poor galaxies}

Normal giant galaxies are found both in clusters and low density environments in the 
field. In particular giant ellipticals are strongly clustered and
are predominantly found near the centres of rich clusters, while spiral 
galaxies are less clustered. The difference in the galaxy population in clusters
and the field is an important observable, but not easy to interpret in view
of the various  effects that may affect galaxies in different
environments: merging, interactions, harassment, dynamical disruption, stripping, 
cooling flows,  pressure confinement by hot gas etc.

Dwarf elliptical galaxies are found predominantly in clusters or as companions 
of giant field galaxies (Binggeli et al. 1990) and there appears to be a lack of 
isolated field dEs. The last point may in part be due to selection effects, since low
luminosity dEs have low surface brightness and may therefore have been missed by 
most local surveys. New dE members of the Local Group are still being discovered
(Armandroff et al. 1998, Karachentsev and Karachentseva 1999) illustrating the 
point that our view of even local dEs are largely incomplete. Dwarf irregulars 
follow the same structures as those outlined by massive galaxies (Thuan et al. 1987; 
Comte et al. 1994).

The LSBGs are less clustered than 
``normal'' (giant) galaxies (e.g. Bothun et al. 1986a, 1993; Mo et al. 1994), 
in the sense that they tend 
to avoid clusters, and are not found close to field galaxies. However, this
is based on comparison with galaxy catalogues that are badly incomplete for
dwarf galaxies, and especially for LSBGs and dEs. Therefore not much can be said
from these studies about how LSBGs cluster with other LSBGs and faint dwarfs.
Large volume limited samples of LSBGs are required for this purpose. Taylor
(1997) finds from a VLA study of their environments, that about one quarter of 
the LSBGs appear to have H{\sc i} companions, with a detection limit
 of $\sim 10^7 M_{\odot}$.
These H{\sc i} companions tend to have faint optical counterparts (Taylor, private communication), 
and are thus likely to be LSBGs themselves. This suggests that LSBGs need not to  be extremely isolated. It is
however clear that LSBGs avoid massive giant galaxies, which is understandable since LSBGs close to giant 
galaxies have a high probability to interact which
would lead to increase their star formation rate, thus transforming the LSBG into
a high surface brightness galaxy.

Iovino et al. (1985) presented results suggesting that H{\sc ii} galaxies 
are less clustered than normal giants.
Salzer (1989) investigated the spatial distribution of 
ELGs in the UM catalogue, finding that the ELGs follow in most parts the 
structures outlined by bright normal galaxies, but tend to avoid the regions 
with the highest galaxy density. On the other hand Comte et al. (1994) showed that KISO ultraviolet 
excess galaxies were distributed in a similar manner to ``normal'' galaxies.
The spatial distribution of the SBS sample
was investigated by Pustilnik et al. (1995) who found similar results, but
in addition a significant fraction ($\sim 20 \%$) was found in voids.
Most H{\sc ii} galaxies/BCGs seem to be rather isolated when compared to existing 
galaxy catalogues and redshift surveys (Campos-Aguilar et al. 1993, Telles and 
Terlevich 1995). But these results mainly show that BCGs avoid giant
galaxies, since these constitute the catalogues used for comparison. No constraint
can be imposed  on the clustering with other dwarf galaxies, e.g. LSBGs. The
latter have indeed been found to be common companions to H{\sc ii} 
galaxies from
the studies by Taylor and collaborators (cf. Taylor 1997). A recent study (Telles and Maddox 
1999) attempts to address also the dwarf-dwarf clustering by comparing BCGs
with APM (automatic plate measuring machine) catalogues, which contains more low luminosity galaxies, and comes to
similar conclusions  as previous studies in the sense that BCGs mainly populate
environments less dense than normal galaxies. However, even the APM
is badly incomplete for LSBGs, which are the likely companions in view of
Taylor' results.

Given the apparent tendency of BCGs to avoid dense environments, such as 
rich clusters, it has been speculated whether  the voids may be 
filled with BCGs and other faint dwarfs such as LSBGs. If so the current view of the 
baryonic matter distribution in the Universe would be very biased, and a large mass fraction 
would  have been missed. This would be in agreement with  ``biased''
galaxy formation theories (e.g. Dekel and Silk 1986) where dwarfs arise from low density 
peaks in the primordial density fluctuation spectrum. There have been some studies 
addressing this question: Pustilnik et al. (1995) found that 20\% of BCGs may
reside in voids. Popescu et al. (1997) finds some void galaxies, but show 
that the voids are not filled by an undiscovered population of BCGs. 
Similar results are reached by Lindner et al. (1999). BCGs in or near voids
are predominately found near the borders (Lindner et al. 1996).
The cases for other types of dwarfs, e.g. LSBGs, are less clear, but there is
presently nothing that points toward a large density of any galaxy type in 
void, although this should be further investigated.

Dwarf irregular galaxies seem to be abundant in most environments, both in
rich clusters and as pure field galaxies (Binggeli et al. 1980). However there are 
few studies addressing the dIs directly, and we have already seen that this class generally
encompasses many different kinds of low mass irregular galaxies, including BCGs
and LSBGs.

In conclusion, metal-poor BCDs and LSBGs avoid rich cluster environments, but may 
have neighbours of the LSBG type. Dwarf ellipticals are found in clusters or 
nearby  luminous galaxies, while dIs are found in most environments (Binggeli et al. 1990).

%%%%%%%%%%%%%%%%%%%%%%%%%%%%%%%%%%%%%%%%%%%%%%%%%%%%%%%%%%%%%%%%%%%%%%%%%%%%%%%%%%%%%%%%
%%%%%%%%%%%%%%%%%%%%%%%%%%%%%%%%%%%%%%%%%%%%%%%%%%%%%%%%%%%%%%%%%%%%%%%%%%%%%%%%%%%%%%%%
%%%%%%%%%%%%%%%%%%%%%%%%%%%%%%%%%%%%%%%%%%%%%%%%%%%%%%%%%%%%%%%%%%%%%%%%%%%%%%%%%%%%%%%%

\np
\section{Global relations and evolutionary links}

\subsection{The metallicity--luminosity relation, and other empirical relations}

As we saw in sections 3.1 and 3.2, the metallicity of dEs and dIs correlate 
positively with their luminosity (Aaronson 1986, Lequeux et al. 1979, Kinman and Davidson 1981, Skillman et al. 1989). Whether they follow the same relation is a
subject of debate due to the uncertain scaling between [O/H] and [Fe/H],
(see  Sect. 4.2.1;  and Richer and McCall 1995). 
It would be instructive 
to examine the location of BCGs and LSBGs in the metallicity--luminosity diagram. 
The luminosity is usually represented by the B-band absolute magnitude which might 
be significantly affected by ongoing star formation, especially in BCGs, and thus 
is not necessarily a good estimator of the  stellar mass. Indeed, NIR magnitudes
would be a better choice, but data are rather scarce.
Earlier work indicates that BCGs follow such a metallicity--luminosity relation, but with considerable 
scatter (Kunth 1986, Campos-Aguilar et al. 1993). 

In Fig. 10, we show the 
oxygen abundance vs. luminosity diagram for dIs (crosses), BCGs (filled symbols) and 
LSBGs (open symbols), based on data collected from the literature, with available 
abundances and integrated B magnitudes. 
In addition we show the location of candidate tidal dwarfs and dEs with measured 
nebular oxygen abundances. For the BCGs we have divided the sample into morphological
subtypes according to the classification schemes by Telles et al. (1997) and Loose and 
Thuan (1986a). Galaxies with regular outer isophotes, classified as Type II according to 
Telles et al. (1997) or as iE or nE according to Loose and Thuan (1986a), will be referred to
as Type II, while ``irregular'' ones will be referred to as Type I 
 following Telles et al. (1997).
If there is no classification in the literature, but published images are available,
we classified the galaxies according to this scheme. The solid line shows the 
$M_B - Z$ relation for dIs by Richer and McCall (1995), while the short and long 
dashed lines shows the relation for dEs (see caption).

At first sight many BCGs do not appear extreme when compared to the normal dIs. 
Indeed some BCGs are more metal rich than dIs at a given luminosity, while the opposite 
would be expected if BCGs were bursting dIs. These BCGs  may be in a post burst stage 
and the fresh metals may have become ``visible'' already. Secondly, some  ``extreme
BCGs'' appear  much more metal-poor at a given luminosity. These extreme BCGs (XBCGs) 
are 3 magnitudes brighter or more at a given metallicity, or equivalently 0.5 dex less 
metal rich at a given luminosity  as compared to the dI relation. 
There is a tendency for the  Marlowe et al. (1999) blue amorphous 
galaxies  to lie above the dI relation.
Galaxies of Type~II  follow the dI relation in a broad sequence, while 
Type I  have a tendency to fall below the dI relation. The same phenomenon 
is seen for four galaxies  from the sample 
by \"Ostlin et al. (1999a), of which three have irregular morphology (Type I). The most
metal-poor galaxies   also have irregular 
morphologies and fall far below the dI relation. Thus the BCGs span a factor of 10 in metallicity 
at a given luminosity.
The intriguing XBCGs include IZw18 and SBS~0335-052, 
 ESO~338-IG04 (=Tololo~1924-516), ESO~400-G43, Haro~11(=ESO~350-IG38),
ESO~480-IG12 (\"Ostlin et al. 1999a,b; Bergvall and \"Ostlin 1999); UM~133, UM~448, C~0840+1201  
(Telles and Terlevich 1997), UM~420, UM~469 and UM~382 (Masegosa et al. 1994, and 
Salzer et al. 1989a).
LSBGs occupy locations in the range from dIs to XBCGs (see also Bergvall and R\"onnback 1994,
Bergvall et al. 1998). While enriched galactic winds  
could possibly explain the extreme metal deficiency of the least massive dwarfs, the existence of XBCGs
in general cannot be understood in this way. 
While suggestive of important differences between
different samples and types, these trends should be regarded as preliminary and
should be put on more solid ground.

\begin{figure}
\vspace{-2cm}
\hspace{1cm}
\resizebox{12cm}{!}{\includegraphics{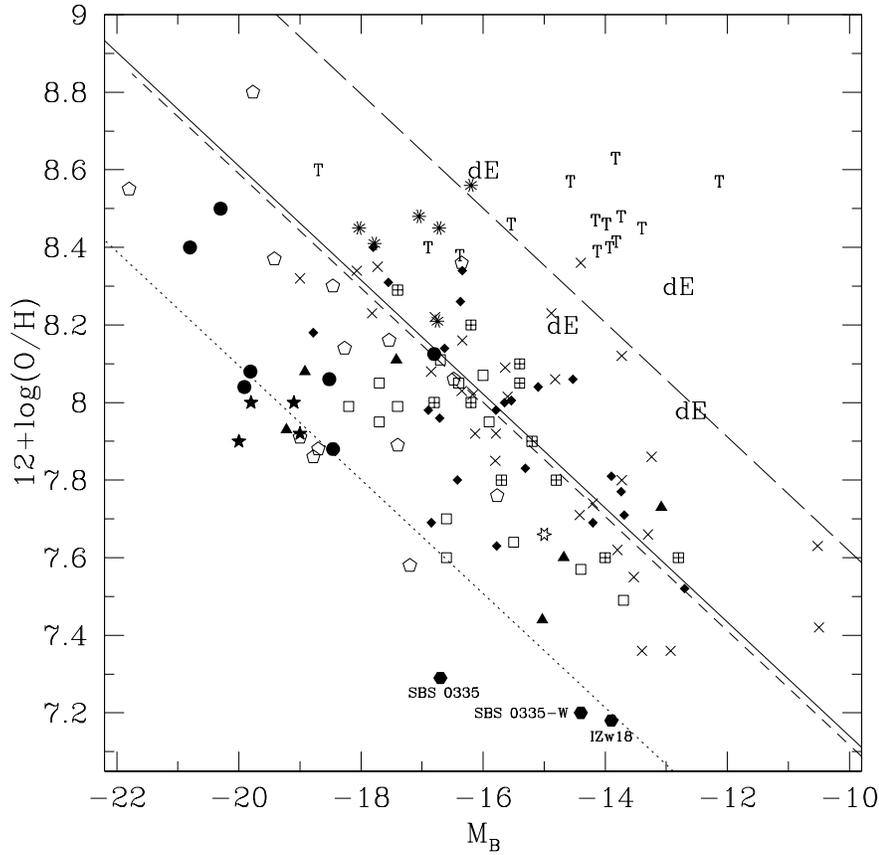}}
\vspace{-0.5cm}
\caption{{\protect\small The luminosity versus metallicity diagram for dwarf galaxies. 
The crosses represent dIs taken from Richer and McCall (1995) and Skillman et al. (1989).
Filled symbols represent galaxies classified as BCGs or H{\sc ii}-galaxies: The small filled
diamonds are ``regular'' galaxies which  can be classified as Type II or iE/nE according 
to Telles et al. (1997) and Loose and Thuan (1986a), respectively, while  filled circles are 
galaxies that can be classified as Type I (see text for description of types). Filled triangles are BCGs
for which no classification or images were available. The three most metal-poor galaxies are
labelled and shown as filled hexagons; their morphology are indicative of Type I. The filled 
stars are luminous BCGs from \"Ostlin et al. (1999a,b), three of which are of Type I.  
The asterisks show the location of ``blue amorphous galaxies'' (Marlowe et al. 1999, except 
for II~Zw40 which is the filled circle falling on the short dashed line). 
LSBGs are shown as open squares (blue LSBGs, Bergvall et al. 1999, R\"onnback and 
Bergvall 1985) and open pentagons (McGaugh 1994, McGaugh and Bothun 1994).
The open star is H{\sc i}~1225+01, the H{\sc i}-cloud in Virgo (Salzer et al. 1991). The boxes with plusses 
inside are quiescent (dI/LSBG) dwarfs from van Zee et al. (1997b,c). Candidate tidal dwarfs (Mirabel et 
al. 1992, Duc and Mirabel 1994, 1998) are shown as ``T'', and dEs  are shown as ``dE'' (data from Mateo 1998). 
The solid line shows the $M_B - O/H$
relation for dIs from Richer and McCall (1995), while the dotted line shows the same relation
offset by 3.5 magnitudes, indicating the location of XBCGs. The short dashed line shows the 
$M_V - Z$ relation for dE/dSph from
Caldwell (1998) assuming $(B-V) = 0.75$ and [O/H] = [Fe/H], while the long dashed line shows
the same relation assuming  [O/H]  = [Fe/H] + 0.5.
When necessary, we have rescaled absolute magnitudes to H$_0=75$km/s/Mpc.}
{\scriptsize Metallicities of BCGs  from:
Izotov and Thuan (1999), Lipovetsky et al. (1999), Bergvall and \"Ostlin (1999), Telles and
Terlevich (1997), van Zee et al. (1998), Kunth and Joubert (1985), Alloin et al. (1978),
Thuan et al. (1996), Masegosa et al. (1994). 
Absolute B-magnitudes for BCGs from: Telles and Terlevich (1997, assuming $B-V=0.5$), 
Papaderos et al. (1996a, 1998), \"Ostlin et al. (1999a), Thuan et al. (1996), Mazzarella and 
Boroson (1993), Salzer et al. (1989a), Schulte-Ladbeck et al. (1998).}}
\end{figure}

Another important relation, between H{\sc I} mass and blue luminosity,
has been addressed by several authors. A positive correlation was found by
Chamaraux (1977) for a sample of Zwicky BCGs. It has been found  (e.g. Staveley-Smith 
et al. 1992) that for gas rich dwarfs, the hydrogen mass to blue luminosity decreases 
with increasing luminosity, i.e. low luminosity galaxies are more H{\sc i} rich, and
have apparently converted a smaller fraction of their neutral gas content into stars. 
Moreover, LSBGs have proportionally more  H{\sc I} than BCG in the sense that 
they have higher $M_{H{\sc I}}/L_B$ values (see e.g. fig 1a in Bergvall et al. 1998). 
BCGs and dIs seem to have similar  $M_{H{\sc I}}/L_B$ ratios.
McGaugh and de Blok (1997) showed that, for a sample of disc galaxies (extending from  normal 
spirals to LSB dwarfs),  $M_{H{\sc I}}/L_B$ increases systematically with decreasing 
surface brightness, and is typically $M_{H{\sc I}}/L_B = 1$~ (solar units) for LSBGs. LSBGs have low
mass densities and have thus been inefficient in converting H{\sc i} to stars and metals.
Thus, to a first approximation, metallicity anticorrelates with the gas mass fraction 
 (cf. Lequeux et al. 1979, Kinman and Davidson 1981, Pagel 1997)
as  expected from closed box models, but this cannot be the whole explanation 
(Matteucci and Chiosi 1983). Dust--to--gas ratios positively correlate with 
metallicity for dIs, while BCGs appear comparably more dust rich (Lisenfeld and Ferrara 
1998), although in many BCGs one can only put upper limits to the dust content.
The general relation between surface brightness and luminosity
for dwarf and late type galaxies (Ferguson and Binggeli 1994), implies that 
dwarfs have low mass densities and/or are inefficient star formers.
It has been argued that this could be a pure selection effect, since faint
low surface brightness would be more difficult to detect (Phillipps et al. 1988).

Terlevich and Melnick (1981) report a positive correlation between the H$\beta$ luminosity,
the H$\beta$ line width, and linear size for giant H{\sc ii} regions and H{\sc ii} galaxies. 
Subsequent work confirmed a dependence also on metallicity (Melnick et al. 1987, se also 
Campos-Aguilar et al. 1993). This also opened up the interesting possibility to use 
H{\sc ii}-galaxies as distance indicators (Melnick et al. 1987, 1988; see Sect. 8.6).
If, as Melnick et al. (1987) argue, the  H$\beta$ line-width
is due to virial motions, this relation reflects an underlying dependence on galaxy mass.
Whereas this relation cannot hold for dwarfs irregulars with low star formation activity,
it is striking that low mass galaxies with strong star formation activity seems to form 
a well defined sequence. This relation may reflect an
 intrinsic upper limit, perhaps regulated 
by  feedback, on the possible star formation rate (directly proportional to the H$\beta$ 
luminosity) as a function of mass or mass density. These galaxies would then represent the most
efficient star formers with respect to their mass. Similarly, Meurer et al. (1997) find
an upper limit on the bolometric surface brightness in starbursts at high and low redshifts,
implying the existence of a physical mechanism limiting the global areal star formation rate 
in galaxies.

\subsection{Evolutionary scenarios and connections}

Like all other galaxies, dwarf galaxies evolve: the extremely gas poor dEs must 
have originally  contained gas to form their observed stellar populations. Moreover,
many BCGs have unsustainable star formation rates and  therefore represent a 
transient stage, unless they form stars with a very different IMF. Thus, evolutionary
connections between different dwarf types must exist, unless some initial conditions 
determined the future evolution of different types of galaxies. Even if 
links exist, there might be several different, physically distinct
channels in the evolution of dwarfs. A discussion  like this necessarily becomes
speculative, and we invite the readers to make their own judgement.

Several evolutionary scenarios that link different types of dwarfs have been
discussed over the years. One can think of two different types of evolution,
either {\it internal} or {\it passive} where the evolution of a galaxy proceeds,
according to its physical (initial) conditions or through {\it external} effects 
such as mergers, or interactions with galaxies 
or intergalactic matter. In the latter case, the environment will be a key parameter.
Possibly, cluster and field galaxies evolve in a similar manner, but the clock runs faster 
in a high density environment. Since dEs are found in clusters
or as companions to field giants, their evolution is likely related to
their environment. 

Lin and Faber (1983) suggested that the dSph satellites of the Milky Way were
dIs that had lost their gas by ram pressure stripping (see also van den Bergh 1994). 
It was later pointed out that this could not be the general explanation since, on average, 
dEs have higher surface brightness than dIs at a given luminosity (Bothun et al. 
1986, Ferguson and Binggeli 1994). However, this is mainly a problem for relatively
luminous dEs (Skillman and Bender 1995). Searle et al. (1973) and Thuan (1985) proposed that BCGs 
originate in low surface brightness dIs, an idea that was further elaborated
by Davies and Phillipps (1988) where they suggested that dIs evolve into dEs
after a number of bursts in the BCG stage. The star formation is regulated by
continuous gas infall from the halo, but this model includes no physics to 
explain the suggested behaviour. One can also 
imagine a cyclic BCG-dE scenario where a starburst in a BCG gives rise to
a superwind, which expells the gas and halts star formation. The expelled gas
later cools and falls back on the galaxy, creating a new starburst. Silk et al. 
(1987) proposed that the BCG phenomenon could be explained by gas expelled from 
dwarfs at high redshift, now accreting on dwarf ellipticals (see also Babul and Rees
1992). However, this can hardly work for the field BCGs since there are seemingly very few
dEs to accrete onto.
Gas loss through supernovae driven winds has been a popular mechanism for 
forming dEs
(Larson 1974, Vader 1986, Dekel and Silk 1986), but although outflows are observed in some 
nearby dwarfs (Sect. 3.3) they do not appear capable of clearing a galaxy of its ISM.
Galaxies that have managed to retain gas until the present epoch are unlikely
to loose it now, and become dEs (Ferrara and Tolstoy 1999).
Skillman and Bender (1995) point out that the majority of Local Group dEs/dSph
formed most of their  stars in an initial burst, while dIs have had more
extended SFHs. However, there has been quite some progress in  recent 
years in  unveiling the SFHs of Local Group dwarfs, and the distinction 
between SFHs of dIs and dEs has been somewhat blurred  (Sect. 4.1, 4.2; Grebel 1998).
Despite being gas poor, most Local Group dEs have significant young or
intermediate populations, and  in active SF phases they would appear as
BCGs or dIs depending on the extent of these episodes. The best example is probably 
Carina (Smecker-Hane et al. 1994) which went through a major SF
episode some Gyrs ago, demonstrating that dI-dE transitions may have occurred fairly recently. 
A thorough discussion on the 
the origin of dEs is given by Ferguson and Binggeli (1994), see also Skillman and 
Bender (1995).
Meurer et al. (1992) 
argued that  BCGs like NCG~1705 could evolve into  nucleated dEs. Similarly, it 
has been suggested that compact galaxies at higher 
redshifts ($z= 0.5 ~{\rm to}~ 1$) are the progenitors of the present day dEs
(Koo et al. 1995). Since outflows seem rather inefficient, the only way for field
BCGs to evolve into dEs (of some kind) is through total gas consumption, which
requires very high star formation efficiencies. Field dEs appear too rare, and field Es
typically too massive to be the successors of most BCGs, but it may apply for
the most massive BCGs, especially those seemingly involved in mergers.

Many dIs appear more simple and  able to form 
stars at more or less continuous rates over long time scales. Although the details
of their star formation activity are not completely understood, there is less 
{\it need} for evolutionary connections since their star formation activity is 
typically sustainable over a Hubble time or more. Given the arguments above, dIs
are not likely to passively evolve into dEs (unless perhaps on timescales 
comparable to the age of the Universe or longer).
However there might be connections to  dEs (especially in clusters) and BCGs (some of which may 
fade into dIs). Given their large number 
and richness in gas,  LSBGs may serve as important fuel for star formation.

\paragraph{Constraints on BCG evolution from photometric structure:}
BCGs have on average higher central surface brightnesses than dIs, even after
the starburst component has been subtracted (Papaderos et al. 1996b, Marlowe et al. 1999),
which argues against  links between BCGs and dEs/dIs,  unless the structure can 
change during the bursting phase. (Note however that this may be an artefact due to insuffiently
deep data, see Sect. 4.4.1). Papaderos et al. (1996b) propose that  in the 
initial phase a BCG may contract because of accretion, while the successors of 
BCGs may expand due to gas loss. However  the required expansion/contraction
necessary to explain structural differences would require at least 50\% of mass 
 loss/gain (Marlowe et al. 1999) which is quite unrealistic given the apparent inefficiency
of winds,  the observed gas mass fractions and large DM content of dIs. Marlowe et al. 
(1999) also note that
the burst, as defined from excess surface brightness over a disc fit, is 
quite modest, and that a similar excess is seen in some dIs (although the underlying 
surface brightness is lower). Then  BCGs may simply be regarded as unusually active 
dIs, because of higher central mass densities (van Zee 1998c). In the non burst phase, they
will  be classified as a high surface brightness dIs or as 
amorphous galaxies  or BCGs with low emission line equivalent widths. 
This may be the case for  many BCGs, especially those close to the dI relation in
Fig. 10, but  hardly for the XBCGs discussed below.

\paragraph{Constraints from luminosity, metallicity, and gas content:} 
Some BCGs  are much more metal-poor at a given luminosity
than the dIs forming the basis of the metallicity--luminosity diagram. We called such objects  
``extreme BCGs'' (XBCGs).
Thus, under the conservative assumption that the current burst  has not yet affected the 
observed nebular abundances, a dI would need to brighten by 3 magnitudes to become
an XBCG. That amount of brightening is not observed from profile fitting
and is unrealistic in view of the observed H{\sc i} masses as the following example illustrates:

Imagine a typical dI with $M/L_B=2$ for its stellar population, on which we add a 10 Myr 
old burst. The burst should have $M/L_B > 0.05$ for a  Scalo IMF (Bruzual and Charlot 2000). 
To accomplish a brightening with 3 magnitudes, the burst has to be 15 times brighter than the 
underlying galaxy, and hence make up more than one third of the total stellar mass. This would 
require, for realistic star formation efficiencies of say 10\%, that the dI precursor had a 
neutral hydrogen mass several times larger than its total stellar mass. The observed gas mass 
fractions are much lower, and therefore  XBCGs cannot originate in bursting dIs (see also Bergvall et al. 1998).
On the other hand, a LSBGs turning on a burst would need to increase its absolute luminosity 
with only $\sim 1.5$ magnitudes to meet the location of XBCGs in the $M_B - Z$ diagram. This would simultaneously
reproduce the $M_{H{\sc I}}/L_B$ values for BCGs (Bergvall et al. 1998). Thus LSBGs are more 
likely to be the precursors of XBCGs than dIs.
Moreover, Telles and Terlevich (1997) found the colours of the underlying 
component in BCGs to be consistent with those of blue LSBGs.
This argument against dI--BCG evolution does however not apply to BCGs which are found 
close to, or above, the metallicity--luminosity relation for dIs. These galaxies must
 definitely have a different history from the XBCGs

\paragraph{Mass vs. metallicity:} 
Another indication that dIs like those in the Local Group are not the progenitors
of the most metal-poor BCGs comes from a comparison with total mass, rather than 
absolute blue magnitude. Data for BCGs are scarce, but taking IZw18 (van Zee et al. 1998a), 
SBS~0335-052 (Papaderos et al. 1998) and luminous BCG from the \"Ostlin (1999a,b)
sample and comparing them to Local Group dIs (data from Mateo 1998) it is clear 
that XBCGs appear to be  an order of magnitude more massive than dIs at a
given metallicity. Thus the scatter in the metallicity--luminosity diagram is not
primarily due to various degrees of star formation affecting the B-luminosity.
Interestingly, metal-poor LSBGs (like ESO146-IG14, Bergvall and R\"onnback 1995;
 and some  galaxies in the sample of de Blok et al. 
1996 and McGaugh 1994) fall on the XBCG mass-metallicity relation. However, three BCGs studied by van 
Zee et al. (1998) and NGC1705 (Meurer et al. 1998) fall in the dI range.
Quiescent dwarfs (dI/LSBG) from van Zee et al. (1997,1997) overlap with BCGs.
The BCGs overlapping with dIs are of the regular type II.
This comparison indicates that different types  may indeed have a different origin, but
 we caution that 
this excursion into studying the mass versus metallicity was based on inhomogeneous 
data  from various sources, using various methods. Thus, it should not be taken at 
face value, but as a motivation for further studies.

\paragraph{Dwarf mergers?} 
Several investigators have noted the wide range of morphologies displayed by BCGs
(see Sect 4.4.1). Telles et al. (1997) note that luminous BCGs (Type I) on average have a
more perturbed morphology and Telles and Terlevich (1997) speculate that this may be 
related to mergers or interactions, but they note that merging is disfavoured, 
based on the clustering properties of BCGs. As we discussed in Sect. 6.3 this 
conclusion might be too pessimistic since the present investigations may be 
incomplete for dwarfs, especially LSBGs.
Taylor et al. (1993, 1995, 1996a, 1996b) investigated the H{\sc i} environment of BCGs and LSBGs
finding that 60 \% and 30 \% of them respectively have H{\sc i} companions with faint optical counterparts,  with a typical
detection limit of $10^7 M_{\odot}$.  From the higher fraction of H{\sc i} companions around  BCG, 
Taylor (1997) argue that LSBGs are not the likely progenitors of BCGs. However, this may be a
selection bias in the sense that the more frequent are the H{\sc i} companions, the more
probable mergers will be.

From a kinematical and morphological study \"Ostlin et al. (1999a,b) found  that a sample of
luminous BCGs (most of them XBCGs) were likely the product of mergers, involving gas 
rich dwarfs or H{\sc i}-clouds. 
A merger can provide the necessary change of the structure of the host and offers a mean for 
changing the kinematics of galaxies, from rotating (like dIs) to essentially non-rotating 
(like E and dE).  A critical point, especially if some BCGs are to evolve into E/dE,
is how efficient the gas consumption may be? Galaxies with a large gaseous disc
at moderate column densities are not likely to be able to consume this gas on 
a short timescale. In mergers, the star formation efficiency might be
higher, especially if globular clusters are able to form (Goodwin 1997). Van Zee 
et al. (1998c) point out that BCGs have higher central H{\sc i} densities than normal
dIs. This is a natural explanation for the high star formation rates, but some
mechanism is needed to put the gas where it is, since the gas consumption time 
scale is rather short. If gas consumption is less efficient, perhaps because gas 
settles into a rotating disc, the merger remnant would rather evolve into a dI
or an amorphous galaxy. If a merging gas rich dwarf contains a rather unpolluted 
HI halo, this could dilute the metal enriched ISM
and lower the observed abundances, to produce XBCGs. 
Clearly, mergers would be able to explain many properties of BCGs,
but of course this does not imply that merging is the general mechanism, especially
concerning those with regular morphology.  The absence 
of morphological perturbations in Type II BCGs may however be an observational effect. 
Since these systems are on the average fainter, their underlying components have low 
surface brightnesses, and thus morphological irregularities such as tails will have low 
surface brightnesses and be difficult to detect.

\medskip

We have fast outlined some ideas about evolutionary connections. It is  clear 
 that the overall picture is not yet 
well understood. There is probably no unified scheme that can explain all dwarfs. 
The origin and evolution of dE/dSph galaxies is probably related to their environment, and
it cannot be ruled out that some dE/dSphs are stripped dIs, although dIs are not likely 
to passively evolve into dEs. 
Many BCG-like galaxies may simply be extreme dIs. 
Some BCGs, especially the XBCGs,
may be triggered by mergers involving gas-rich galaxies such as LSBGs. 
Of course if some BCGs are truly young, their progenitors must be pure gas clouds
and not visible as galaxies. Such gas clouds have however not been found, and although 
the existence of genuinely young BCGs at the current epoch cannot be ruled out, we do not 
consider that there is any BCG where there is an unambiguous evidence for pure youth 
(cf. Sect. 4.4.2 and 5).

%%%%%%%%%%%%%%%%%%%%%%%%%%%%%%%%%%%%%%%%%%%%%%%%%%%%%%%%%%%%%%%%%%%%%%%%%%%%%%%%%%%%%%%%%%%%%
%%%%%%%%%%%%%%%%%%%%%%%%%%%%%%%%%%%%%%%%%%%%%%%%%%%%%%%%%%%%%%%%%%%%%%%%%%%%%%%%%%%%%%%%%%%%%
%%%%%%%%%%%%%%%%%%%%%%%%%%%%%%%%%%%%%%%%%%%%%%%%%%%%%%%%%%%%%%%%%%%%%%%%%%%%%%%%%%%%%%%%%%%%%
\np

\section{Metal-poor galaxies, cosmology and the early Universe}

\subsection{The primordial helium abundance}

According to conventional wisdom, helium is  mainly produced during the first few
minutes of our observed Universe and subsequent stellar nucleosynthesis has
added at most 10 to 20 \% to its primordial value. An accurate measurement
of the primordial Helium is critical for our understanding of the origin of
the Universe because, under the assumption of the standard hot Big Bang model
of nucleosynthesis, it will eventually provide a fundamental parameter in
cosmology, that is, the nucleon density in the early Universe.
 Up to now, the most accurate method of measuring helium abundance consists
of observing diffuse gas which has been ionised by the ultraviolet photons from
hot stars in star forming dwarf galaxies, in particular BCGs. The goal being 
to measure $Y$, the helium mass fraction present in ionised gas of low heavy 
element abundance. Early studies have pioneered a method by which one can
extrapolate the He/H versus O/H relation to the primordial abundance, $Y_p$, as the
oxygen abundance tends to zero (Peimbert \& Torres-Peimbert 1974;
Lequeux et al. 1979). The extrapolation is rendered easier by choosing
extreme metal-poor galaxies. Extensive
searches are still underway to select blue low mass galaxies with even lower
oxygen abundance and of highest excitation. This last point is crucial, since
a good oxygen and helium abundance determination is only possible if the
temperature of the gas can be obtained from the very few critical forbidden
line ratios which can be properly measured (Kunth and Sargent 1983).
Because useful constraints can only be achieved if the He abundance is 
measured to better than 5\% accuracy this exercise needs care at
each step that from the observations leads to the final derivation.
 Problems
connected with fluorescent and collisional excitation affecting the helium
lines must be addressed as well as the effect induced by electron
temperature fluctuations. Another irritating question is  the amount of
 neutral helium to correct for, because it
remains undetected (Masegosa et al. 1994). The stellar population, due to the 
most recent burst of star
formation, produces underlying absorption in the H and He nebular lines (Izotov and Thuan 1998a).
To circumvent this difficulty, high resolution spectroscopic data are
needed.
Despite improved observations, the scatter in the $Y$ vs $Z$ diagram has
remained large (see e.g. Kunth 1996). Is this due to systematic errors or an intrinsic cosmic
dispersion? It has been argued that galaxies with high $Y$ might be affected
by localised pollution from Wolf-Rayet (WR) stars (Pagel et al. 1986, 1992), and
hence that galaxies showing (WR) features should be excluded from the analysis.
However, this has been disputed by Esteban and Peimbert (1995) and Kobulnicky 
and Skillman (1996) who find no evidence for localised pollution in WR galaxies.

The problem of the recombination coefficients for 
helium has been a subject of controversy after Smits (1991) published new 
values that differed from those of Brocklehurst (1972), but these calculations
were  discovered later on to be in error by Smits himself (1996). A new grid of emissivities
improving the predictions for He emission lines are now available (Benjamin et al. 1999).  
Problems inherent to the observations and reduction processes have also been
considered and have led to at least two distinct approaches. One method (Skillman et al. 1994)
involves repeated observations with different systems (telescopes, standard
stars, reduction procedures, etc.) while
Thuan and Izotov (1998a) have favoured an
approach by which the same procedure (telescope, etc.) is used to ensure
more homogeneity of the data. Reviews of some problems in connection with deriving
the primordial He abundance can be found in Terlevich et al. (1996) and Kunth (1996).

The last estimate of the 
primordial helium abundance (Izotov and Thuan, 1998a) using a  "self-consistent"
treatment, gives $Y_p =0.244 \pm 0.002$~ (see Fig. 11), a value slightly higher than that of 
Pagel et al. (1992), i.e. more in line with the predictions of the 
Standard Big Bang nucleosynthesis. 
 Perhaps equally important, is a
considerable scatter in the helium values at a given metallicity: an
indication of possible differences in the chemical evolution history
of these dwarf galaxies. This scatter remains, although 
 observational errors have been considerably reduced over the last
decade with the advent of linear detectors.
 The new Izotov and Thuan (1998a) data are available in a way 
that  offers the
possibility to  check various 
pending observational problems 
(such as a grating  ``second order''  contamination that is possibly
present beyond 7200\AA\ and may affect redshifted objects in the 7065\AA\ HeI
line) and  perform independent analysis. Their work however still leaves room for 
further investigation in particular to check the importance of underlying  stellar 
 absorption that may  artificially weaken some HeI lines. It is for this 
unfortunate
 reason that region NW in the most deficient metal-poor IZw18 is difficult to use 
and has been alternatively rejected and reintroduced by Izotov and Thuan (1998a,b) 
in the final derivation of $Y_p$. 
 The advent of 2D spectroscopy will also help to better take into account 
 stratification effects in the nebula and localise possible contaminations
 due to WR stars and supernovae.

\begin{figure}
\hspace{1.0cm} \resizebox{12cm}{!}{\includegraphics{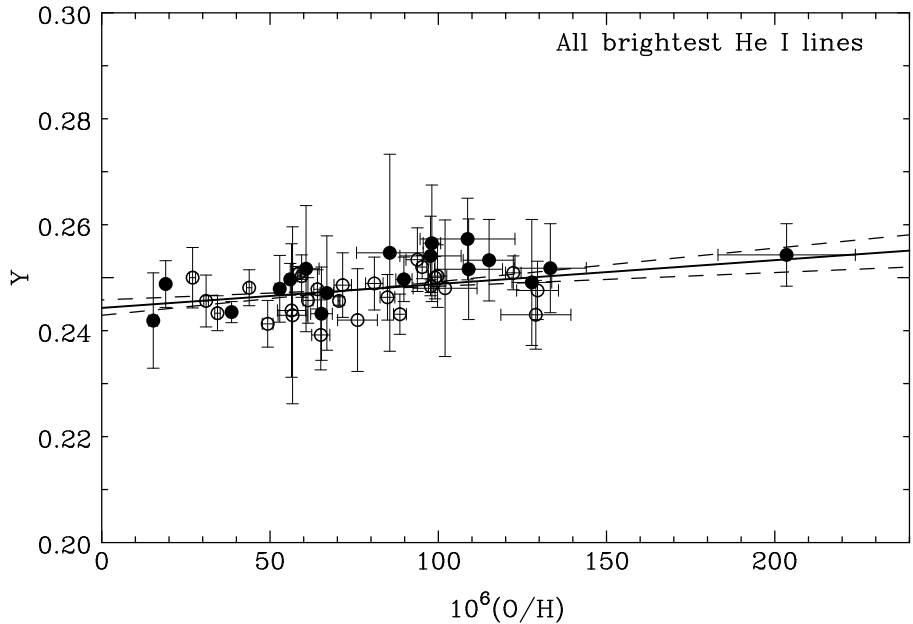}}
\caption{{\protect\small 
Example of recent derivation of the primordial He abundance through regression analysis.
The meaning of filled vs. open symbols is described in Izotov and Thuan (1998). 
(Courtesy T.X. Thuan.)}}
\end{figure}

\subsection{QSO-absorption line systems}

Spectra of high redshift QSOs display absorption lines from intervening 
HI clouds, arising when the QSO continuum is absorbed at the local 
rest wavelength of the resonant Ly$\alpha$ line. The QSO absorption systems
come with a range of column densities from below 10$^{14} {\rm cm}^{-2}$, the
Ly$\alpha$ forest, to more than 10$^{20} {\rm cm}^{-2}$, the Damped Ly$\alpha$
systems (DLAs).

Metal absorption lines associated with QSO absorption line systems  can be used 
to study the metallicity and its temporal evolution in interstellar/intergalactic gas. This 
field of research has experienced rapid progress recently thanks to the new 
generation of 10m class telescopes. Now even low density  Ly$\alpha$ forest
systems can be studied (but not those with too low $N_H$) and have been found to
 contain heavy elements, on average [C/H]$\sim -2.5$,
although the metallicities are still very uncertain due to the limited understanding
of the physical condition in these systems (Pettini 1999). 

The DLAs are better understood in terms of physics though their true nature is 
still not clear, and are on average more metal rich. At redshifts above ~$z=1.5$~
Pettini et al. (1997) and Prochaska and Wolfe (1999) determined an average zinc
(where problems with dust depletion are expected to be small) abundance of
[Zn/H]$=-1.15$. There  is a large scatter at all redshifts, and no strong trend in
metallicity with redshift (but see also Lu et al. 1996). It is generally assumed 
 that DLAs are associated with 
galaxies at an early stage of evolution.  It is not
clear whether all remote galaxies at redshift larger than 2 are
chemically young systems or whether damped  Ly$\alpha$ lines
preferentially pick out unevolved systems.

Pettini et al. (1995) explore the possibility to measure the 
N/O ratio at a metallicity lower than those of the most metal-poor dwarf galaxies 
known. They find that relative to the Sun, N is more underabundant than O by at 
least a factor of 15 in agreement with Lu et al. (1996, 1998). Izotov and Thuan (1999) 
attribute this low N/O to uncertainties related to unknown physical conditions in 
the interstellar medium of high-redshift galaxies that make the correction
factors for unseen low-ionisation species difficult to assess. On the other hand 
Pilyugin (1999) manages to account for the discrepant behaviour of DLAs and low 
metallicity BCGs, if assuming (among other things) that DLAs have had their 
last star formation event less than 1 Gyr ago, whereas the observed metals
in the most metal-poor BCGs have been produced in a previous event, more than
1 Gyr ago.

Vladilo (1998) proposes that the abundance pattern in DLAs is more consistent
with them being dwarf galaxies rather than disc galaxies. By dwarf galaxy,
an LMC like galaxy was considered, i.e. much more massive than the most metal
poor dwarfs which have narrow H{\sc i} profiles, but consistent with the most luminous
BCGs and dIs. However it is not clear whether this agrees with the observed 
kinematics of DLAs (Prochaska and Wolfe 1997). Observations of galaxies associated 
with lower redshift DLAs present a wide variety of morphologies (Le Brun et al. 1997). 
Thus it may be the case that DLAs are not linked with a specific kind of galaxy,
but with all sorts of galaxies.

Although the nature of DLA galaxies and their relation to galaxies seen at
low redshift is far from clear (Pettini 1999), they are among the most metal
poor galaxies known. With a dozen 10m class telescopes available in the beginning
of the next decade, there is good hope that our understanding of these systems will
significantly improve.

\subsection{Star forming galaxies at high redshift}

It has been conjectured that primeval galaxies  
would be nearly dust free and devoid of metals at their early stages ( but see
Puget and Guiderdoni 1999).
The search for ``normal'' galaxies (i.e. where light is generated by stars) at
high redshifts has seen a tremendous progress in the last years thanks 
to programs employing the Lyman break technique (cf. Steidel et al. 1995, 1996, 1999), 
and  projects like the Hubble Deep Field (HDF, Williams et al. 1996).    
Steidel and collaborators (1996) have shown that on the order of a few percent
of the galaxies at faint optical magnitudes (${\cal R} \sim 25$) are actively
star forming galaxies  at redshift greater than 3. 
With the assumption that the bulk of  emission is produced by photoionisation of 
stars, their star formation rate is found to range from a few up to a thousand solar masses
per year, hence much larger than in local metal-poor, low mass, galaxies such as IZw18. 
Morphologically  many appear to be star-forming spheroids smaller than the bulge of 
a spiral galaxy. They may be part of the reservoir from
which many of today's luminous galaxies were formed through hierarchical
merging.  Presently, the true mass and the metallicity of these galaxies is
poorly known.  Only a few  have been followed up using near-infrared
spectroscopy (Pettini et al. 1998) showing that their dust content (as
obtained from rest-frame visible light) does not require a major
reassessment for the star formation rates derived from the rest frame UV. 
Masses obtained from line-widths lead to masses of $10^{10}M_{\odot}$, hence  
larger than that of the local dwarf galaxies we have reviewed throughout this paper, and
furthermore these may even be underestimated (Pettini et al. 1998).
There are indications that  energetic outflows are taking place,
similar to what is now observed in local starbursts (Kunth et al. 1998). 
Broadened interstellar lines are most likely the result of mechanical input
from winds and supernova. Metallicity estimates using synthetic stellar-wind
profiles  calibrated against a local star-forming galaxy sample suggest
Magellanic Cloud-like composition (Leitherer 1999).
Although many galaxies are found to be extremely dusty even at large redshift (Puget and 
Guiderdoni 1999), it is possible that there exists an intrinsically fainter population of dust-free 
star-forming galaxies that could represent the earliest phases of galaxy formation. 
New Ly$\alpha$ emitters are now found at high-redshift from surveys using large 
telescopes with narrow-band filters (Hu et al. 1998, Pascarelle  et al. 1998). Limits down 
to a few $10^{-18}$erg/cm$^2$/s are now reachable and give access in principle, to galaxies 
with fainter continuum magnitudes than the Lyman break galaxies (see Fig. 12). 
On the other hand  local starbursts (Kunth et al. 1998) indicate
 that an unknown 
fraction of the youngest galaxies may not be Ly$\alpha$ emitters.
The systematic search and discovery of this kind of objects should offer the 
opportunity of studying  processes 
of star and galaxy formation and evolution at a substantial cosmological look-back time. 
Unfortunately even the brightest emission lines  objects are difficult to follow up, 
and only  the advent of 8m class telescopes with near-IR spectrographs
will allow  to tackle the difficult task of measuring metal abundances.

Deep optical galaxy counts have shown a  strong excess of faint blue
galaxies, which however is absent in near-infrared $K$-band surveys (Tyson
1988). Redshift surveys find that the bulk of this 
faint population is at intermediate redshifts ($z<0.6$; see e.g. 
Broadhurst et al. 1988; Colles et al. 1990; Lilly, Cowie \& Gardner 
1991). It has been proposed (see also Cowie et al. 1991, Babul and Rees 1992),
that the faint blue excess could be explained by a population of 
star bursting dwarf galaxies at intermediate redshifts. It has been demonstrated that
the blue excess can be identified with a population of irregular/peculiar
galaxies (e.g. Glazebrook et al. 1995, see  Ellis 1997 for a review).

Spectroscopic follow up of compact galaxies in the  flanking fields around the
Hubble Deep Field (HDF) 
have shown these to be active star forming galaxies at $0.4 < z < 1.4$ with 
narrow emission line. In  many respects these objects are similar to
the most luminous local BCGs and H{\sc ii} galaxies (e.g. Guzm\'an et al. 1997).
Their metallicities are largely unknown but do not seem to be very low.
Similar high redshift blue compact galaxies were found in the Canada-France
redshift survey (Schade et al. 1995, Hammer et al. 1997).
Kobulnicky and Zaritsky (1999) studied rather luminous emission line galaxies
at redshift 0.1 to 0.5, and found oxygen abundances from $1/5 Z_{\odot}$
to  $Z_{\odot}$, and N/O values consistent with local galaxies of similar
metallicity.

If, as in the local Universe, low metallicity galaxies at high redshift would 
be dwarf galaxies of low or moderate luminosity, they would be missing in existing 
surveys. Indeed, the luminosity 
function of LBGs at faint magnitudes is very steep, indicating a large population
of intrinsically fainter objects lurking below the current detection threshold
(Steidel et al. 1999)

\subsection{Metal production at high redshift}

The work of Steidel and  collaborators (e.g. Steidel et al. 1996)
has confirmed  a substantial population of star-forming galaxies at ~$z \sim 3$,  
with a  comoving number density of roughly 10 to 50 \% that of present
day luminous ($L \ge L^{\star}$) galaxies. From these data it has been possible to 
sketch - although with large uncertainties and questions - the star-forming history 
of the Universe at high redshifts (Madau et al. 1996). From this work, 
and other similar samples, e.g. from the HDF,  the overall SFR of the galaxy population seems to 
increase from ~$z=0$~ to ~$z=2$, during which time a significant fraction of the heavy 
elements in the Universe are formed, while it tails off again towards higher redshift.
Recently, Steidel et al. (1999) questioned the decrease in the number of
star forming galaxies at high redshift, and suggest that the cosmic star formation
rate as inferred from Lyman break galaxies has been constant from $z= 1$~ to ~$z= 4$~.
Recent results from sub-mm observations of redshifted dust emission with the SCUBA
bolometer array (e.g. Hughes et al. 1998) and from the ISO satellite (Aussel et al. 1999; 
Puget and Guiderdoni1999) suggest that a significant amount of 
high redshift star formation may be dust obscured, meaning that estimates of the cosmic
metal production rate from optical observations might be too low.
The cosmic metal production history may also be constrained by observations
of the extragalactic background at different wavelengths (see Pagel 1997 for
a discussion).

Numerical simulations by Cen and Ostriker (1999) predict the evolution of the 
metal content of the Universe as a function of density, by incorporating star 
formation and its feedback on the IGM. At a given density (corresponding e.g. 
to a rich cluster, a disc galaxy, dwarf etc.) their models predict an evolution 
with redshift, but more importantly, the metallicity is found to be a strong
function of density. At low redshift, low density environments will still be 
very metal-poor as suggested by the presence of very metal-poor gas-rich dwarf
galaxies, and Ly$\alpha$ absorption systems (Shull et al. 1998). Ferrara and 
Tolstoy (1999) find in their simulations of the feedback from star formation 
on the ISM in dwarf galaxies, that low mass galaxies could 
be the main contributor to  metals in the IGM.

The idea of so called  population III objects has been discussed for quite a while.
These hypothetical objects would be (or host) the first generation of stars, appearing 
before the main epoch of galaxy formation, and thus be a source of pre enrichment in the 
Universe. The subject remains speculative, but the existence of Pop III stars
cannot be ruled out (Cayrel 1996).

\begin{figure}
\hspace{2.0cm} \resizebox{10cm}{!}{\includegraphics{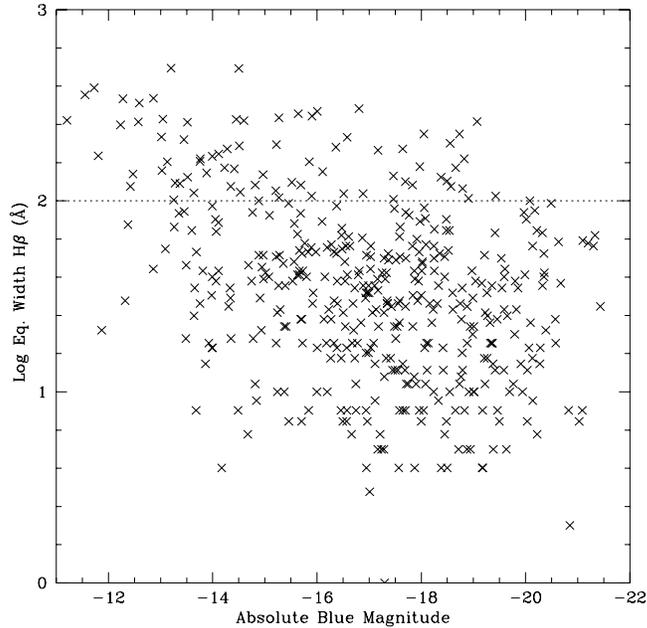}}
\caption{{\protect\small 
The relation between absolute blue magnitude and equivalent with of H$\beta$
for a sample of H{\sc ii} galaxies (Terlevich et al. 1991). The apparent trend 
may in part be due to selection effects. (Courtesy R. Terlevich)}}
\end{figure}

\subsection{The r\^ole of dwarfs in hierarchical structure formation}

What r\^ole do metal-poor dwarf galaxies play in the hierarchical
build-up of galaxies  in the early Universe? Observations 
suggest high merger frequencies at high redshift, but what is merging? 
Certainly not just normal spirals but more likely dwarfs that  may or
may not have counterparts in the local Universe.

If galaxies form  by the cooling and condensation of gas within dark
matter halos, the ability to detect them depends on the sizes of the pieces
out of which galaxies are assembled and on the time interval during  which this
takes place.  There have been suggestions that the principal galaxy building blocks
are to be found throughout the redshift range of about ~$z = 2$~ to ~$5$. In any case, such young
galaxies would, in their earliest phase, be expected to have strong emission lines but 
faint continua and would hence fall far below the magnitude threshold of current Lyman
break surveys. In fact there is a trend for local galaxies (see for
instance the Terlevich et al. 1991, catalogue of H{\sc ii} galaxies, and Fig. 12) with very
strong emission lines to have fainter magnitudes. Moreover, if galaxies are
more inclined to be in groups and clusters at the time of formation they
offer an interesting challenge for existing CDM models. HST images obtained
by Pascarelle et al. (1998) have revealed emitting objects in sheets or
clustered in ways that fit the theoretical picture of Rauch et al. (1997).
Observations suggest high merger frequencies at high redshift, but again the nature
of the merging blocks  (masses and metallicity) remains to be investigated.

Have today's giant galaxies (big spirals and ellipticals) ever been metal-poor?
The intuitive answer would of course be -- yes. But, if massive galaxies are 
built up by successive merger of small galaxies, it is 
possible that they were already rather metal rich when reaching a total mass
comparable to that of an $L^{\star}$ galaxy. In that case, the only metal-poor
galaxies at any redshift will be low mass galaxies,  difficult to
study at cosmological distances.

A long standing problem in galaxy formation models is the so called 
overcooling problem (White and Rees 1978), leading to an overproduction of dwarf 
galaxies at the expense of massive galaxies, in disagreement with observations of 
the luminosity function of galaxies. This problem still remains in current
galaxy formation models (cf. Somerville and Primack 1998). Various mechanisms,
have been proposed such as SN feedback and a high UV background keeping dwarf sized clouds ionised, 
preventing them to cool (Efstathiou 1992). The problem may be 
somewhat relaxed if most dwarfs have merged into giants or have been disrupted. 
If some low mass galaxies formed before the Universe was re-ionised and subsequently 
experience a blow-out of their ISM, they could  represent population 
III objects, responsible for  cosmic metal enrichment at very early epochs (see 
Ferrara and Tolstoy 1999 and references therein).

\subsection{H{\sc ii} galaxies as distance indicators}

It is established that giant extragalactic H{\sc ii} regions (GEHRs) display
a  correlation between their intrinsic luminosities and the width of 
their emission lines, the $\sigma$--$L({\rm H}\beta )$ relation (Terlevich \& 
Melnick 1981). Melnick et al. (1987,1988) have shown that the 
relation found for GEHRs holds also for H{\sc ii} galaxies. The scatter in the 
$\sigma$--$L({\rm H}\beta )$ relation is small enough that it can be used to 
determine  distances.

Recent work with HIRES at the Keck (Koo et al. 1995, Guzm\'an et al. 1996) has shown that a 
large fraction of the numerous compact galaxies found at intermediate redshifts have kinematical 
properties similar to those of luminous local H{\sc ii} galaxies. They
exhibit fairly narrow emission line widths ($ \sigma = 30$ ~ to~  $160 $~km/s ) rather than 
the 200 km/s typical for galaxies of similar luminosities. In particular
galaxies with $\sigma < 65$ km/s seem to follow the same relations 
in $\sigma$, $M_B$ and  $L({\rm H}\beta )$ as the local ones. Recent infrared spectroscopy
of Balmer emission lines in a few Lyman break galaxies at  $z=3$ 
 (Pettini et al. 1998) suggests 
that these adhere to the same relation, although this has to be confirmed for a larger 
sample. This opens the important possibility of applying the distance 
estimator and map the Hubble flow up to extremely high redshifts and 
simultaneously to study 
the behaviour of starbursts of similar luminosities over a huge 
redshift range. It may prove to be a useful  method to measure  $q_0$ because
the redshift interval of present day applicability considering the most luminous H{\sc ii} galaxies 
is  larger than for methods involving SNe, allowing 
a good discrimination between deceleration ($q_0$) and curvature ($\Lambda$). 
Possible complications concern the treatment of the effects of metallicity and extinction 
in these systems.

%%%%%%%%%%%%%%%%%%%%%%%%%%%%%%%%%%%%%%%%%%%%%%%%%%%%%%%%%%%%%%%%%%%%%%%%%%%%%%%%%%%%%%%%%%
%%%%%%%%%%%%%%%%%%%%%%%%%%%%%%%%%%%%%%%%%%%%%%%%%%%%%%%%%%%%%%%%%%%%%%%%%%%%%%%%%%%%%%%%%%
%%%%%%%%%%%%%%%%%%%%%%%%%%%%%%%%%%%%%%%%%%%%%%%%%%%%%%%%%%%%%%%%%%%%%%%%%%%%%%%%%%%%%%%%%%

\np

\section{Summary  and Conclusions}

 Extreme properties are often sought for in astronomy as one way to sharpen
our understanding of  fundamental concepts. In such a context,
metallicity proves to be of crucial importance. Because metals build up
as a function of time after the release of nucleosynthesis products, the
most metal-poor galaxies help us to understand the primordial Universe and the
subsequent formation and evolution of galaxies. 

We have discussed in this paper the reason why local
metal-poor galaxies are found amongst the dwarfs. Whether this property remains
true at high redshift where proto-galaxies begin to form, is certainly a
challenging question. If the hierarchical model of galaxy formation is correct
and if building block galaxies are similar to the well-studied dwarfs we
discussed in this review they may be unreachable observationally, even with
the advent of the most recent large telescopes. On the other hand, large,
massive and gas-rich proto-galaxies may have some properties similar to those of
the most unevolved objects we see today. The same applies to the distant
starbursts that  have been found using Lyman break techniques or  Ly$\alpha$
emission searches.

Evolution versus formation has been a clear issue in order to characterise the
few known extreme star-forming dwarfs like IZw18. Most of the properties of these
galaxies: blue colours, gas and dust content, and extreme metal
deficiency once led to the belief that, although rare, small condensations were able
to produce genuinely young galaxies at the current cosmic epoch. This question 
has indeed received a lot of attention in the recent years but we consider it as 
somewhat ``pass\'e''. In most cases - but for a
few intriguing ones - an old underlying stellar population has been found,
revealing one or several previous bursts of star formation. The same seems to
apply to dwarf gas-poor galaxies as one can tell from the study of their colour
magnitude diagrams. The problem of the existence of local young galaxies may not
be completely solved yet but its importance is less acute, since  in any case, there is
no  significant population of dwarf galaxies still in the process of formation. 
Evidences that most of the galaxies were formed at ealier epochs come from a more 
direct view of what the Universe looks like at high redshift and also  the  hint 
that the metal production rate in the Universe has passed its peak. 
The question of present cosmological interest would rather be to determine the
epoch when the bulk of dwarfs were formed. Was the formation of dwarfs confined 
to very high redshifts, or was it still going on at intermediate redshifts
where the faint blue galaxies emerge and the Universe was about 2/3 its present age?

 Nevertheless, for years, local extremely metal poor galaxies turned out to be our best test
laboratory. We have learned a lot about the properties of their
massive stars (formation and evolution, appearance of WR stars), the evolution
of the dynamics of the gas in the gravitational potential of the parent galaxy
as a superbubble evolves and the chemical enrichment of the interstellar medium
after the fresh products are well mixed. We have discussed  their
dynamical properties and interactions with companions in order to understand the
triggering mechanism that ignites their bursts of star formation. These galaxies
have been modelled to constrain the scenario of the starbursts, with the major
motivation of  describing how various classes of dwarfs  could be linked
together into one or several evolutionary scenarios. This is not completely
settled yet and a lot of work is still needed both theoretically and
observationally in this area. Surveys are aimed to  build  luminosity
functions that will help in answering some of the above questions and also
produce new targets that will open the possibility to investigate  the
properties of metal-poor galaxies at much larger redshift. We clearly enter a
new area in which metal-poor galaxies  and the sub-class of metal-poor dwarf
galaxies will observationally begin to be considered in cosmology. They enter in the study of
the cosmic metal enrichment, the possibility to establish new distance
indicators, the study of the yield of heavy elements and the derivation of the
primordial helium, and finally the r\^ole these galaxies play in the hierarchial buildup of
structures in the Universe.
 
Will we ever find a metal-free galaxy? In our local universe we concur that 
young galaxies (hence possible metal free candidates) are unlikely to be found.
In fact we have stressed that the most metal poor known dwarf galaxies are orders of
magnitudes more metal rich than the most extreme halo stars in our Galaxy (Cayrel 1996). 
The metallicity distribution of the Galactic disc stars  and the so-called 
G-dwarf problem (Pagel 1997) indicate pre-enrichment before the disc formed.
The oldest known systems in our Galaxy (and the Universe), the halo globular clusters,  have 
metallicities similar to local dwarf spheroidal galaxies. Hence, the oldest
globular clusters were self enriched or formed from pre-polluted material.  At high 
redshift the challenge is to pick up galaxies at the epoch of formation. If our Galaxy 
is a representative case we do not expect pristine (or nearly pristine, with
abundances much lower than the most metal-poor local dwarf galaxies) matter to be 
found anywhere but in dwarfs or sub galactic clumps.

%%%%%%%%%%%%%%%%%%%%%%%%%%%%%%%%%%%%%%%%%%%%%%%%%%%%%%%%%%%%%%%%%%%%%%%%%%%%%%%%%%%%%%%%%%%%%%%%%%%
%%%%%%%%%%%%%%%%%%%%%%%%%%%%%%%%%%%%%%%%%%%%%%%%%%%%%%%%%%%%%%%%%%%%%%%%%%%%%%%%%%%%%%%%%%%%%%%%%%%
%%%%%%%%%%%%%%%%%%%%%%%%%%%%%%%%%%%%%%%%%%%%%%%%%%%%%%%%%%%%%%%%%%%%%%%%%%%%%%%%%%%%%%%%%%%%%%%%%%%

\np
 
\setlength{\parindent}{0em}

\small

{\large \bf Acknowledgements}

\medskip
We first of all thank L. Woltjer who inspired this review paper. 
We thank  N. Bergvall, S. Charlot, T. Karlsson, B. Lanzoni, F. Legrand,  J. Masegosa, 
N. Prantzos, J. Salzer, 
D. Schaerer, E. Skillman,  C. Taylor, R. Terlevich, T.X. Thuan, M. Tosi and J. Vilchez 
for useful discussions.  In particular we thank R. Guzm\'an for comments on an early draft
version of this paper, and James Lequeux for several careful readings of the manuscript.
We are grateful to F. Legrand, R. Terlevich and T.X. Thuan, for providing us with figures used in
this review. S. Charlot is thanked for making  the most recent update of the  Bruzual and Charlot
spectral evolutionary synthesis models available to us. 
G. \"Ostlin acknowledges support from The Swedish Foundation for International Cooperation
in Research and Higher Education (STINT).  
In writing this review we made frequent use of the NASA astrophysics data system (ADS) 
and the NASA/IPAC extragalactic database (NED), which is operated by the Jet Propulsion 
Laboratory, California Institute of Technology, under contract with the National Aeronautics 
and Space Administration. 
We acknowledge the use of NASA's SkyView facility
          (http://skyview.gsfc.nasa.gov) located at NASA Goddard
          Space Flight Center.
HST-images of I~Zw18 and SBS~0335-052 were obtained from the archive at the Space Telescope 
Science Institute, which is operated by the association of Universities for Research in 
Astronomy, Inc., under NASA contract NAS5-26555.

\bigskip

{\bf Appendix: A list of some used abbreviations and acronyms:}

\begin{tabbing}
BCD:~~~~~ \= blue compact dwarf galaxy \\
BCG:    \> blue compact galaxy \\
C:	\> carbon \\
CDM:	\> cold dark matter \\
dE:	\> dwarf elliptical galaxy \\
dI:	\> dwarf irregular galaxy \\
dSph:	\> dwarf spheroidal galaxy \\
DLA:    \> damped Lyman alpha \\
E:	\> elliptical galaxy \\
ELG:    \> emission line galaxy \\
FBS:    \> first Byurakan survey \\
FOS:	\> Faint Object Spectrograph (HST istrument) \\
Fe:	\> iron \\
GC:     \> globular cluster \\
H:	\> hydrogen \\
He:	\> helium \\
HST:	\> Hubble Space Telescope \\
IGM:    \> intergalactic medium \\
IMF:    \> stellar initial mass function \\
IR:	\> infrared \\
IRAS:	\> infrared astronomical satellite \\
ISM:    \> interstellar medium \\
ISO:	\> infared space observatory \\
$L^{\star}$:	\> the characteristic galaxy luminosity in the galaxy luminosity funtion \\
LBG:	\> Lyman break galaxy \\
LF:	\> luminosity function (of galaxies) \\
LMC:	\> Large Magellanic Cloud \\
LSBG:	\> low surface brightness galaxy \\
N: 	\> nitrogen \\
O:	\> oxygen \\
QSO:	\> quasar, quasi stellar object \\
S:	\> Sulphur \\
SBS:    \> second Byurakan survey \\
SF:	\> star formation \\
SFR:    \> star formation rate \\
SFH:    \> star formation history \\
SMC:	\> Small Magellanic Cloud \\
SSC:    \> super star cluster \\
UV:	\> ultraviolet \\
VLA:	\> Very Large Array \\
WR:     \> Wolf-Rayet \\
XBCG:	\> extreme blue compact galaxy \\
Y:	\> helium abundance \\
Z:      \> metallicity, heavy element  abundance \\
z:	\> redshift \\

\end{tabbing}

{\large \bf References}

\bigskip

\setlength{\baselineskip}{0.75em}
\setlength{\parindent}{-1.5em}

Aaronson M., 1986, in ``Star-forming dwarf galaxies and related objects'', Eds. Kunth D., 
Thuan T.X., Tran Thanh Van J., \'Editions Fronti\`eres, p. 125 \\ 

Allen D.A, Wright A.E, Goss W.M. 1976, MNRAS, 177, 91  \\

Alloin D., Bergeron J., Pelat D., 1978, A\&A 70, 141  \\

Aloisi A., Tosi M., Greggio L., 1999, AJ 118, 302 \\

Aparicio A., Herrero A., S\'anchez F., Eds., ``Stellar Astrophysics for the Local Group'',
Cambridge University press, 1998 \\ 

Armandroff T.E., Davies J.E., Jacoby G.H., 1998, AJ 116, 2287 \\ 

Arnault P., Kunth D., Casoli F., Combes F., 1988,  A\&A 205, 41 \\

Arnault Ph., Kunth D., Schild H. 1989, A\&A, 224, 73  \\

Arp H., 1965, ApJ 142, 383  \\

Arp H., O'Connell R.W., 1975, ApJ 197, 291 \\

Arp H., Sandage A., 1985, ApJ 90, 1163  \\

Augarde R., Chalabaev A., Comte G., Kunth D., Maehara H., 1994, A\&AS 104, 259 \\

Aussel H., Cesarsky C.J., Elbaz D., Starck J.L., 1999, A\&A 342, 313 \\

Babul A., Rees M.J., 1992, MNRAS 255, 346 \\

Balkowski C., Chamaraux P., Weliachew L., 1978, A\&A 69, 263 \\

Barbuy B., 1988, A\&A 191, 121 \\

Barth A.J., Ho L.C., Fillipenko A.V., Sargent W.L., 1995, AJ 110, 1009  \\

Bautista M.A., Pradhan A.K., 1995, ApJ 442, L65 \\

Bell E.F.,  Bower R.G., de Jong R.S., Hereld M., Rauscher B.J., 1999, MNRAS 302, L55 \\

Benjamin R.A., Skillman E.D., Smits D.P., 1999, ApJ 514, 307 \\

Bergvall N. 1985, A\&A 146, 269  \\

Bergvall N., J\"ors\"ater S., 1988, Nature 331, 589  \\

Bergvall N.,  \"Ostlin G.,  1999, in preparation  \\

Bergvall N., R\"onnback J.,  1994, in ``ESO/OHP worshop on dwarf galaxies'', Eds. Meylan G., 
Pruginel P., ESO Conference and Workshop proceedings No. 49, p. 433 \\

Bergvall N., R\"onnback J.,  1995, MNRAS 273, 603  \\

Bergvall N., \"Ostlin G., Pharasyn A., R\"onnback J., Masegosa J., 1998, 
in Highlights in Astronomy, Ed. Andersen J., Vol. 11A, p. 103\\

Bergvall N., R\"onnback J., Masegosa J., \"Ostlin G.,  1999 A\&A 341, 697  \\

Binggeli B., 1994, in ``ESO/OHP worshop on dwarf galaxies'', Eds. Meylan G., 
Prugniel P., ESO Conference and Workshop proceedings No. 49, p. 13 \\

Binggeli B., Sandage A., Tarenghi M., 1984, AJ 89, 64  \\

Binggeli B., Sandage A., Tammann G.A., 1988, ARA\&A 26, 509 \\

Binggeli B., Tarenghi M., Sandage A., 1990, A\&A 228, 42  \\

Boroson T.A., Salzer J.J., Trotter A., 1993, ApJ 412, 524  \\

Bothun G.D.,  Caldwell C.N., 1984, ApJ 280,528  \\

Bothun G.D., Beers T.C., Mould J.R., Huchra J.P., 1986a, ApJ 308, 510 \\

Bothun G.D., Mould J.R., Caldwell N., MacGillivray H.T., 1986b, AJ 92, 1007  \\ 
Bothun G.D., Impey C.D., Malin D.F., Mould J., 1987, AJ 94, 23 \\

Bothun G., Schombert J., Impey C., Schneider S., 1990, ApJ 360, 427  \\

Bothun G.D., Schombert J.M., Impey C.D., Sprayberry D., McGaugh S.S., 1993, AJ 106, 530 \\

Bothun G., Impey C., McGaugh S., 1997, PASP 109, 745  \\

Bottinelli L., Gougenheim L., Heidmann J., 1973, A\&A 22, 281  \\

Bottinelli L., Duflot R., Gougenheim L., Heidmann J., 1973, A\&A 22, 281  \\

Briggs F.H., 1997, Pub.Ast.  Soc of Australia 14,31 \\

Broadhurst T.J., Ellis R.S., Shanks T., 1988, MNRAS 235, 827 \\

Brocklehurst M., 1972, MNRAS 157, 211 \\

Bruzual G., Charlot S. 2000, in preparation \\

Caldwell, Amandroff, Da Costa, Seitzer, 1998, AJ 115, 535  \\

Campbell A., Terlevich R., Melnick J., 1986, MNRAS 223, 811  \\

Campos-Aguilar A., Moles M., Masegosa J., 1993, AJ 106, 1784 \\

Carignan C., Beaulieu S., C\^ot\'e S., Demers S., Mateo M., 1998, AJ 116, 1690  \\

Cayrel R., 1996, A\&AR 7, 217 \\

Cen R., Ostriker J.P., 1999, ApJ 519, L109  \\

Chamaraux P., 1977, A\&A 60, 67  \\

Chamaraux P., Heidmann J., Lauqu\'e R., 1970, A\&A 8, 424  \\

Chengalur J.,N., Giovanelli R., Haynes M.P., 1995, AJ 109, 2415 \\

Cervi\~no M. 1998, Ph. D. Thesis, Universidad Complutense de Madrid \\

Cervi\~no M., Mas-Hesse J.M., 1994, A\&A, 284, 749  \\

Colless M., Ellis R.S., Taylor K., Hook R.N., 1990, MNRAS 244, 408 \\

Comte G., 1998, Astrophysics, 41, 89  \\

Comte G., Augarde R., Chalabaev A., Kunth D., Maehara H., 1994, A\&A 285, 1  \\

Combes F., 1986, in ``Star-forming dwarf galaxies and related objects'', 
Eds. Kunth D., Thuan T.X., Tran Thanh Van J., \'Editions Fronti\`eres, p. 307; \\

Conti P. 1991, ApJ, 377, 115  \\

Conti P., Vacca W.,  1994, ApJ, 423, L97  \\

Corbelli E., Salpeter E.E, 1993a, ApJ 419,94 \\

Corbelli E., Salpeter E.E, 1993b, ApJ 419,104 \\

C\^ot\'e S., Freeman K.C., Carignan C., Quinn P.J., 1997, AJ 114, 1313  \\

Cowie  L.L., Songaila A.A., Hu E.M., 1991, Nature 354, 460 \\

Da Costa G.S., Armandroff T.E., 1990, AJ 100, 162  \\

Davies J.I., Phillipps S., 1988, MNRAS 233, 553  \\

Davies J., Impey C., Phillipps S., eds.,  ``The Low Surface Brightness Universe'', 
proc. IAU Col. 171, pub. PASP conf ser. No. 170 \\

de Blok W.J.G., van der Hulst J.M.,  1998, A\&A 335, 421  \\

de Blok W.J.G., McGaugh S.S., van der Hulst J.M.,  1996, MNRAS 283, 18  \\

de Blok W.J.G.,  van der Hulst J.M., Bothun G.D., 1995, MNRAS 274, 235 \\

Dekel A., Silk K., 1986, ApJ 303, 39 \\

De Mello D.F., Schaerer D., Heldmann J., Leitherer C. 1998, ApJ, 507,199  \\

de Vaucouleurs G., de Vaucouleurs  A.; Corwin J.R.,  Buta R.J.,  Paturel  G., Fouque P., 
Third reference catalogue of bright galaxies (RC3), version 9, 1991, New York : Springer-Verlag \\

de Young D.S., Heckman T.M., 1994, ApJ 431, 598  \\

Disney M., Phillipps S., 1983, MNRAS 205, 1253  \\

Djorgovski S., 1990, AJ 99, 31 \\

Doublier V., Comte G., Petrosian A., Surace C., Turatto M., 1997, A\&AS 124, 405 \\

Doublier V., Kunth D., Courbin F., Magain P, 1999, A\&A accepted (astroph/9902294) \\

Duc P.-A., Mirabel I.F., 1994, A\&A 289, 83  \\
 
Duc P.-A., Mirabel I.F., 1998, A\&A 333, 813 \\

Dufour R.J., Hester J.J., 1990, ApJ 350, 149 \\

Dufour R.J., Garnettt D.R., Shields G.A., 1988, ApJ 332, 752 \\

Dufour R.J., Garnettt D.R., Skillman E.D., Shields G.A., 1996, in ``From Stars to Galaxies: 
The Impact of Stellar Physics on Galaxy Evolution'', Eds. Leitherer C., Fritze-von-Alvensleben U., 
Huchra J., ASP Conf. Ser. 98, p.358 \\

Dunlop J.S., Hughes D.H., Rawlings S., Eales S.A., Ward M.J., 1994, Nature 370, 347 \\

Edmunds M.G., 1990, MNRAS 246,678  \\

Edvardsson B., Andersen J., Gustafsson B., Lambert D.L., Nissen P.E., Tomkin J., 1993, 
A\&A 275, 101   \\

Efstathiou G., 1992, MNRAS 256, P43 \\

Eggen O.J., Lynden-Bell D., Sandage A.R., 1962, ApJ 136, 748 \\

Ellis R.S., 1997, ARA\&A 35, 389  \\

Esteban C., Peimbert M., 1995, A\&A 300, 78 \\

Fanelli M.N., O'Connell R.W., Thuan T.X., 1988, ApJ 334, 665 \\

Ferguson H., Binggeli B., 1994, A\&AR 6, 67  \\

Ferrara A., Tolstoy E., 1999, MNRAS submitted, (astroph/9905280) \\

French H.B., 1980, ApJ 240, 41  \\

Gallagher J.S., Hunter D.A., 1987, AJ 94, 43  \\

Gallagher J.S., Littleton J.E., Matthews L.D., 1995, AJ 109, 2003 \\

Gallego J., Zamorano J., Rego M., Vitores A.G., 1997, ApJ 475, 502  \\

Garnett D.R., 1990, ApJ, 363, 142  \\

Garnett D.R.,  Skillman E.D., Dufour R.J., Peimbert M., Torres-Peimbert S., Terlevich R.J.,
Terlevich E.,  Shields G.A., 1995, ApJ 443, 64  \\

Garnett D.R.,  Skillman E.D., Dufour R.J., Shields G.A., 1997, ApJ 481, 174 \\

Garnett D.R., Shields G.A., Peimbert M., Torres-Peimbert S., Skillman E.D., Dufour R.J.
Terlevich E., Terlevich R.J., 1999, ApJ 513, 168  \\

Gerola H., Seiden P.E., Schulman L.S., 1980, ApJ 242, 517 \\

Gilmore G., Wyse, R.F.G., 1991, ApJ 367, L55  \\

Giovanelli R., Haynes M.P., 1989, ApJ 346, L5 \\

Giovanelli R., Scodeggio M., Solanes J.M., Haynes M.P., Arce H., Sakai S., 1995, AJ 109, 1451 \\ 

Glazebrook K., Ellis R.S., Santiago B., Griffiths R., 1995,  MNRAS 275, L19 \\

Gondhalekar P.M., Johansson L.E.B., Brosch N., Glass I.S., Brinks E., 1998, A\&A 335, 152 \\

Gordon D., Gottesman S.T., 1981, AJ 86, 161 \\ 

Goodwin S.P., 1997, MNRAS  284, 785 \\

Gorgas J., Pedraz S., Guzm\'an R., Cardiel N., Gonz\'ales J., 1997, ApJ 481, L19  \\

Gottesman S.T., Weliachew L., 1972, Astrophys. Letters. 12, 63 \\

Grebel E., 1998, in ``The Stellar Content of Local Group Galaxies'', Eds. Whitelock P. and 
Cannon R., IAU symposium 192, Publisher: Astron. Soc. of the Pacific., p.17  \\

Gustafsson B., Karlsson T., Olsson E., Edvardsson B., Ryde N., 1999, A\&A 342, 426  \\
%(The origin of carbon, investigated by solar type stars in the Galactic disc) \\

Guzm\'an R., Koo D.C., Faber S.M., Illingworth G.D., Takamiya M., R.G. Kron, Bershady M., 
1996, ApJ 460, L5 \\

Guzm\'an R., Gallego J., Koo D.C., Phillips A.C., Lowenthal J.D., Faber S.M., 
Illingworth G.D., Vogt N.P., 1997, ApJ 489, 559  \\

Hammer F., Flores H., Lilly S.J., Crampton D., Le F\`evre O., Rola C., 
Mallen-Ornelas G., Schade d., Tresse L., 1997, ApJ 481, 49 \\

Haro G., 1956, Bol Obs. Tonantzintla y Tacubaya 2, 8 \\

Haser S., Pauldrach A., Lennon D., Kudritzki R.-P., Lennon M., Puls J., Voels S., 1998, A\&A 330, 285  \\

Hazard C., 1986, in ``Star-forming dwarf galaxies and related objects'', Eds. Kunth D., 
Thuan T.X., Tran Thanh Van J., \'Editions Fronti\`eres, p. 9  \\

Held E.V., Mould J.R., 1994, AJ 107, 1307  \\

Hensler G., Rieschick A., 1998, in Highlights in Astronomy, Ed. Andersen J., Vol.  11A, p 139, (Kluwer) \\

Hilker M., Richtler T., Gieren W., 1995, A\&A 294, 648  \\

Hidalgo-G\'amez A.M., Olofsson K., A\&A 334, 45 \\

Ho L.C., Filippenko A.V., 1996, ApJ 472, 600  \\

Hu E.M., Cowie L.L. \& McMahon R. G., 1998,  ApJ 502, L99  \\

Hughes D.H., Serjeant S., Dunlop J., et al. Nature 394, 241 \\

Hunter D.A., Thronson H.A. Jr, 1995, ApJ 452, 238 \\

Ibata R.A., Gilmore G., Irwin M.J., 1994, Nature 370, 194 \\

Impey C., Bothun G., Malin D., Stavely-Smith L., 1990, ApJ 351, L33 \\

Iovino  A., Melnick J., Shaver P., 1988, ApJ 330, L17  \\

Israel F.P., van Driel W., 1990, A\&A 236, 323 \\

Israel F.P., Tacconi L.J., Baas F., 1995, A\&A 295, 599  \\

Israelian G., Lopez R.J.G., Rebolo R., 1998, ApJ 507,805 \\

Izotov Y.I., Thuan T. X., 1998a, ApJ, 500, 188  \\

Izotov Y.I., Thuan T.X.,  1998b, ApJ, 497, 227  \\

Izotov Y.I.,  Thuan T.X., 1999 ApJ 511, 639   \\

Izotov Yu.I., Guseva N.G., Lipovetskii V.A., Kniazev A.Yu, Stepanian J.A., 1990, Nature 343, 238 \\ 

Izotov Y.I., Lipovetsky V.A., Guseva N.G., Kniazev A.Y., Stepanian J.A., 1991, A\&A 247, 303 \\

Izotov Y.I., Foltz C.B., Green R.F., Guseva N.G., Thuan T.X. 1997a, ApJ, 487, L37   \\

Izotov Y., Lipovetsky V.A., Chaffee F.H., Foltz C.B., Guseva N.G., Kniazev A.Y., 1997b, ApJ 476, 698 \\

Izotov Y.I., Papaderos P., Thuan T.X., Fricke K.J., Foltz C.B., Guseva N.G., 1999, 
A\&A submitted, (astroph/9907082)  \\

Izotov Y.I., Chaffee F.H., Foltz C.B., Green R.F., Guseva N.G.,  Thuan T. X., 1999b,
ApJ accepted, (astroph/9907228) \\

James P., 1991, MNRAS 250, 544  \\

James P., 1994, MNRAS 269, 176  \\

Jerjen H., Binggeli B., 1997, in ``The second Stromlo symposium: the nature of elliptical 
galaxies'' eds. Arnaboldi M., Da Costa G.S., Saha P., ASP conf. ser. Vol 116, p. 239  \\

Karachentsev I.D.,  Karachentseva V.E., 1999, A\&A 341, 355 \\

Kennicutt R.C.Jr, 1989, ApJ 344, 685 \\
 
Kinman T.D., Davidson K., 1981, ApJ, 243, 127  \\

Kinman T.D., Hintzen P., 1981,  PASP 93, 405 \\

Kniazev A. Yu., Pustilnik S.A., Ugryumov A.V., 1998, Bull. Spec. Astrophys. Obs. 4, 23 \\

Kniazev A. Yu., et al. 1999, in preparation \\

Kobulnicky H.A., 1998, in ``Abundance Profiles: Diagnostic Tools for Galaxy History'', 
Eds. Friedli D., 
Edmunds M., Robert C., Drissen L., ASP Conf. Ser. Vol. 147, p. 108 \\

Kobulnicky H.A., Skillman E.D., 1996, ApJ 471, 211  \\

Kobulnicky H.A., Skillman E.D., 1998, ApJ 497, 601  \\

Kobulnicky H.A., Skillman E.D., Roy J.-R., Walsh J.R., Rosa M.R., 1997, ApJ 477, 679 \\

Kobulnicky H.A., Kennicutt R.C. Jr, Pizagno J.L., 1999 ApJ 514, 544 \\

Kobulnicky H.A., Zaritsky D., 1999, ApJ 511, 118 \\

Kormendy  J., Bender R., 1994, in ``ESO/OHP worshop on dwarf galaxies'', Eds. Meylan G.,  
Prugniel P., ESO Conference and Workshop proceedings No. 49, p. 161 \\

Koo D.C., Guzm\'an R., Faber S.M., Illingworth G.D., Bershady M.A., Kron R.G., Takamiya M., 
1995, ApJ 440, L49 \\ 

Kudritzki R.P., 1998, in ``Stellar astrophysics for the Local Group'', eds. Aparicio A., Herrero A.,
Sanchez F., Cambridge university press  \\

Kunth D., 1986, in ``Star-forming dwarf galaxies and related objects'', Eds. Kunth D., Thuan T.X., 
Tran Thanh Van J., \'Editions Fronti\`eres, p. 183  \\

Kunth D.,  1996, in ``The Sun and Beyond'',  eds. Tran Thanh Van J., Celnikier L., Chong Trong H., 
\'Editions Fronti\`eres, p. 357 \\

Kunth D., Joubert M. 1985, A\&A, 142, 411  \\

Kunth D., Sargent W.L.W. 1981, A\&A, 101, L5  \\

Kunth D., Sargent W.L.W., 1983, ApJ 273, 81  \\

Kunth D., Sargent W.L.W., 1986, ApJ 300, 496  \\

Kunth D., Schild H. 1986, A\&A, 169, 71  \\

Kunth D., Sargent W.L.W., Kowal C., 1981, A\&AS 44, 229  \\

Kunth D., Maurogordato S., Vigroux L., 1988, A\&A 204, 10  \\

Kunth D., Lequeux J., Sargent W.L.W., Viallefond F.,  1994 A\&A 282, 709 \\

Kunth D., Matteucci F., Marconi G. 1995, A\&A, 297, 634  \\

Kunth D., Mas-Hesse J.M., Terlevich E., Terlevich R., Lequeux J., Fall S.M., 1998, A\&A 334,11  \\

Lacey C., Guiderdoni B., Rocca-Volmerange B., Silk J., 1993, ApJ 402, 15 \\   

Lanzetta K.M., Bowen D.B., Tytler D., Webb J.K., 1995, ApJ 442,538 \\

Lauberts A., Valentijn E.A., 1989, ``The surface photometry catalogue of the ESO-Uppsala galaxies'', ESO \\

Lauqu\'e R., 1973, A\&A 23, 253  \\

Larson R.B., 1974, MNRAS 169, 229 \\

Le Brun V., Bergeron J., Boiss\'e P., Deharveng J.M., 1997, A\&A 321, 733  \\

Lee M.G., Freedman W., Mateo M., Thompson I., Roth M., Ruiz M.-T.,
1993, AJ 106, 1420  \\

Lee H., McCall M.L., Richer M.G., 1998, BAAS 193, 5304  \\

Lee H., McCall M.L., Richer M.G., 1998 in ``Abundance profiles: diagnostic tools for galaxy history'',
eds. Friedli D., Edmunds M., Robert C., Drisen L., ASP conf. ser. 147, p. 313 \\

Legrand F., 1998, PhD Thesis, Institut d'Astrophysique de Paris \\

Legrand F., Kunth D., Mas-Hesse J.M., Lequeux, J. 1997a, A\&A 326, 929  \\

Legrand F., Kunth D., Roy J.-R., Mas-Hesse J.M., Walsh J.R.,  1997b, A\&A 326, 17  \\

Legrand F., et al., 1999, in preparation  \\

Lehnert M.D., Bell R.A., Hesser J.E., Oke J.B., 1992, ApJ 395, 466  \\

Leitherer C., 1999, in ``Chemical evolution from zero to high redshift'', Eds. Walsh J., Rosa M.,  
Lecture notes in Physics (Berlin:Springer) in press. (STSCI preprint 1312)  \\

Leitherer C., Robert C., Drissen L., 1992, ApJ 401, 596 \\

Lequeux J., Viallefond F., 1980, A\&A 91, 269  \\

Lequeux J., Peimbert M., Rayo J.F., Serrano A., Torres-Peimbert S., 1979, A\&A 80, 155  \\

Lequeux J., Maucherat-Joubert M., Deharveng J.M., Kunth D., 1981, A\&A 103, 305 \\

Lequeux J., Kunth D., Mas-Hesse J.M., Sargent W.L.W., 1995, A\&A,301,18  \\

Lilly S.J., Cowie L.L., Gardner J.P., 1991, ApJ 369, 79 \\

Lin D.N.C., Faber S.M., 1983, ApJ 266, L21 \\

Lindner U., Einasto M., Einasto J., Freudling W., Fricke K., Lipovetsky V., Pustilnik S., Izotov Y.
Richter G., 1996, A\&A 314, 1 \\

Lindner U., Fricke K.J., Einasto J., Einasto M., 1998, in Highlights of Astronomy, Ed. Andersen
J., Vol. 11A, p 111 \\

Lisenfeld U., Ferrara A., 1998, ApJ 496, 145 \\

Lipovetsky V.A., Chaffe F.H., Izotov Y.I., Foltz C.B., Kniazev A.Y., Hopp U., 1999, ApJ 519, 177 \\

Lo K.Y., Sargent W.L.W., 1979, ApJ 227, 756 \\

Lo K.Y., Sargent W.L.W., Young K.,  1993, AJ 106, 507 \\

Loose H.-H., Thuan T.X., 1986a, in ``Star-forming dwarf galaxies and related objects'', 
Eds. Kunth D., Thuan T.X., Tran Thanh Van J., \'Editions Fronti\`eres, p. 73 \\

Loose H.-H., Thuan T.X., 1986b, ApJ 309, 59  \\

Lu L., Sargent W.L.W., Barlow T.A., Churchill C.W., Vogt S.S.,  1996, ApJS 107, 475  \\

Lu L., Sargent W.L.W., Barlow T.A., 1998, ApJ 115, 55  \\

MacAlpine G.M., Smith S.B., Lewis D.W., 1977, ApJSS 34, 95 \\

MacLow M.-M., Ferrara ., 1999, ApJ 513, 142  \\

Macri L.M., Huchra J.P., Stetson P.B., et al., 1999, ApJ 521, 155 \\

Madau P., Ferguson H.C., Dickinson M.E., Giavalisco M., Steidel C.C., Fruchter A., 1996, MNRAS 283, 1388 \\

Madore B.F., Freedman W.L., Silbermann N., et al., 1999, ApJ 515, 29 \\

Maeder, A. 1992, A\&A, 264,105  \\

Maiz-Appelaniz J., Mas-Hesse J.M., Munoz-Tunon C., Vilchez J.M., Castaneda H.O., 1998, A\&A 329, 409\\

Marconi G., Matteucci F., Tosi M., 1994, MNRAS 270, 35 \\

Markarian B.E., 1967, Afz 3,24  \\

Markarian B.E., Stepanian D.A., 1983, Afz 19, 639 \\

Marlowe A.T., Heckman T.M., Wyse R.F.G., Schommer R.,  1995, ApJ, 438, 563  \\

Marlowe A.T., Meurer G.R., Heckman T.M., Schommer R., 1997, ApJS 112, 285 \\

Marlowe A.T., Meurer G.R., Heckman T.M., ApJ accepted, (astroph/9904089)\\

Martin C.,  1996, ApJ 465, 680  \\

Martin C.,  1998, ApJ 506, 222  \\

Masegosa J., Moles M., Campos-Aguilar A., 1994, ApJ 420, 576  \\
 
Mas-Hesse J.M., Kunth D., 1999, A\&A accepted (astroph/9812072)\\

Massey P., Armandroff T., Pyke R., Patel K., Wilson C., 1995, AJ 110, 2715  \\

Mateo  M., 1998, ARA\&A 36, 435  \\

Mathewson D.S.,  Ford V.L., 1996, ApJS 107, 97 \\

Matteucci F., 1996, Fund. Cosm. Phys. 17, 283 \\

Matteucci F., Chiosi C., 1983, A\&A 123, 121  \\

Matthews L.D., Gallagher J.S., 1996, AJ 111, 1098 \\

Maza et al. 1999, in preparation \\

Mazzarella J.M., Boroson T.A., 1993, ApJS 85, 27 \\

McGaugh S., 1994, ApJ 426, 135  \\

McGaugh S.S., Bothun G.D., 1994, AJ 107, 530 \\

McGaugh S.S., de Blok W.J.G., 1997, ApJ 481, 689 \\

McMahon R.G., Irwin M.J., Giovanelli R., Haynes M.P., Wolfe A.M., Hazard C., 1990, ApJ 359, 302 \\

Melnick J., Moles M., Terlevich R., Garcia-Pelayo J.M., 1987, MNRAS 226, 849  \\

Melnick J., Moles M., Terlevich R., 1985a, A\&A 149, L24  \\

Melnick J., Terlevich R., Eggleton P.P., 1985b, MNRAS 216, 255  \\

Melnick J., Terlevich R., Moles M., 1988, MNRAS 235, 297   \\

Melnick J., Heydari-Malayeri M., Leisy P., 1992, A\&A 253, 16 \\

Merlino S., Masegosa J., Kniazev A., Pustilnik S., Ugryumov A., Izotov Y.,  Marquez I.,
1999, in ``Astrophysics with the NOT'', Ed Piirola V., in press. \\

Meurer G.R., Freeman K.C., Dopita M.A., Cacciari C.,  1992, AJ 103, 60 \\

Meurer  G.R., Heckman T.M., Leitherer C., Kinney A., Robert C., Garnett D.R., 1995, AJ 110, 2665 \\

Meurer G.R., Carignan C., Beaulieu S., Freeman K.C., 1996, AJ 111, 1551 \\

Meurer G.R., Heckman T.M., Lehnert M.D., Leitherer C., Lowenthal J., 1997, AJ, 114, 54 \\

Meurer G.R., Staveley-Smith L., Killeen N.E.B., 1998, MNRAS 300, 705 \\

Meynet G. 1995, A\&A, 298, 767  \\

Mihos C.J., McGaugh S.S., de Blok W.J.G., 1997, ApJ 477, L79 \\  

Miller B.,W., 1996, AJ 112, 991  \\

Miller B.W., Hodge P., 1996, ApJ 458, 467  \\

Mirabel I.F., 1989, in ``Structure and Dynamics of the Interstellar Medium, Proceedings of IAU Colloq. 120''
Eds. Tenorio-Tagle G., Moles M., Melnick J., Lecture Notes in Physics (Springer), Vol. 350, p.396  \\

Mirabel I.F., Dottori H., Lutz D., 1992, A\&A 256, L19  \\

Mo H.J., McGaugh S.S., Bothun G.D., 1994, MNRAS 267, 129 \\

Mould J.R., 1978, ApJ 20, 434 \\

Murakami I., Babul A., 1999, MNRAS in press, (astroph/9906084) \\

O'Connell R.W., Gallagher J.S. III, Hunter D.A., Colley W.N., 1995, ApJ 446, L10 \\

Olofsson K., 1995, A\&A 293, 652 \\

\"Ostlin G., 1998, PhD Thesis, Uppsala University \\

\"Ostlin G., 1999a, ApJ submitted  \\

\"Ostlin G., 1999b, in preparation  \\

\"Ostlin G., 1999c, in preparation  \\

\"Ostlin G., Bergvall N., R\"onnback J., 1998, A\&A 335, 85  \\

\"Ostlin G., Amram P., Masegosa J., Bergvall N., Boulesteix J., 1999a, A\&AS 147, 419 \\

\"Ostlin G., Amram P.,  Bergvall N., Masegosa J., Boulesteix J., 1999b, A\&A submitted \\

\"Ostlin G., Bergvall N., et al. 1999c, in preparation  \\

Pagel B.E.J., 1993, in ``New Aspects of Magellanic Cloud Research'', Eds. Bashek B., 
Klare G., Lequeux J., Lecture notes in Physics 416 (Berlin:Springer) p. 330 \\

Pagel B.E.J., 1997, ``Nucleosynthesis and chemical evolution of galaxies'', 
Cambridge University press \\

Pagel B.E.J.,  Edmunds M.G., 1981, ARA\&A,19, 77 \\

Pagel B.E.J.,  Terlevich R.J., Melnick J.,  1986, PASP 98, 1005 \\

Pagel B.E.J., Simonson E.A., Terlevich R.J., Edmunds M.G., 1992, MNRAS 255, 325  \\

Pantelaki I., Clayton D.D., 1987, in ``Starbursts and galaxy evolution'', Eds. 
Thuan T.X., Montmerle T., Tran Thanh Van J., \'Editions Fronti\`eres, p. 145  \\

Papaderos P., Fricke K.J., Thuan T.X., Loose H.-H., 1994, A\&A 291, 13  \\

Papaderos P., Loose H.-H., Thuan T.X., Fricke K.J., 1996a, A\&AS 120, 207  \\

Papaderos P., Loose H.-H., Thuan T.X., Fricke K.J., 1996b, A\&A 314, 59  \\

Papaderos P., Izotov Y.I., Fricke K.J., Thuan T.X., Guseva N.G., 1998, A\&A 338, 43  \\

Pascarelle S.M., Windhorst R.A., Keel W.C., 1998, AJ 116, 2659 \\

Patterson R.J., Thuan T.X., 1996, ApJS 107, 103 \\

Peimbert M., 1967, ApJ 150, 825 \\

Peimbert M., Serrano A., 1980, Rev. Mex. Astron. Astrofis., 5, 9   \\

Peimbert M., Torres-Peimbert S., 1974, ApJ 193, 327 \\

Pe\~na M., Ruiz M.T., Maza J., 1991, A\&A 251, 417  \\

Persic M., Mariani S., Cappi M., Bassani L., Danese L., Dean A.J.,
Di Cocco G., Franceschini A., Hunt L.K., Matteucci F., Palazzi E.,
Palumbo G.G.C., Rephaeli Y., Salucci P., Spizzichino A., 1998,
A\&A 339, L33 \\

Pesch P., Sanduleak N., 1983, ApJS 51, 171  \\
 
Petrosian A.R., Boulesteix J., Comte G., Kunth D., LeCoarer E., A\&A 318, 390  \\

Pettini M., 1999, in ``Chemical evolution from zero to high redshift'', Eds. Walsh J., Rosa M., 
Lecture notes in Physics (Berlin:Springer) in press. (astroph/9902173)  \\

Pettini M., Lipman K., 1995, A\&A 297, 63 \\

Pettini M., Lipman K., Hunstead R.W., 1995, ApJ 451, 100  \\

Pettini M., Smith L.J., King D.L., Hunstead R.W., 1997 ApJ 486, 665  \\

Pettini M., Kellogg M., Steidel C.C., Dickinson M., Adelberger K.L., Giavalisco M., 1998,
ApJ 508, 539 \\
 
Phillipps S., Davies J.I., Disney M.J., 1988, MNRAS 233, 485  \\

Phillipps S., Disney M.J., Davies J., 1993, MNRAS 260, 453  \\

Pilyugin L.S., 1992, A\&A 260, 58   \\

Pilyugin L.S., 1993, A\&A 277, 42  \\

Pilyugin L.S., 1999, A\&A accepted, atro-ph/9904157  \\

Popescu C.C., Hopp U., Els\"asser H., 1997, A\&A 325, 881 \\

Prantzos N., 1998, in ``Abundance Profiles: Diagnostic Tools for Galaxy History'', 
Eds. Friedli D., 
Edmunds M., Robert C., Drissen L., ASP Conf. Ser. Vol. 147, p. 171 \\

Press W.H., Schechter P., 1974, ApJ 187, 425 \\

Prochaska J.X., Wolfe A.M., 1997, ApJ 487, 73  \\

Prochaska J.X., Wolfe A.M.,  1999, ApJS  121, 369  \\

Puget J.L., Guiderdoni B., 1999, in ``Space Infrared Astronomy, today and to-morrow'', eds. F. Casoli, F. David, J. Lequeux, EDP Sciences/Springer-Verlag, in press \\

Pustil'nik S., Ugryumov A.V., Lipovetsky V.A., Thuan T.X., Guseva N., 1995, ApJ 443, 499 \\

Pustilnik S.A., Engels D., Ugryumov A.V., et al., 1999a, A\&AS 137, 299 \\

Pustilnik S.A., Brinks E., Thuan T.X., Lipovetsky V.A., Izotov Y.I., 1999b, in prep \\

Rauch M., Heahnelt M.G., Steinmetz M., 1997, ApJ 481, 601 \\

Renzini A.,  Voli M., 1981, A\&A 94, 175 \\

Richer M.G., McCall M.L., 1995, ApJ 445, 642  \\

Richer M.G., McCall M.L., Arimoto N., 1997, A\&AS 122, 215  \\

Richer M.G., McCall M.L., Stasi\'nska  G., 1998, A\&A 340, 67  \\

R\"onnback J., Bergvall N., 1994, A\&AS 108, 193  \\

R\"onnback J., Bergvall N., 1995, A\&A 302, 353  \\

Roy J.-R, Kunth D., 1995, A\&A 294, 432  \\

Russell S.C., Bessell M.S.,1989, A\&AS 70, 865 \\

Sage L.J., Salzer J.J., Loose H.-H., Henkel C., 1992, A\&A 265, 19  \\

Sage L.J., Welch G.A., Mitchell G.F., 1998, ApJ 507, 726  \\

Salpeter E.E., 1955, ApJ 121, 161 \\

Salzer J.J., 1989, ApJ 347, 152 \\

Salzer J.J. 1999, in "Dwarf Galaxies and  Cosmology", eds. Thuan et al., Editions Frontieres in press\\

Salzer J.J., MacAlpine G.,M., Boroson T.A., 1989a, ApJSS 70, 447  \\

Salzer J.J., MacAlpine G.,M., Boroson T.A., 1989b, ApJSS 70, 479  \\

Salzer J.J., di Serego Alighieri S., Matteucci F., Giovanelli R., Haynes M.P., 1991, AJ 101,1258 \\

Salzer J.J., Moody J.W., Rosenberg J.L., Gregory S.A., Newberry M.V., 1995, AJ 109, 2376  \\

Sandage A., Brucato R., 1979, AJ 84, 472  \\

Sanders D.B., Soifer B.T., Elias J.H., Madore B.F., Matthews K., Neugebauer G., Scoville N.Z., 
1988, ApJ 325, 74 \\

Sargent W.L.W., Searle L., 1970, ApJ, 162, L155  \\

Scalo J., 1990, in ``Windows on Galaxies'', Eds. Fabbiano G., Gallagher J.S., Renzini A.,  
Dordrecht:Kluwer, p. 125 \\

Scalo J., 1998, in ``The stellar initial mass function'', Eds. Gilmore G., Howell D., 
ASP Conf. Ser. 142, p. 201  \\

Schade D., Lilly S.J., Crampton D., Hammer F., Le F\`evre O., Tresse L., 1995, ApJ 451, L1 \\

Schaerer D. 1996, ApJ, 467, L17  \\

Schaerer D., Vacca W.D. 1998, ApJ, 497, 618  \\

Schaerer D., Contini,Th., Kunth D., Meynet G. 1997, ApJ 481, L75  \\

Schaerer D., Contini,Th., Kunth D. 1999, A\&A, 341, 399  \\

Schechter P., 1976, ApJ 203, 297 \\

Schmidt M., 1968, ApJ 151, 393  \\

Schmutz W., Vacca W.D., 1999, New Astronomy, submitted  \\

Schneider S.E., 1989, ApJ, 343, 94 \\

Schneider S.E., Thuan T.X., Mangum J.G., Miller J., 1992 ApJS 81, 5 \\

Schneider S.E., Spitzak J.G., Rosenberg J.L., 1998, ApJ 507, L9 \\

Schulte-Ladbeck R.E.,  Crone M.M., Hopp U., 1998, ApJ 493, L23 \\

Schulte-Ladbeck R.E., Hopp U., Crone M.M., Greggio L., 1999, 525, 709 \\

Schweizer F., 1978, in ``Structure and properties of nearby galaxies (IAU symposium 77)'' 
eds. Berkhuijsen E.M.,  Wielebinski R., Reidel Dordrecht, p. 279 \\ 

Searle L., Sargent W.L.W., 1972, ApJ 173, 25  \\

Searle L., Sargent W.L.W., Bagnuolo W.G., 1973, ApJ 179, 427   \\

Searle L., Zinn R., 1978, ApJ 225, 357 \\

Serrano  A., Peimbert M., 1983, Rev. Mex. Astron. Astrofis.,8,117; \\

Shetrone M.D., Bolte M., Stetson P.B., 1998, AJ 115, 1888  \\

Shull M.J., Penton S.V., Stocke J.T., Giroux M.L., van Gorkom J.H., Lee Y.H., Carilli C., 
1998, ApJ 116, 2094 \\

Silk J., Wyse R.F.G., Shields G.A., 1987, ApJ 322, L59 \\

Skillman E.D., 1996, in ``The Minesota lectures on extragalactic neutral hydrogen'', ed. 
Skillman E.D.,  ASP Conf. Ser. 106, p. 208  \\

Skillman E.D., 1997, Rev. Mex. Astron. Astrofis. Ser. Conf. 6, 36 \\

Skillman E.D., 1998, in ``Stellar astrophysics for the Local Group'', eds. Aparicio A., Herrero A.,
Sanchez F., Cambridge university press \\

Skillman E.D., Bender R., 1995, Rev. Mex. Astron. Astrofis. Ser.  Conf., 3, 25 \\

Skillman E.D., Kennicutt R.C., 1993, ApJ 411, 655 \\

Skillman E.D., Kennicutt R.C., Hodge P.W., 1989, ApJ 347, 875  \\

Skillman E.D., Terlevich R.J., Kennicutt R.C., Garnett D.R., Terlevich E., 1994, ApJ 431, 172 \\

Smecker-Hane T.A., Stetson P.B., Hesser J.E., Lehnert M.D., 1994, AJ 108, 507  \\

Smits D.P., 1991, MNRAS 251, 316 \\

Smits D.P., 1996, MNRAS 278, 683  \\

Somerville R.S., Primack J.R., 1999, MNRAS accepted (astroph/9802268) \\

Spite M., Cayrel R., Francois P., Richtler T., Spite F., 1986, A\&A 168,197 \\

Staveley-Smith L., Davies R.D., Kinman T.D., 1992, MNRAS 258, 334 \\

Steidel C.C., Pettini M., Hamilton D., 1995, AJ, 110, 2519 \\

Steidel C.C., Giavalisco M., Pettini M., Dickinson M., Adelberger K.L., 1996, ApJ 462, L17 \\

Steidel C.C., Adelberger K.L., Giavalisco M., Dickinson M., Pettini M., 1999, ApJ 519, 1 \\

Suntzeff N.B., Mateo M., Terndrup D.M., Olszewski E.W., Geisler D., Weller W., 1993, ApJ 418, 208\\

Szomoru A., Guhathakurta P., van Gorkom J.H., Knapen J.H., Weinberg D.H., Fruchter A.S., 
1994, AJ  108, 491 \\

Takase B., Miyauchi-Isobe N., 1984, Tokyo Astr. Observ. Annals, 2nd Ser.,Vol 19, 595  \\

Tayler R.J., 1976, MNRAS 177, 39 \\

Taylor C.L., 1997, ApJ 480, 524  \\

Taylor C.L., Brinks E., Skillman E.D., 1993, AJ 105, 128 \\

Taylor C.L., Brinks E., Grashuis R.M., Skillman E.D., 1995, ApJS 99, 424 \\

Taylor C.L., Brinks E., Grashuis R.M., Skillman E.D., 1996a, ApJS 102, 189 (erratum)  \\

Taylor C.L., Thomas D.L., Brinks E., Skillman E.D., 1996b, ApJS 107, 143  \\

Telles E., Maddox S., 1999, MNRAS submitted (astroph/9903037) \\

Telles E., Terlevich R., 1995, MNRAS 275, 1 \\

Telles E., Terlevich R., 1997, MNRAS 286, 183 \\ 

Telles E., Melnick J., Terlevich R., 1997, MNRAS 288, 78 \\

Tenorio-Tagle G., 1996, AJ 111, 1641 \\

Tenorio-Tagle G., Silich S.A., Kunth D., Terlevich E., Terlevich R., 1999, MNRAS in press,
(astroph/9905324) \\

Terlevich R.,  Melnick J., 1981, MNRAS 195, 839  \\

Terlevich R.,  Melnick J., Masegosa J., Moles M., Copetti M.V.F., 1991, A\&AS 91, 285 \\

Terlevich E., Skillman E.D., Terlevich R., 1996 in ``The interplay between massive star 
formation, the ism and galaxy evolution'', Eds. Kunth D., Guiderdone B., Heydari-Malayeri M., 
Thuan T.X., \'Edition Fronti\`eres, p. 395,   \\

Thuan T.X., 1983, ApJ 268, 667  \\

Thuan T.X., 1985, ApJ 299, 881  \\

Thuan T.X., 1986a, in ``Star-forming dwarf galaxies and related objects'', 
Eds. Kunth D., Thuan T.X., Tran Thanh Van J., \'Editions Fronti\`eres, p. 105 \\

Thuan T.X., Martin G.E., 1981, ApJ 247, 823  \\

Thuan T.X., Seitzer P.O., 1979, ApJ 231, 680 \\

Thuan T.X., Gott J.R, Schneider S.E., 1987 ApJ 315, L93  \\

Thuan T.X., Izotov Y.I., Lipovetsky V.A., 1996, ApJ 463, 120  \\

Thuan T.X., Izotov Y.I., Lipovetsky V.A., 1997, ApJ 477, 661 \\

Thuan T.X., Sauvage M., Madden S., 1999, ApJ 516, 783 \\

Tikhonov N.A.,  Karachentsev I.D.,  1993, A\&AS 275, 39 \\

Tolstoy E., Gallagher J.S., Cole A.A., Hoessel J.G., Saha A., Dohm-Palmer R.C., 
Skillman E.D., Mateo M., Hurley-Keller D., 1998, AJ 116, 1244 \\

Torres-Peimbert S., Peimbert M., Fierro J., 1989, ApJ 345, 186 \\

Tosi M., 1988, A\&A 197, 47  \\

Tully B.R., Boesgaard A.M., Schempp W.V., 1980 in ``Photometry, Kinematics and Dynamics of Galaxies'',
Ed. Evans D.S., Knudsen, p. 325 \\

Tyson J.A., 1988, AJ 96, 1\\

Tyson N.D., Scalo J.M., 1988, ApJ 329, 618 \\

Ugryumov A.V., Engels D., Lipovetsky V., et al., 1999, A\&AS 135, 511  \\

Vacca W.D., Conti P.S.,  1992, ApJ 401, 543 \\

Vader J.P., 1986, ApJ 305, 669 \\

van den Bergh S., 1994, ApJ 428, 617 \\

van der Hoek L.B., Groenewegen M.A.T., 1997, A\&AS 123, 305 \\

van der Hulst J.M., Skillman E.D., Smith T.R., Bothun G.D., McGaugh S., de Blok W.J.G., 
1993, AJ 106, 548   \\

van Zee L. Haynes M.P., Salzer J., Broeils A.H., 1996, 112, 129 \\

van Zee L., Haynes M.P., Salzer J., Broeils A.H.,  1997a, AJ 113, 1618   \\

van Zee L., Haynes M.P., Salzer J., 1997b, AJ 114, 2479  \\

van Zee L., Haynes M.P., Salzer J., 1997c, AJ 114, 2497  \\

van Zee L.,  Westpfahl D., Haynes M.P., Salzer J., 1998a, AJ 115, 1000 \\

van Zee L.,  Salzer J., Haynes M.P., 1998b, ApJ 497, L1  \\

van Zee L.,  Skillman E.D., Salzer J.,  1998c, AJ 116, 1186 \\

Viallefond F., Lequeux J., Comte G., 1987, in ``Starbursts and galaxy evolution'', Eds. 
Thuan T.X., Montmerle T., Tran Thanh Van J., \'Editions Fronti\`eres, p. 139  \\

V\'ilchez J.M., 1995, AJ 10, 1090  \\

V\'ilchez J.M., 1999, in ``Chemical evolution from zero to high redshift'',
eds. Walsh J., Rosa M., Lecture Notes in Physics, Berlin:Springer, in press \\

V\'ilchez J.M., Iglesias-P\'aramo J., 1998, ApJ 508, 248 \\

Vitores A.G., Zamorano J., Rego M., Alonso O., Gallego J.,  1996, A\&AS 118, 7 \\

Vladilo G., 1998, ApJ 493, 583  \\

Walsh J.R., Roy J.-R., 1993, MNRAS 262, 27 \\

Walsh J.R., Dudziak G., Minniti D., Ziljstra A.A., 1997, ApJ 487, 651 \\

Wasilewski A.J., 1983, ApJ 272, 68  \\

White S.D.M., Rees M.J., 1978, MNRAS 183, 341 \\

Whitmore B.C., Schweizer F.,  1995, AJ 109, 960 \\

Williams R.E., Blacker B., Dickinson M., et al., 1996, AJ 112, 1335 \\

Worthey G., 1994, ApJS 95, 107 \\

Worthey G., Faber S.M., Gonzalez J.J., 1992, ApJ 398, 69  \\

Young C.K., Currie M.J., 1998, A\&AS 127, 367 \\

Young J.S., Kenney J.D., Tacconi L., Claussen M.J., Huang Y.-L., Tacconi-Garman L., 
Xie S., Schloerb F.P.,   ,1986 ApJ 311, L17 \\

Young L., Lo K., 1996, ApJ 462, 203  \\

Young L., Lo K., 1997a, ApJ 476, 127  \\

Young L., Lo K., 1997b, ApJ 490, 710  \\

Zinnecker H., Cannon R.D., 1986, in "Star forming dwarf galaxies
and related objects", Eds. Kunth D., Thuan T.X., Tran Thanh Van J., \'Editions Fronti\`eres, p. 155  \\

Zwicky F., 1956, Ergebnisse der Exakten Naturwissenschaften 29, 344 \\

Zwicky F., 1965, ApJ 142, 1293   \\

Zwicky F., 1966, ApJ 143, 192  \\

Zwicky F., 1971, Catalogue of selected compact galaxies and of post-eruptive galaxies \\

\end{document}